\documentclass[namedreferences]{solarphysics}
\usepackage[optionalrh]{spr-sola-addons} 
\usepackage{graphicx}        
\usepackage{color}           
\usepackage{url}             

\newcommand{\etal}{{\it et al.}}

\usepackage[pdfborder={0 0 0},urlcolor=blue,breaklinks]{hyperref}
\ifx \doiurl  \undefined \def \doiurl#1{\href{http://dx.doi.org/#1}{\url{#1}}}\fi
\ifx \adsurl  \undefined \def \adsurl#1{\href{http://adsabs.harvard.edu/abs/#1}{\url{#1}}}\fi

\begin{document}

\begin{article}

\begin{opening}

\title{Propagating Disturbances in Coronal Loops: A Detailed Analysis of
Propagation Speeds\\}

\author{G.~\surname{Kiddie}$^{1}$\sep
        I.~\surname{De Moortel}$^{1}$\sep
        G.~\surname{Del Zanna}$^{2}$\sep
        S.W.~\surname{McIntosh}$^{3}$\sep
        I.~\surname{Whittaker}$^{1}$      
       }
\runningauthor{G.Kiddie \etal}
\runningtitle{Propagating Disturbances in Coronal Loops}

   \institute{$^{1}$ School of Mathematics and Statistics, University of St
Andrews, North Haugh, St Andrews, Fife, KY16 9SS, Scotland, UK.
                      \href{mailto:gregk@mcs.st-and.ac.uk}{gregk@mcs.st-and.ac.uk}
                      \href{mailto:ineke@mcs.st-and.ac.uk}{ineke@mcs.st-and.ac.uk}email:
 \href{mailto:icw@mcs.st-and.ac.uk}{icw@mcs.st-and.ac.uk}
            $^{2}$ Department of Applied Mathematics and Theoretical Physics, Wilberforce Road, Cambridge, CB3OWA
                  email:\href{mailto:G.Del-Zanna@damtp.cam.ac.uk}{G.Del-Zanna@damtp.cam.ac.uk}                                 
            $^{3}$ High Altitude Observatory, National Center for Atmospheric Research, P.O. Box 3000, Boulder, CO 80307
                   email: \href{mailto:mscott@ucar.edu}{mscott@ucar.edu}\\   
                               }

\begin{abstract}
 Quasi-periodic disturbances have been observed in the outer solar atmosphere for many years.  Although first interpreted as upflows (Schrijver {\it et al.} Solar Physics.187,261), they have been widely regarded as slow magneto-acoustic waves, due to their observed velocities and periods.  However, recent observations have questioned this interpretation, as periodic disturbances in Doppler velocity, line width, and profile asymmetry were found to be in phase with the intensity oscillations (De Pontieu and McIntosh, Astrophysics. J. 722,1013 (2010), Tian, McIntosh, and De Pontieu, Astrophysics, J.Lett. 727,L37 (2011)), suggesting that the disturbances could be quasi-periodic upflows.  Here we conduct a detailed analysis of the velocities of these disturbances across several wavelengths using the {\it Atmospheric Imaging Assembly} (AIA) onboard the {\it Solar Dynamics Observatory} (SDO).   We analysed 41 examples, including both sunspot and non-sunspot regions of the Sun. We found that the velocities of propagating disturbances (PDs) located at sunspots are more likely to be temperature dependent, whereas the velocities of PDs at non-sunspot locations do not show a clear temperature dependence.  This suggests an interpretation in terms of slow magneto-acoustic waves in sunspots but the nature of PDs in non-sunspot(plage) regions remains unclear.   We also considered on what scale the underlying driver is affecting the properties of the PDs.  Finally, we found that removing the contribution due to the cooler ions in the 193~\AA\ wavelength suggests that a substantial part of the 193~\AA\ emission of sunspot PDs can be attributed to the cool component of 193~\AA.

\end{abstract}

\end{opening}

\section{Introduction}
     \label{S-Introduction} 

Since the launch of SOHO and TRACE, low-amplitude quasi-periodic disturbances
have been found at loop foot points  ({\it e.g.} \inlinecite{demoortel09} for a review).  The first observations of propagating
disturbances (PDs) were found along coronal plumes by \inlinecite{ofman97} using SOHO/UVCS,
and again by \inlinecite{deforest98} using SOHO/EIT. These were observed as intensity
perturbations travelling at approximately the local sound speed.  This led to
their classification as slow magneto-acoustic waves (see reviews by \inlinecite{demoortel09} and \inlinecite{banerjee11}). Propagating disturbances of
a similar nature (in active region loops) were
observed by \inlinecite{berghams99} using SOHO/EIT. 
\inlinecite{schrijver99}, \inlinecite{nightingale99}, \inlinecite{demoortel00} found similar
disturbances using TRACE, while \inlinecite{berghams01} found them using {\it Yokhoh}/SXT. 
These perturbations usually have small amplitudes of the order of a few percent
of the background.  They were found to have velocities of approximately 100
km$s^{-1}$ and periods of two\,--\,ten minutes \cite{mcewan06}.  There has also been
substantial work done in theoretical modelling of these disturbances
\cite{nakariakov00,tsiklauri01,demoortel04,demoortel04(ii)}.  These authors have looked at a variety of aspects
under the assumption that these disturbances are slow magneto-acoustic waves and found that the observed amplitude decay could be explained in terms of thermal conduction.  The quasi-periodic nature of these waves has been attributed  to the leakage of {\it p-}modes into the solar atmosphere \cite{depontieu05,demoortel07,malins07}. 
\inlinecite{marsh09} inferred a coronal temperature using EIS and found a temperature
that agrees  with the seismologically calculated temperature found by
\inlinecite{marsh09b}.  They suggested that this agrees with the interpretation of
the disturbances as slow magneto-acoustic waves.

Although this interpretation was widely accepted for several years, in the last
few years it has been questioned again, as spectroscopic observations from
{\it Hinode}/EIS have shown the situation is not so straightforward.  These
observations still show quasi-periodic intensity perturbations which are correlated ({\it i.e.} in phase) with perturbations in Doppler velocity, line width, and line asymmetry.  This has led to an alternative interpretation as  high-velocity upflows as this coherent behaviour is hard to explain with a slow wave scenario \cite{depontieu10,tian11,nish11}. 
\inlinecite{sakao07} found faint up flows in spectra of transition-region and coronal-loop foot points.  \inlinecite{depontieu09} discovered that these up flows are
ubiquitous in foot points of coronal loops.  A link between small blue-ward
asymmetries in spectra of loop foot points and the propagating disturbances was found \cite{depontieu10, tian11, tian11b}. These were found by fitting the lines with a double Gaussian model and using a red-blue asymmetry analysis \cite{depontieu10,tian11b,bryans10}.  Other studies that use the interpretation as flows include  \inlinecite{doschek07}, \inlinecite{delzanna08}, \inlinecite{he10}, \inlinecite{peter10}, \inlinecite{harra08}, \inlinecite{warren11}, \inlinecite{marsch08}, \inlinecite{hara08b}, \inlinecite{tian11c}, \inlinecite{murray10}, \inlinecite{brooks11}, and \inlinecite{young12}.  This has not closed the debate though. 
\inlinecite{verwichte10} showed that these periodic line asymmetries could be explained
by slow magneto-acoustic waves.  There are many other studies that still show a
preference for the slow wave interpretation \cite{marsh09b,wang09,kitagawa10,banerjee09,mariska09,prasad11}.  It has been suggested that these  PDs can have a close connection with type II spicules \cite{depontieu09,depontieu11,voort09} and they have also been linked with the mass cycle of the solar wind  \cite{mcintosh10b,tian11c}.  Due to their ubiquitous nature, they could have a significant effect on the coronal energy budget.  Recent work by \inlinecite{mcintosh12} shows a slow, downflow of coronal material, which could be the return component of the up flow. Other articles that consider downflows include \inlinecite{kamio11}  and \inlinecite{urra11}.
  
This so called ``Flows versus Waves'' debate has been argued for several years now, with a definitive answer yet to be decided (if there is one).  In this article we 
study propagating disturbances found at loop foot points, using the {\it Atmospheric Imaging Assembly} (AIA) onboard the {\it Solar Dynamics
Observatory} (SDO).   We are going to look
at the velocities of these disturbances over a range of different wavelengths
and temperatures.  We are going to consider the velocities of these disturbances
in different bands, which are dominated by lines formed in a range
of temperatures (\inlinecite{odwyer10}, \inlinecite{delzanna11}).
In particular, we consider the 131, 171, and 193~\AA\ bands.
As shown by Del Zanna \etal (2011), for the active-region loops considered here,
these three bands are dominated by Fe \textsc{viii}, Fe \textsc{ix}, and a range of
ions (Fe \textsc{vii}\,--\,Fe \textsc{xii}) respectively.
In ionisation equilibrium, these three bands show emission
from plasma formed in a broad range of temperatures, centred around
0.4, 0.7, and 1.6 MK \cite{dere09}.
The other AIA bands are more multi-thermal or
lack atomic data \cite{delzanna11}.

The outline of this article is as follows: Section \ref{S-Observations} describes the two primary data sets studied in this article.  Section \ref{S-Propspeed} describes how the velocities of the observed  PDs change with temperature by looking at 131, 171, and 193~\AA\ observations of both data sets.  A description of how the properties of the PDs change across an active region is described in Section \ref{S-Active}.  In Section \ref{S-Cool} we describe a method to remove the contribution due to cool ions in the 193~\AA\ passband and  the effect this has on the properties of the observed PDs.  The discussion and conclusions are presented in Section \ref{S-Conclusions}. 

\section{Observations and Analysis}
     \label{S-Observations}

The two primary data sets investigated are AIA observations of active regions
AR11236 on 22 June 2011 at 15:13UT and AR11301 on 22 September 2011 at 15:01UT.
 Both data sets have a duration of 40 minutes.  We will focus on the 131,171,
and 193~\AA\ passbands. Each passband has a cadence of 12 seconds and exposure times of
2.9, 2.0, and 2.0s respectively.  Each data set has been cleaned and co-aligned using the SolarSoft IDL command {\sf AIA\_PREP} and these are then de-rotated.

For each example a 150 $\times$ 150 pixel subsection is chosen to contain a loop footpoint.  A loop is then outlined by two arcs, and divided
into cross sections.  This is a very similar technique to that of \inlinecite{demoortel00}.  Consecutive images are summed to increase the signal to noise
ratio and then we calculate a running-difference by subtracting from each image the one taken 96s previously.  This is done to highlight the oscillations since we expect their periods to be approximately three\,--\,5 minutes.

\section{Propagation Speeds In Multiple Wavelengths}
     \label{S-Propspeed}

In this section we look at the two data sets outlined in Section
\ref{S-Observations}.  We identify a loop and then create running-difference images to identify PDs along this loop.  If
the PDs  are slow magneto-acoustic waves then their velocity is expected to be the local sound speed.   The sound speed scales
with temperature in the following way; $c^2_s = \gamma\frac{p}{\rho}$ and
$p=\frac{\rho k_B T}{\mu_m}$, and hence  $c^2_s = \alpha T$.  The constant is defined as
$\alpha= \gamma\frac{k_B}{\mu_m}$, where $k_B$ is the Boltzmann constant and $\mu_m$
is
the reduced mass, {\it i.e.} the average mass of all particles in the plasma.  Therefore, the slow-wave propagation speed (which is closely related to the sound speed) is expected to be proportional to the square root of the temperature.  

\subsection{22 September 2011 [Non Sunspot]}
    \label{S-22/09/11}

\begin{figure}[!h]    
   \centerline{\hspace*{0.015\textwidth}
               \includegraphics[width=0.400\textwidth,clip=]{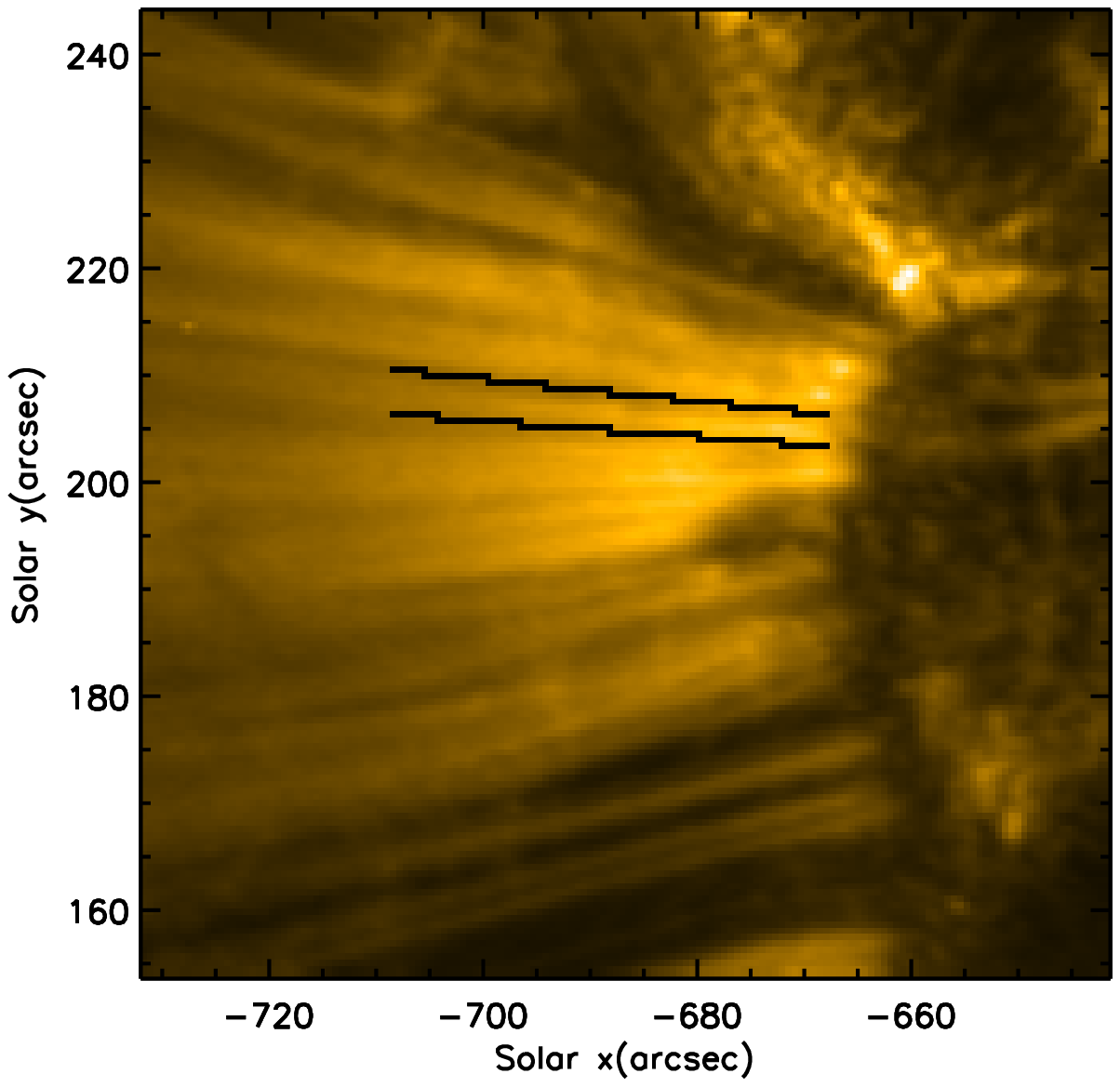}
               \hspace*{-0.03\textwidth}
               \includegraphics[width=0.400\textwidth,clip=]{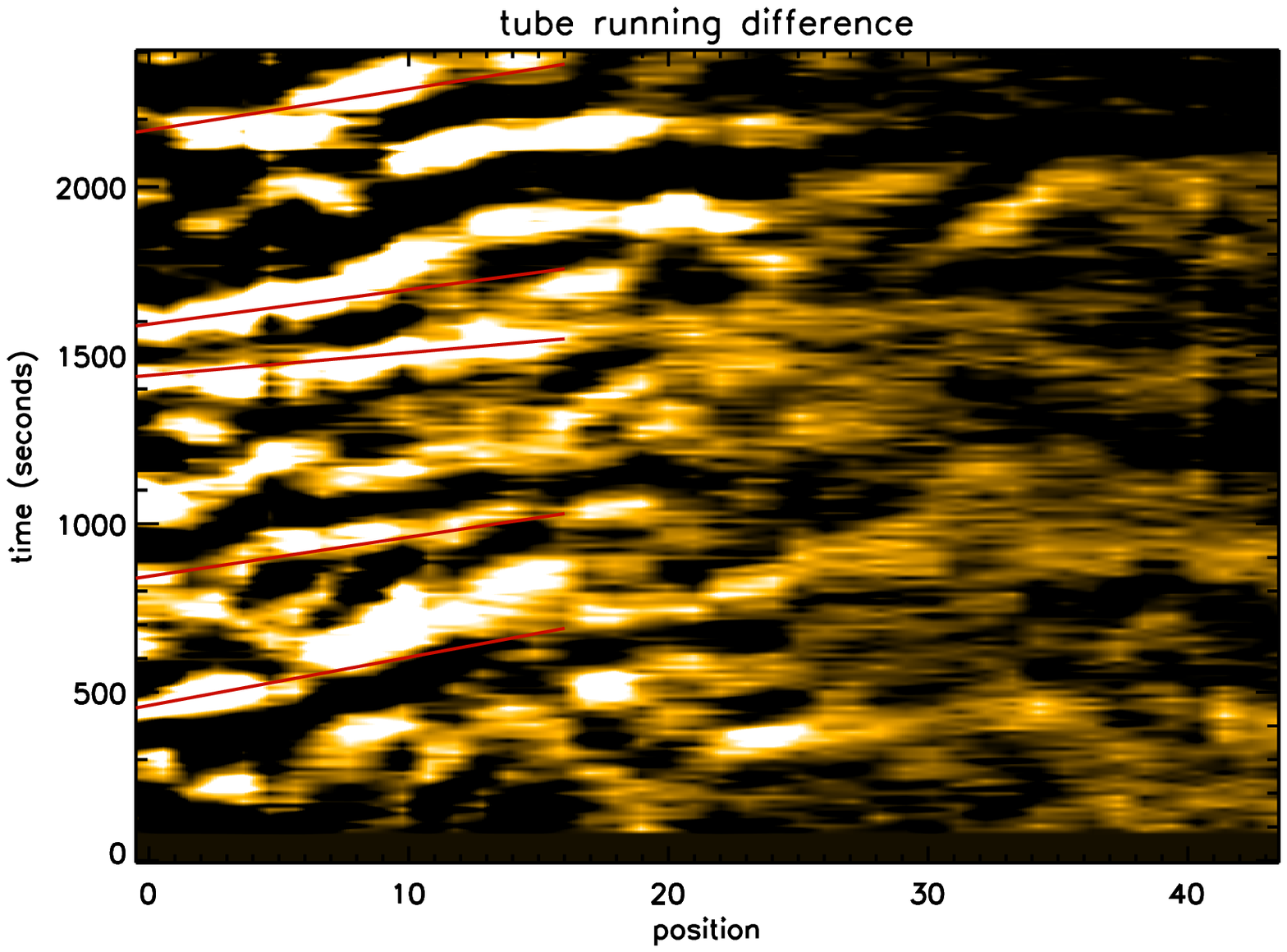}
              }
     \vspace{-0.32\textwidth}   
     \centerline{\Large \bf     
      \hspace{0.0 \textwidth}  \color{black}{\small(a)}
      \hspace{0.400\textwidth}  \color{black}{\small(b)}
         \hfill}
     \vspace{0.31\textwidth}    
   \centerline{\hspace*{0.015\textwidth}
               \includegraphics[width=0.400\textwidth,clip=]{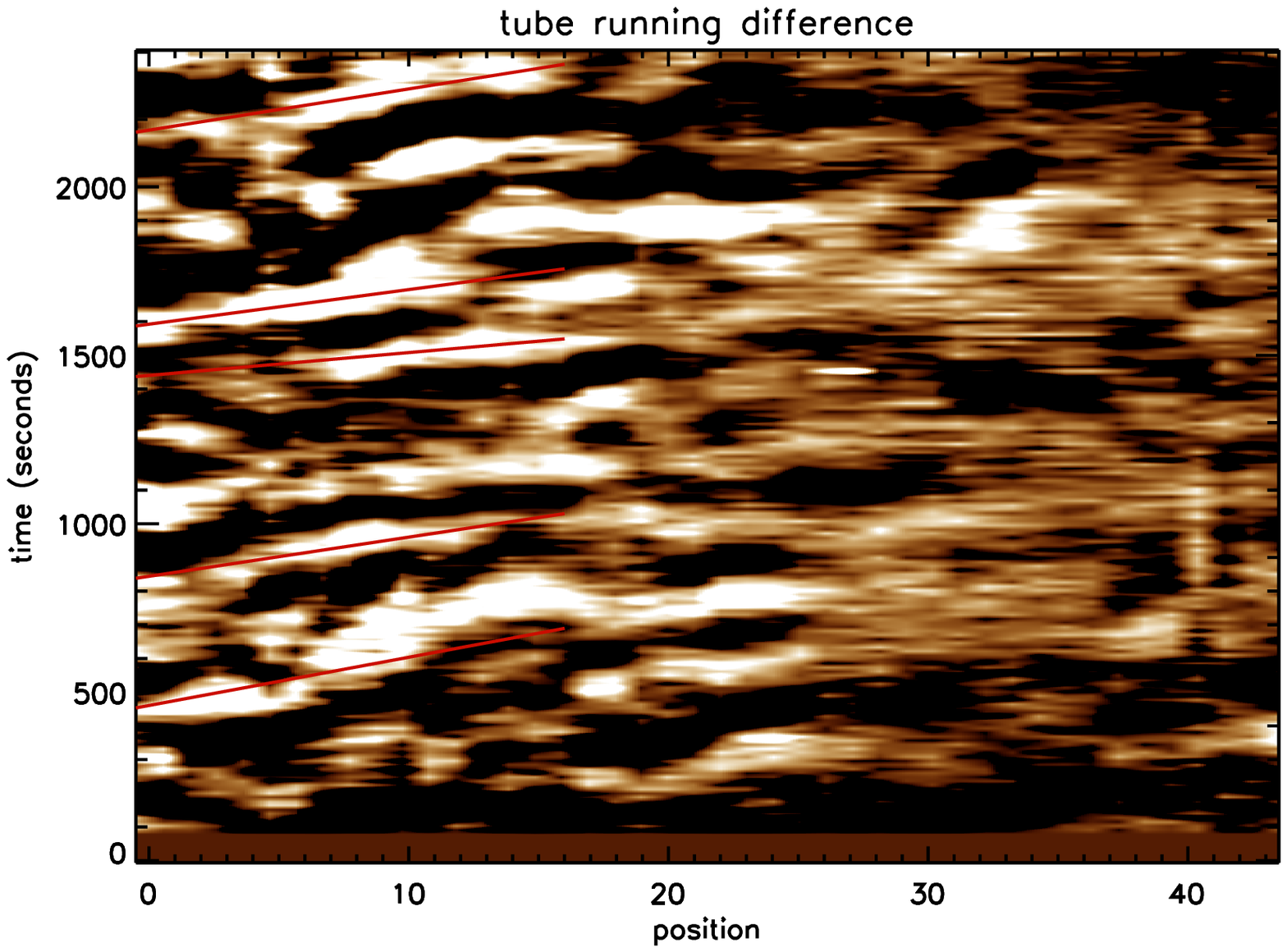}
               \hspace*{-0.03\textwidth}
               \includegraphics[width=0.400\textwidth,clip=]{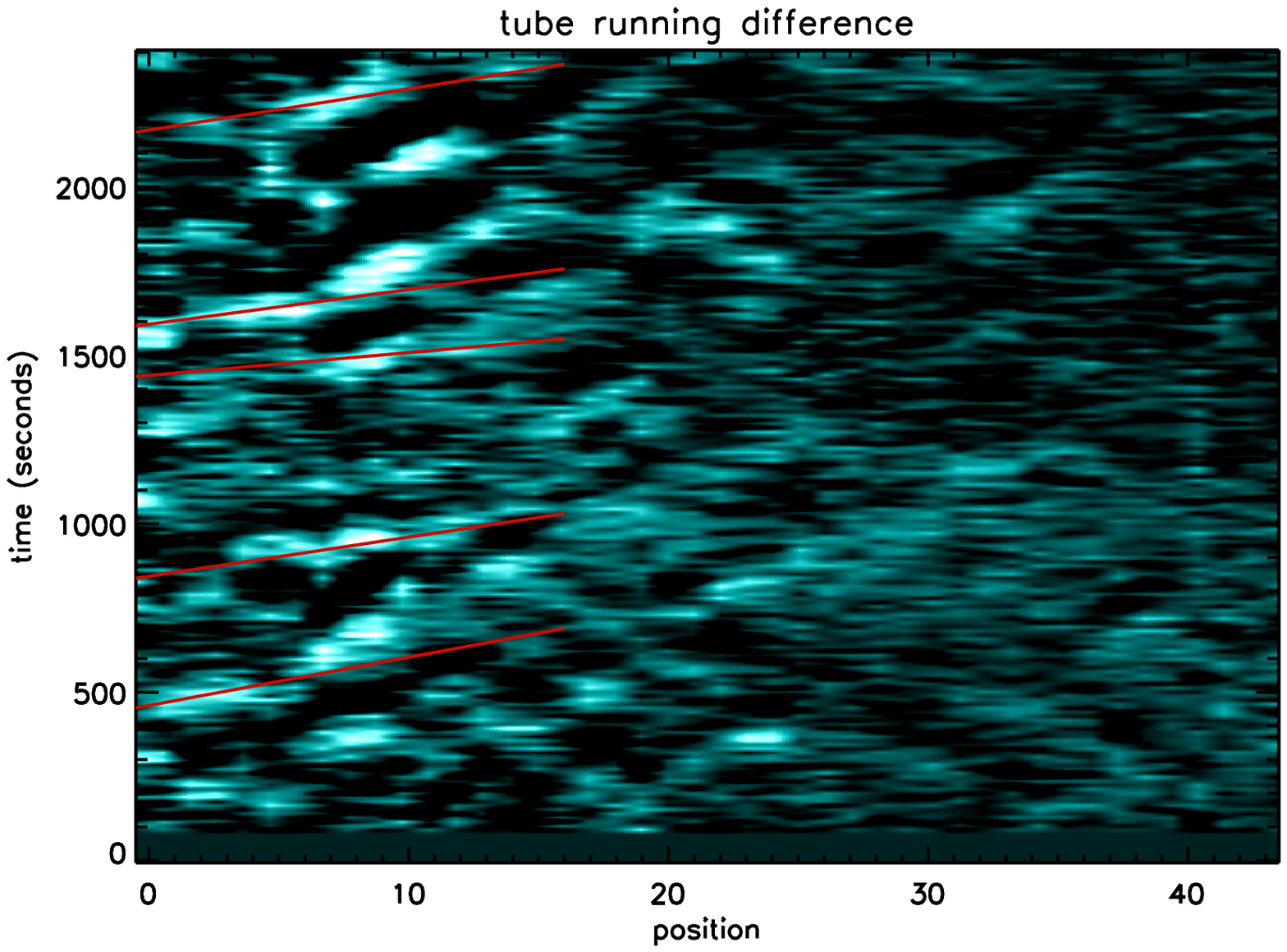}
              }
     \vspace{-0.32\textwidth}   
     \centerline{\Large \bf     
      \hspace{0.0 \textwidth} \color{black}{\small(c)}
      \hspace{0.400\textwidth}  \color{black}{\small(d)}
         \hfill}
     \vspace{0.31\textwidth}    
              
\caption{(a) 171~\AA\ intensity image with the analysed loop outlined by the black lines. (b)\,--\,(d) Running-difference images for the 171 ~\AA, 193 ~\AA, and 131 ~\AA\ passbands, respectively.  The red lines correspond to the gradient of the intensity bands from the 193 ~\AA\ running-difference.}
   \label{F-run_diff_22/09}
   \end{figure}

The first AIA observation is of active region AR11301 on 22 September 2011
as described in Section \ref{S-Observations}. The two arcs that outline the strands
are shown on the top-left plot of Figure \ref{F-run_diff_22/09}.  The loop has
solar coordinates of (-670,204) arseconds at 00.35UT.  To visualise the oscillations, a 
running-difference is made by subtracting the image 96 seconds earlier from each image.  The
running-difference images are shown in Figure
\ref{F-run_diff_22/09}(b)\,--\,(d) for the three different wavelengths.  The overplotted lines in the running-difference images (b)\,--\,(d)
represent the gradient of the intensity bands estimated in the 193~\AA\ passband. 
They are overplotted as a visual aid to compare velocities of the PDs across
the three wavelengths.

The velocities of the PDs are calculated from the inverse of the gradient of
the intensity bands.  For this example we identified five intensity bands in each
wavelength and calculated a range of velocities as in \inlinecite{demoortel00};  the range of possible velocities is estimated from the range of slopes within a given intensity band.    The velocities are displayed in Table \ref{T-One}. Band 1 is the intensity band closest to the bottom of the running-difference images and band 5 is closest to the top.

\begin{table}[!h]
\caption{Average velocities associated with running-difference for all
wavelengths (in
km$s^{-1}$) for 22 September 2011.  The velocities in the brackets show the lower and higher estimates.}
\label{T-One}
\begin{tabular}{c c c c}
\hline
Band & 131~\AA\ & 171~\AA\ & 193~\AA\ \\
\hline
1 & 48(27\,--\,190) & 31(24\,--\,43) & 49(34\,--\,90) \\
2 & 71(41\,--\,243)  & 63(39\,--\,128) & 47(30\,--\,113)\\
3 & 86(53\,--\,228)  & 72(43\,--\,216) & 82(44\,--\,533) \\
4 & 43(29\,--\,87) & 36(26\,--\,56) & 63(37\,--\,261) \\
5 & 42(29\,--\,75) & 43(27\,--\,103) & 50(35\,--\,87) \\ 
\hline
\end{tabular}

\end{table}

\begin{figure}[!h]    
   \centerline{\hspace*{0.015\textwidth}
               \includegraphics[width=0.400\textwidth,clip=]{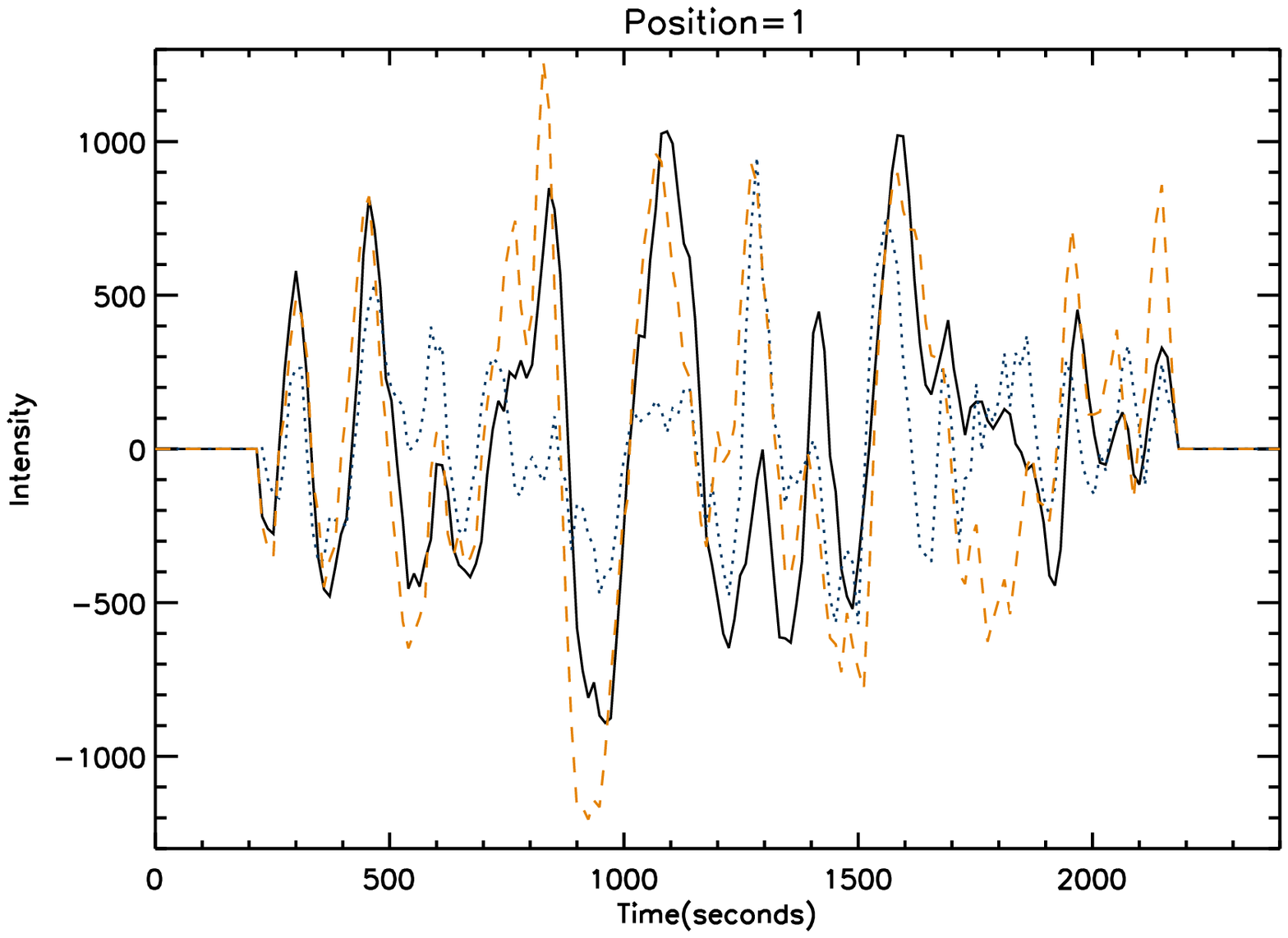}
               \hspace*{-0.03\textwidth}
               \includegraphics[width=0.400\textwidth,clip=]{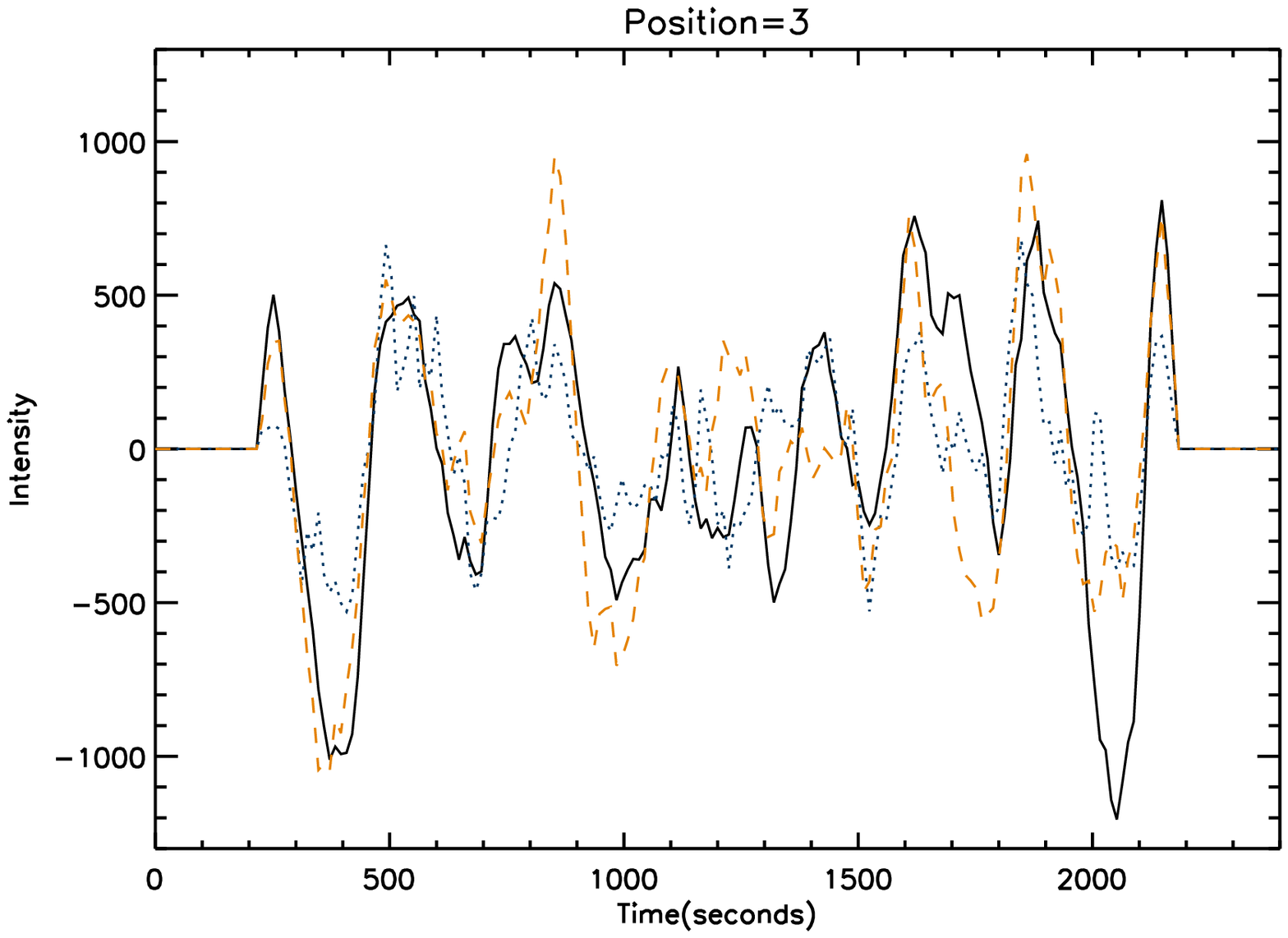}
              }
     \vspace{-0.32\textwidth}   
     \centerline{\Large \bf     
      \hspace{0.0 \textwidth}  \color{black}{\small(a)}
      \hspace{0.400\textwidth}  \color{black}{\small(b)}
         \hfill}
     \vspace{0.31\textwidth}    
   \centerline{\hspace*{0.015\textwidth}
               \includegraphics[width=0.400\textwidth,clip=]{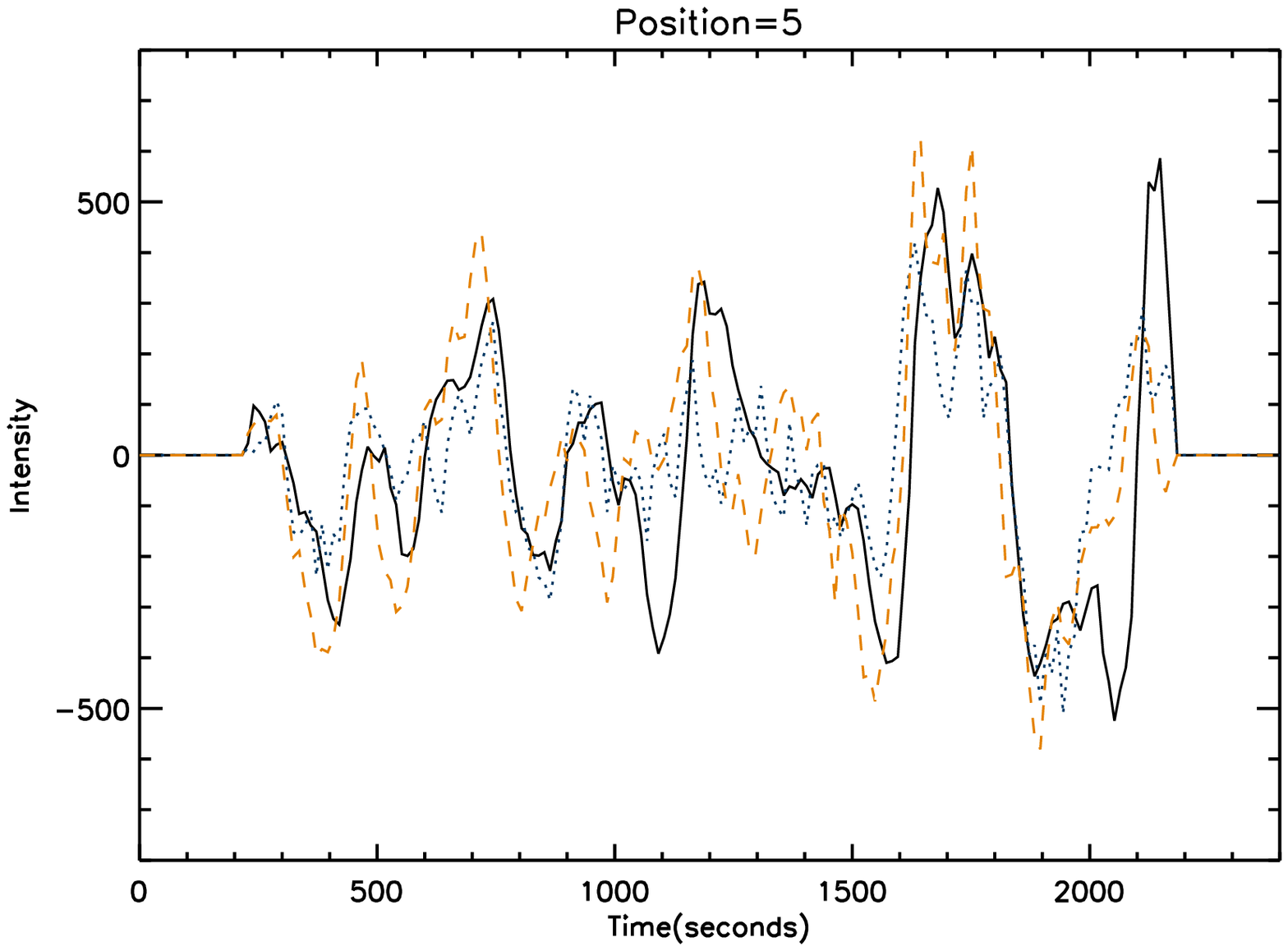}
               \hspace*{-0.03\textwidth}
               \includegraphics[width=0.400\textwidth,clip=]{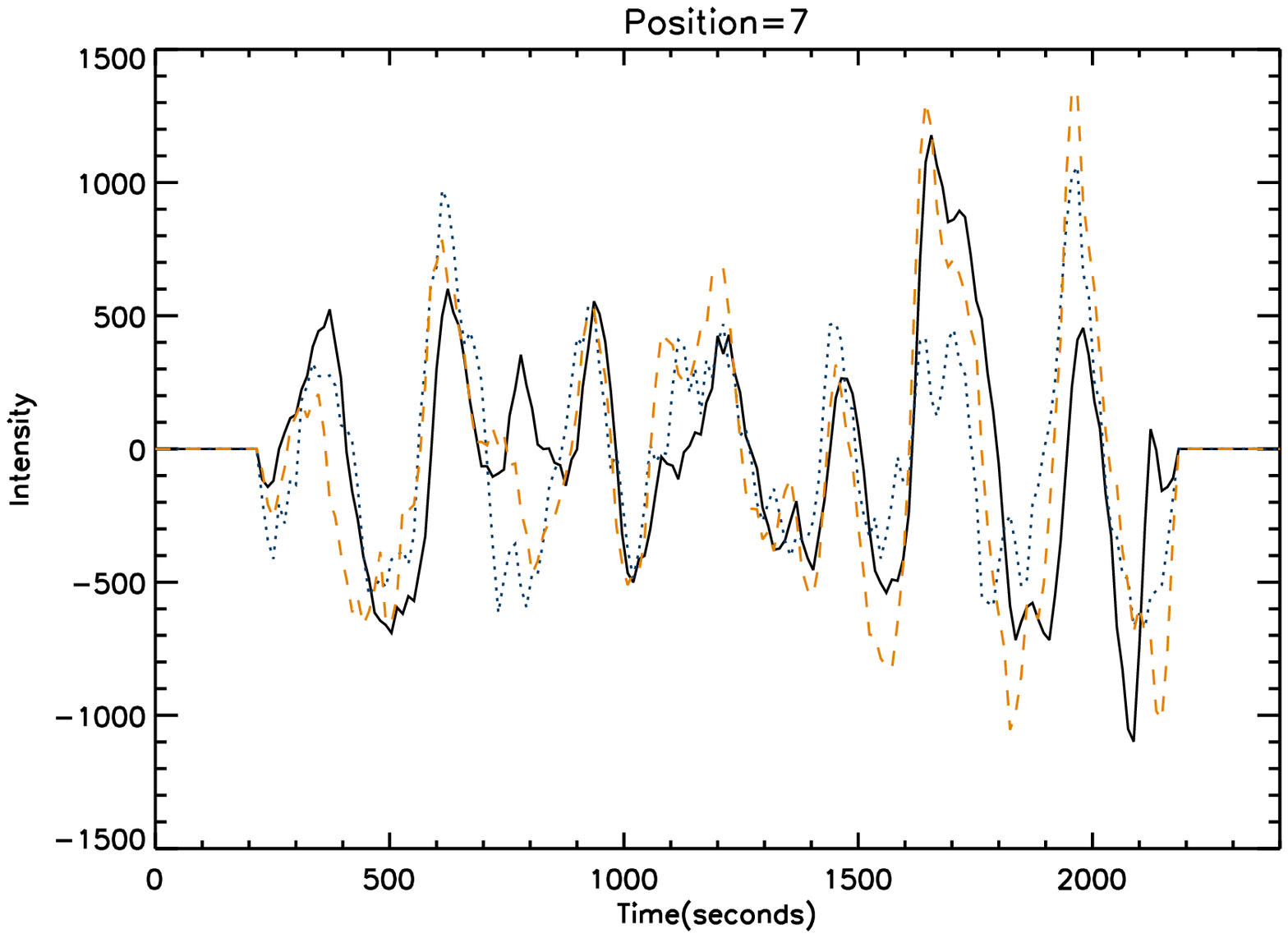}
              }
     \vspace{-0.32\textwidth}   
     \centerline{\Large \bf     
      \hspace{0.0 \textwidth} \color{black}{\small(c)}
      \hspace{0.400\textwidth}  \color{black}{\small(d)}
         \hfill}
     \vspace{0.31\textwidth}    
              
\caption{Cuts through the intensities for  171~\AA\ (black solid), 193~\AA\ (orange dashed) and 131~\AA\
(blue dotted) for different positions along the tube, for 22 September 2011. }
   \label{F-cuts_22/09}
   \end{figure}

From Table \ref{T-One} it is clear that the velocities of the PDs do not vary drastically across the wavelengths, with the largest
range across wavelengths only 23km$s^{-1}$.  To gain a greater insight in how the PDs in each wavelength are related, we take cuts at fixed positions along the loop.  Note that we have subtracted  an eight minute running average from the intensities to highlight the PDs as in \inlinecite{tian11b},  which implies that periods greater than eight minutes will be suppressed.  The PDs are no longer distinguishable by approximately position 12 along the
loop, so we have taken cuts at position 1, 3, 5, and 7, which are displayed in Figure
\ref{F-cuts_22/09}.

The black solid lines in Figure \ref{F-cuts_22/09} correspond to cuts through the
171~\AA\ intensity, the orange dashed lines are cuts through 193~\AA\ and the blue dotted lines 131~\AA.  It should be
noted that the 193~\AA\ (orange dashed) and the 131~\AA\ (blue dotted) lines
have been multiplied by arbitrary constants to make them comparable in size to the
171~\AA\ (black solid) line. In this figure, (a) shows cuts at position 1, which is near the bottom of the loop, (b) shows position 3, (c) position 5, and (d) position 7. The three lines match well at position 1; the 193~\AA\ and 171~\AA\ lines are almost
exactly in phase for the entire time with the 131~\AA\ line also in phase for the
majority of the time.  This trend continues into positions 3 and 5. At position
7 the three lines are still approximately in phase although less so.

\begin{figure}[!h]
 \centering
\scalebox{0.4}{\includegraphics{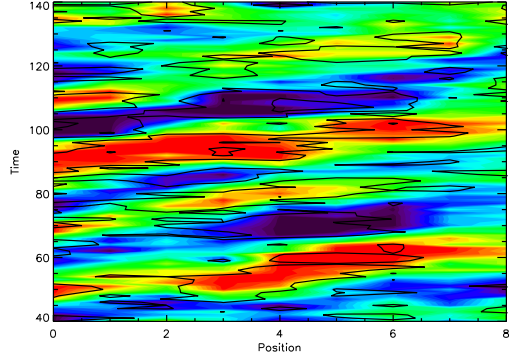}}
\caption{Contour plot of the 193~\AA\ running-difference with 131~\AA\ overplotted (thick black lines) for 22 September 2011.}
\label{F-contour2_22/09}
\end{figure}

The final analysis we will use on this example is to consider a contour plot of
the 193~\AA\ running-difference with the 131~\AA\ contours overplotted.   Figure
\ref{F-contour2_22/09} shows this contour plot for a subsection of the running-difference images shown in Figure \ref{F-run_diff_22/09}.  It is clear from
Figure \ref{F-contour2_22/09}  that the  131~\AA\ contours match the 193~\AA\
contours quite well.  The rough shapes of the bands are well outlined by the 131~\AA\
contours and the gradient of the bands are approximately the same.

The analysis performed on this data set suggests that the velocities
of the PDs do not change considerably with the temperature and do not show the temperature dependence expected for a slow magneto-acoustic wave.

\subsection{22 June 2011 [Sunspot]}
\label{22_06_2011}

\begin{figure}[!h]    
   \centerline{\hspace*{0.015\textwidth}
               \includegraphics[width=0.400\textwidth,clip=]{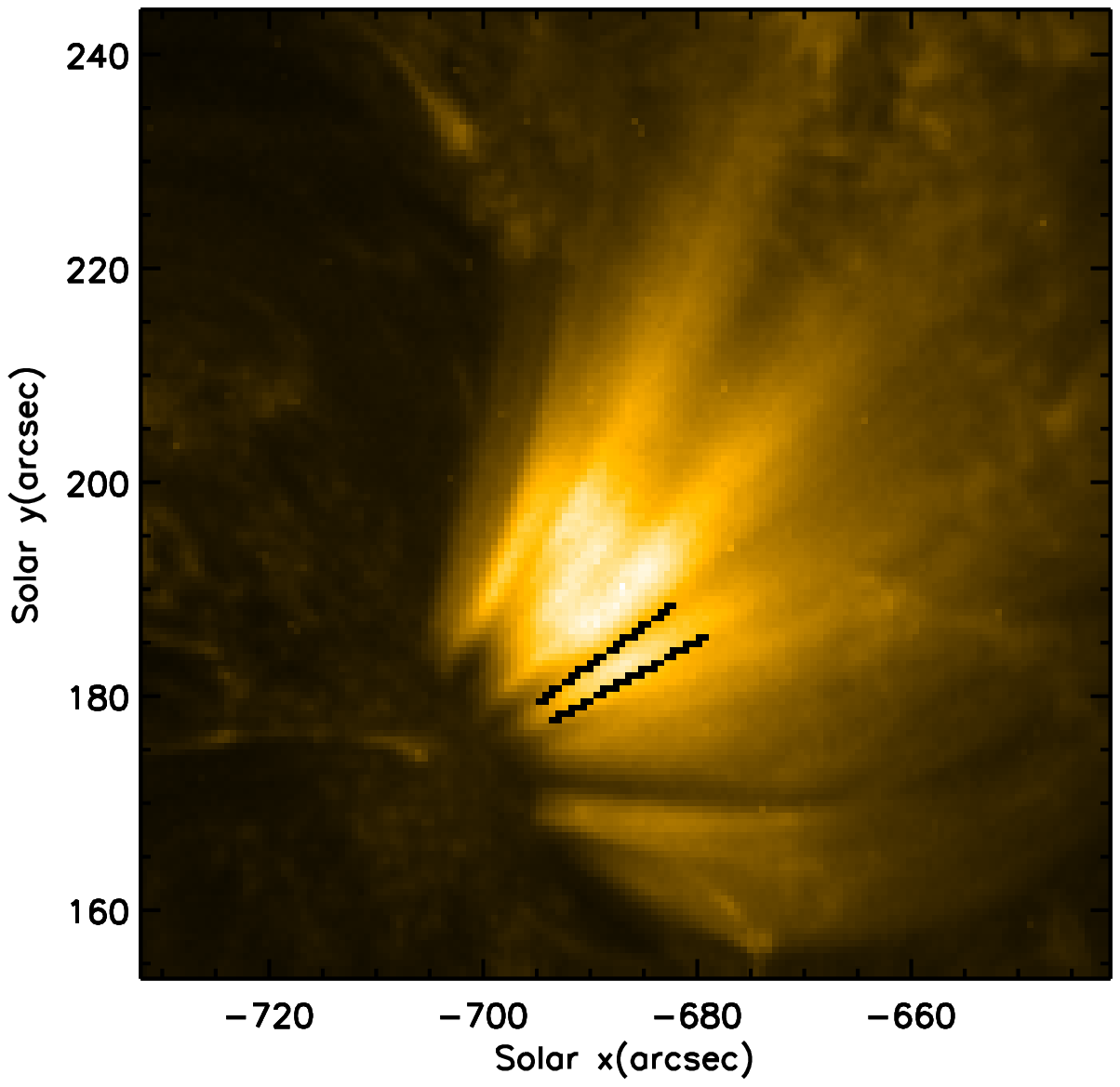}
               \hspace*{-0.03\textwidth}
               \includegraphics[width=0.400\textwidth,clip=]{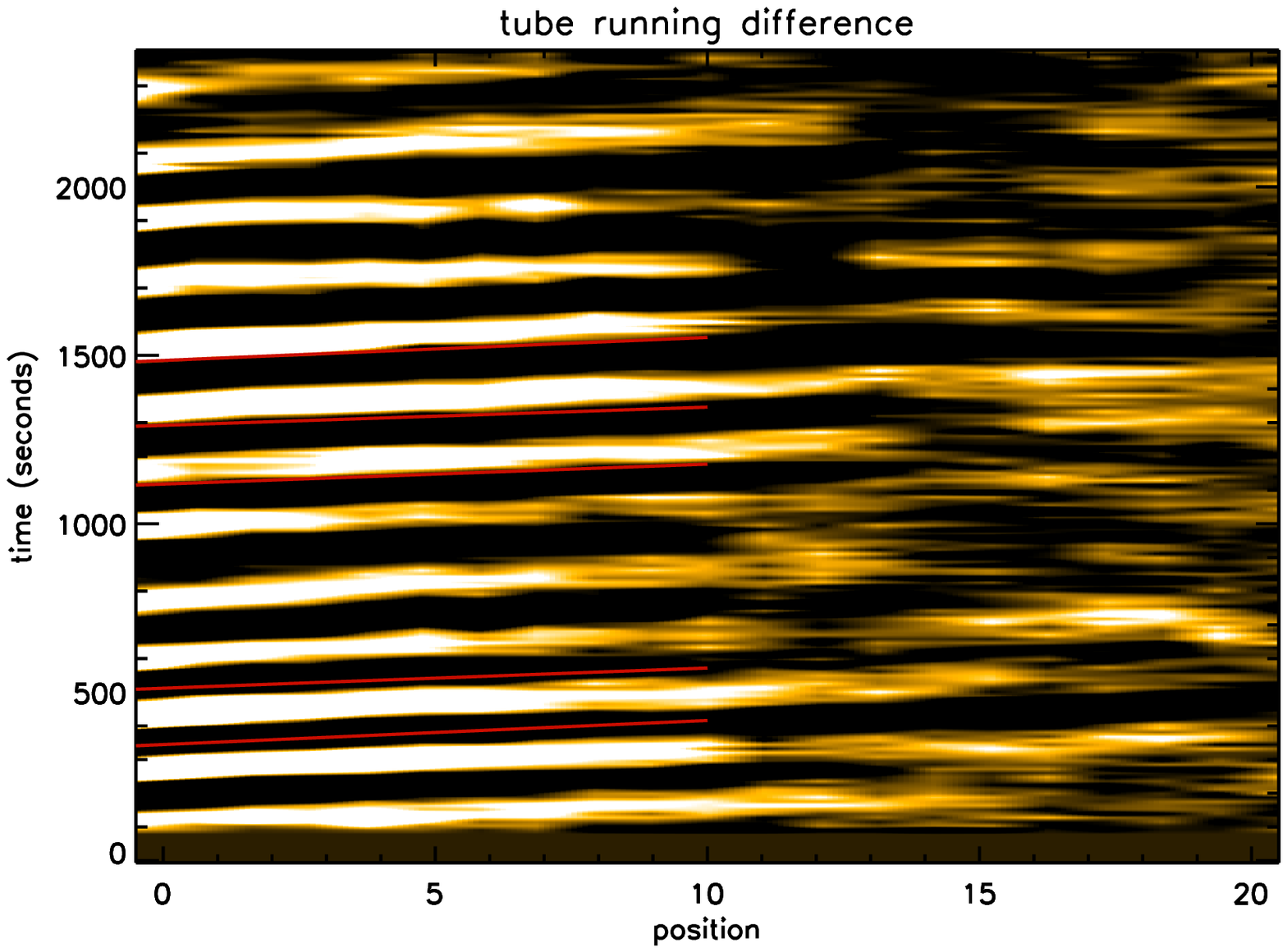}
              }
     \vspace{-0.32\textwidth}   
     \centerline{\Large \bf     
      \hspace{0.0 \textwidth}  \color{black}{\small(a)}
      \hspace{0.400\textwidth}  \color{black}{\small(b)}
         \hfill}
     \vspace{0.31\textwidth}    
   \centerline{\hspace*{0.015\textwidth}
               \includegraphics[width=0.400\textwidth,clip=]{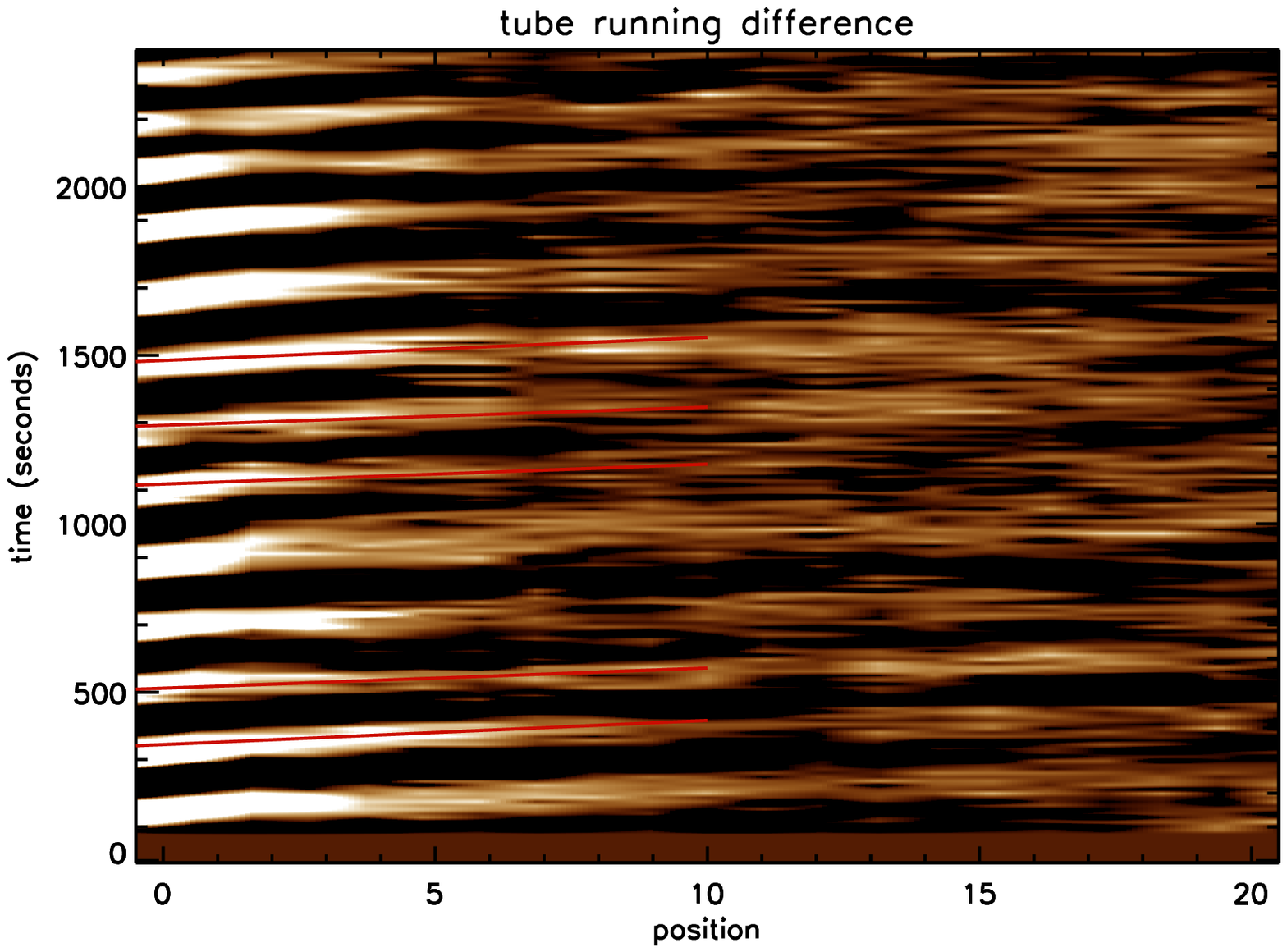}
               \hspace*{-0.03\textwidth}
               \includegraphics[width=0.400\textwidth,clip=]{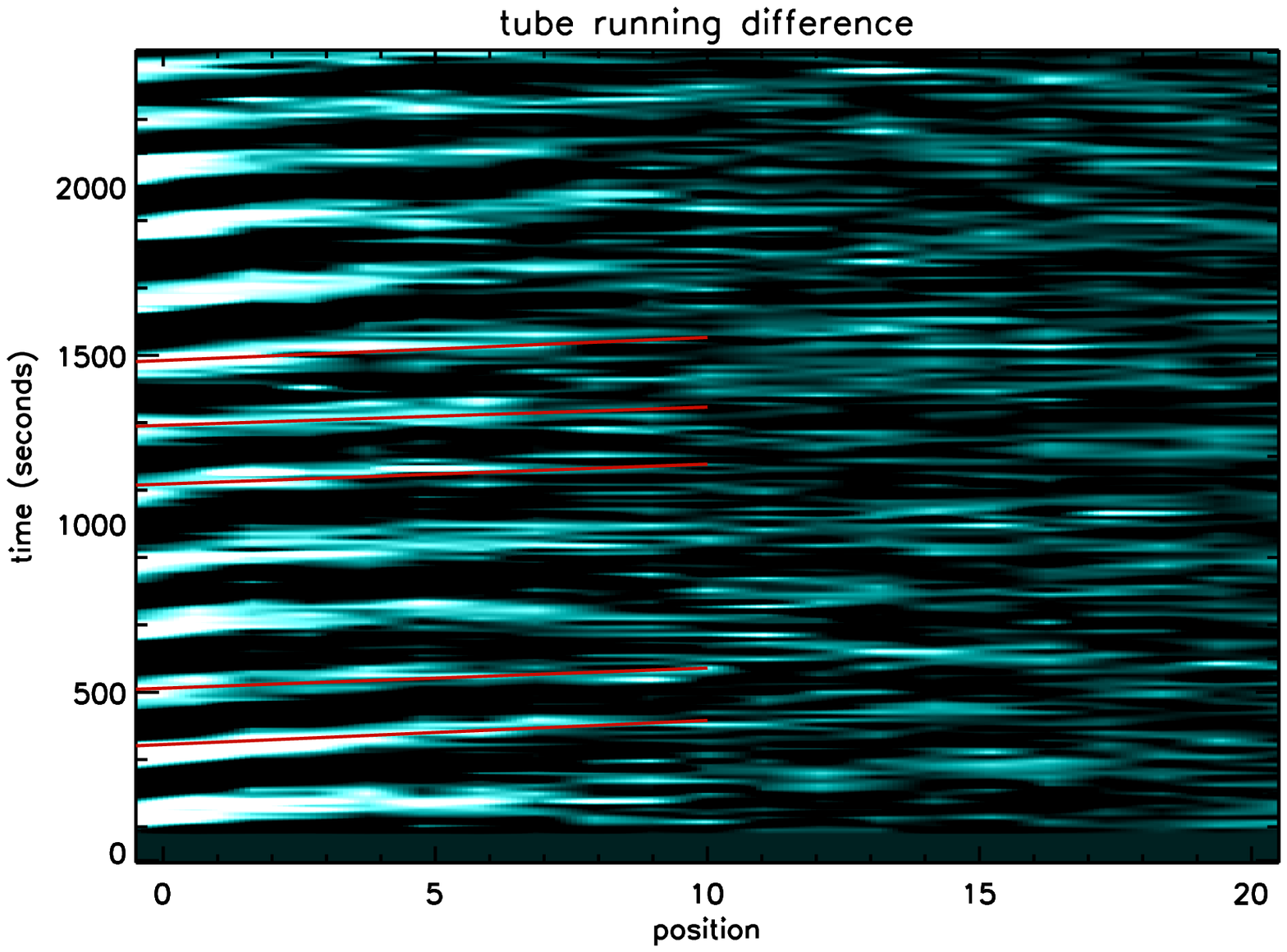}
              }
     \vspace{-0.32\textwidth}   
     \centerline{\Large \bf     
      \hspace{0.0 \textwidth} \color{black}{\small(c)}
      \hspace{0.400\textwidth}  \color{black}{\small(d)}
         \hfill}
     \vspace{0.31\textwidth}    
              
\caption{(a) 171~\AA\ intensity image with the analysed loop outlined by the black lines. (b)\,--\,(d) Running difference images for the 171~\AA, 193~\AA, and 131~\AA\ passbands, respectively for 22 June 2011.  The red lines correspond to the gradient of the intensity bands from the 193 running-difference.}
   \label{F-run_diff_22/06}
   \end{figure}

The second AIA observation is of active region AR11236 on 22 June 2011. The two arcs that outline the loop are
shown in Figure \ref{F-run_diff_22/06}(a).  The loop has solar
coordinates of (338,146) arcseconds at 15:13UT.  Running-difference images for the area
outlined by the two arcs in (a) are shown in (b)\,--\,(d) of Figure \ref{F-run_diff_22/06}.  This loop is rooted in a sunspot umbra, unlike the example analysed in Section \ref{S-22/09/11}, which is situated over a plage region.

We can see from Figure \ref{F-run_diff_22/06} that there are clear intensity
bands indicating PDs in the three wavelengths. This could be an indication of their behaviour; for near-harmonic waves (over this relatively short time interval), we would expect the wave amplitude to be approximately constant, the bands to be equally bright, and very straight (as the wave-propagation speed does not depend on its amplitude in the linear regime).  This is what we observe in Figure \ref{F-run_diff_22/06}.  For flows, one could envisage a more random behaviour with variations in the strength of the flows ({\it i.e.} the amplitude of the PDs) and hence the speed ({\it i.e.} the slope of the bands). This could be an explanation for the irregularity seen in the bands in Figure \ref{F-run_diff_22/09}.  Using the overplotted
lines as an aid, the 131~\AA\  bands may have a slightly greater gradient than the
193~\AA.  The 171~\AA\ bands seem to have  approximately the same gradient but it is
difficult to get an accurate measurement due to the fact the bands do not match spatially.  The
velocities are calculated in the same way as in the previous example and are
displayed in Table \ref{T-Two}.   The straight, parallel nature of the bands in Figure \ref{F-run_diff_22/06}(b)\,--\,(d) is reflected in the relatively small range of speeds between the different bands in Table \ref{T-Two}.  For an non-sunspot example, a much greater disparity between different bands was found (Table \ref{T-One}).  

\begin{table}[!h]
\caption{Average velocities associated with running-difference for all
wavelengths (in
km$s^{-1}$) for 22 June 2011.  The velocities in the brackets show the lower and higher estimates.}
\label{T-Two}
\begin{tabular}{c c c c}
\hline
Band & 131~\AA\ & 171~\AA\ & 193~\AA\ \\ 
\hline
1 & 75(33\,--\,298) & 88(44\,--\,288) & 92(40\,--\,277) \\
2 & 62(29\,--\,557) & 100(54\,--\,733) & 111(42\,--\,177) \\
3 & 67(27\,--\,134) & 111(54\,--\,812) & 112(40\,--\,143)\\
4 & 95(38\,--\,191) & 104(45\,--\,365) & 123(45\,--\,163) \\
5 & 86(37\,--\,257) & 88(43\,--\,335) & 97(44\,--\,580) \\ 
\hline
\end{tabular}
\end{table}

From Table \ref{T-Two} we can see  that the average velocities of the PDs generally
increase from 131~\AA\ to 193~\AA\.  If we assume a characteristic temperatures of
0.4MK for the 131 line, 0.8 MK for 171~\AA\ and 1.2 MK for 193~\AA, the sound speed
increases by a factor of 1.187 from 131~\AA\ to 171~\AA\ and by 1.192 from 171~\AA\ to 193~\AA. The
average velocities in the table match these factors reasonably well.

 \begin{figure}[!h]    
   \centerline{\hspace*{0.015\textwidth}
               \includegraphics[width=0.400\textwidth,clip=]{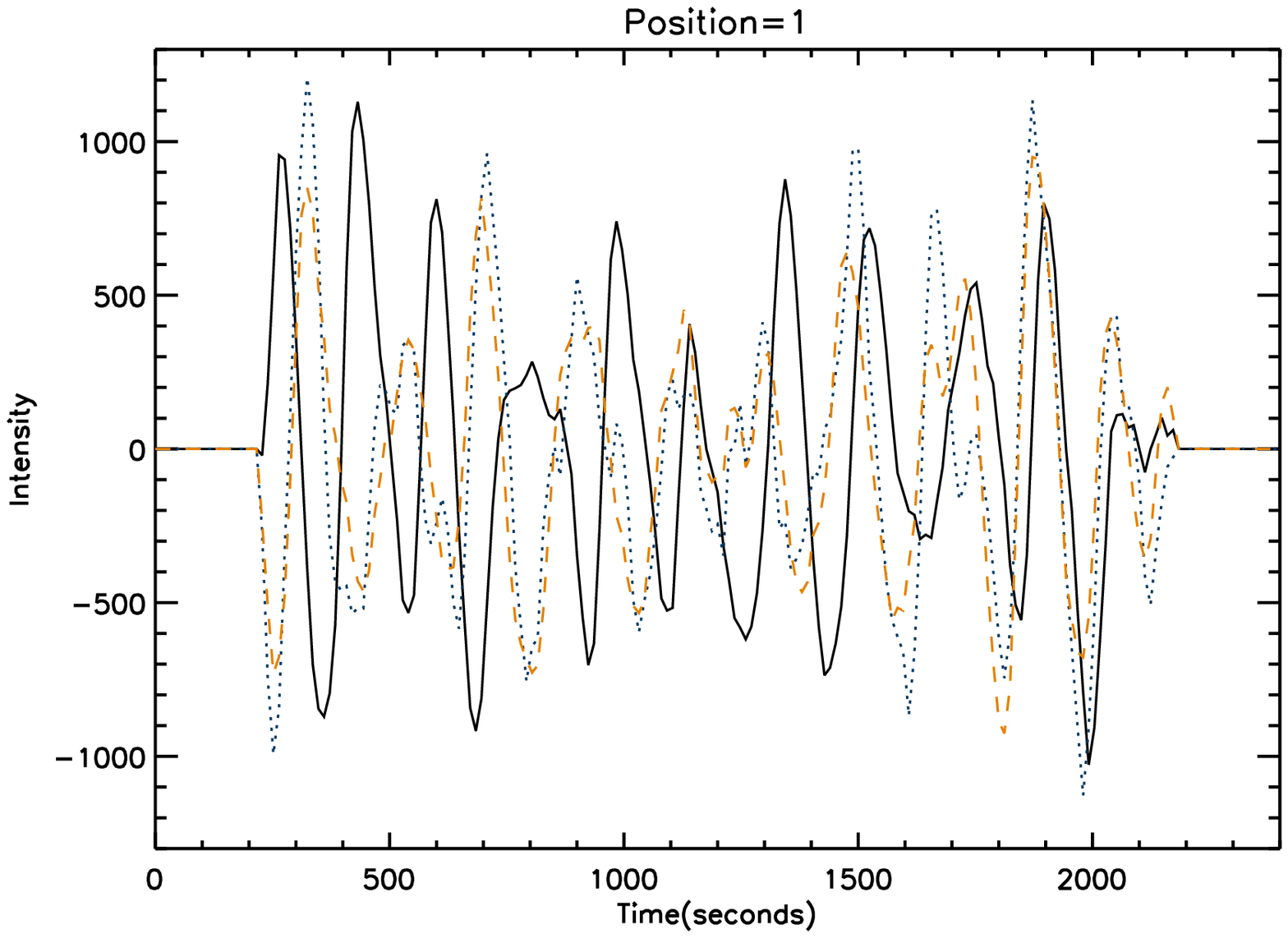}
               \hspace*{-0.03\textwidth}
               \includegraphics[width=0.400\textwidth,clip=]{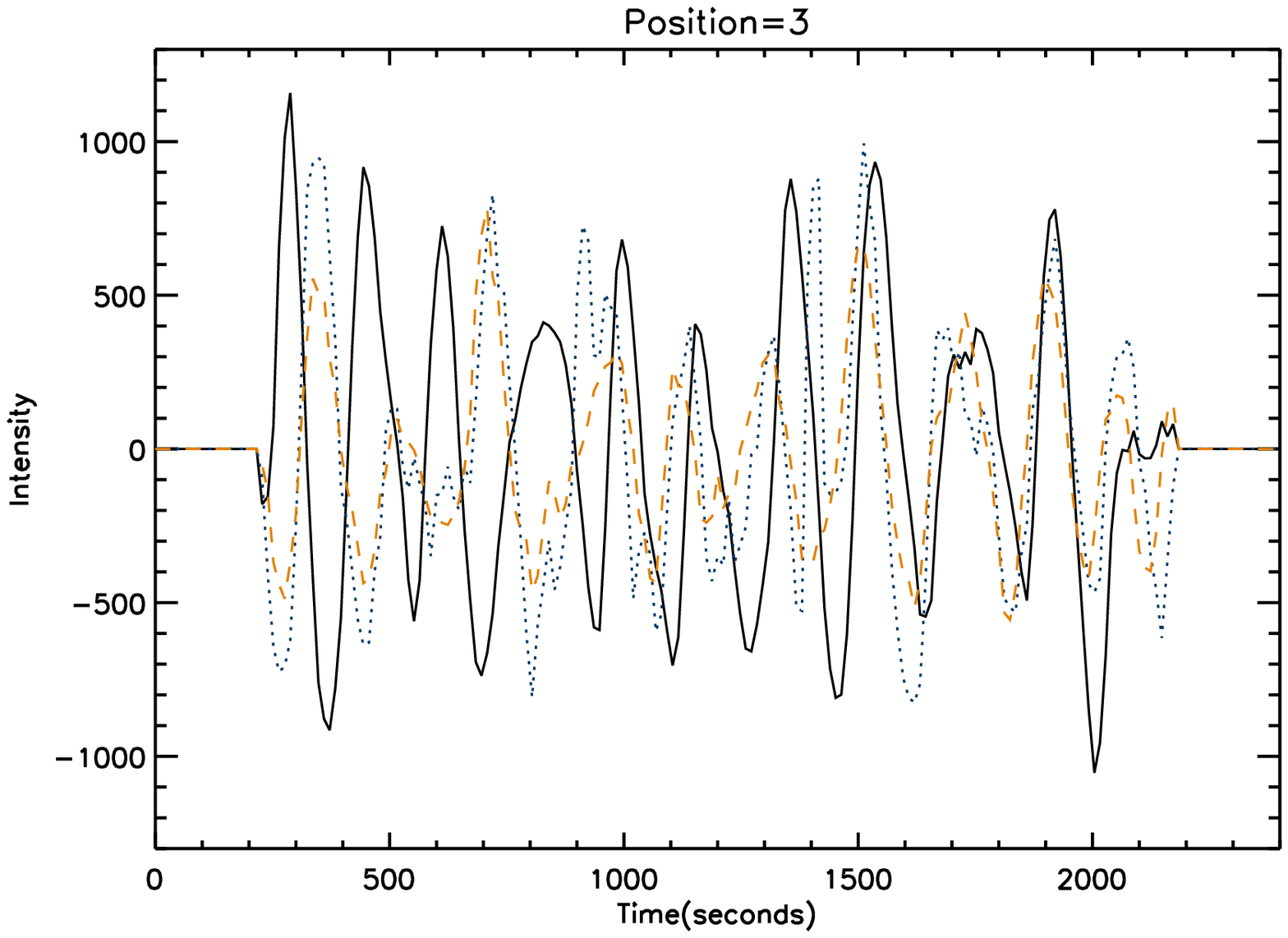}
              }
     \vspace{-0.32\textwidth}   
     \centerline{\Large \bf     
      \hspace{0.0 \textwidth}  \color{black}{\small(a)}
      \hspace{0.400\textwidth}  \color{black}{\small(b)}
         \hfill}
     \vspace{0.31\textwidth}    
   \centerline{\hspace*{0.015\textwidth}
               \includegraphics[width=0.400\textwidth,clip=]{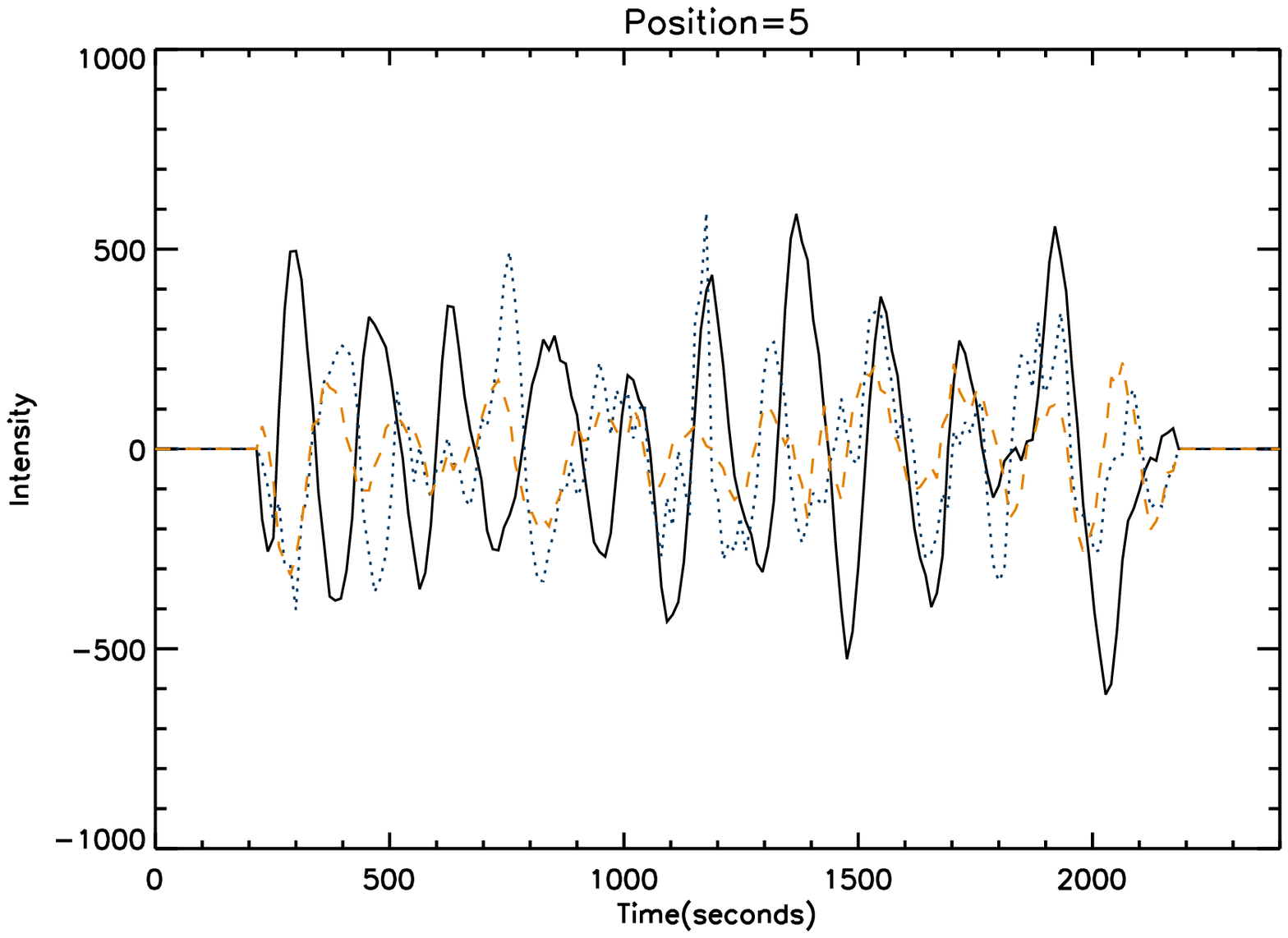}
               \hspace*{-0.03\textwidth}
               \includegraphics[width=0.400\textwidth,clip=]{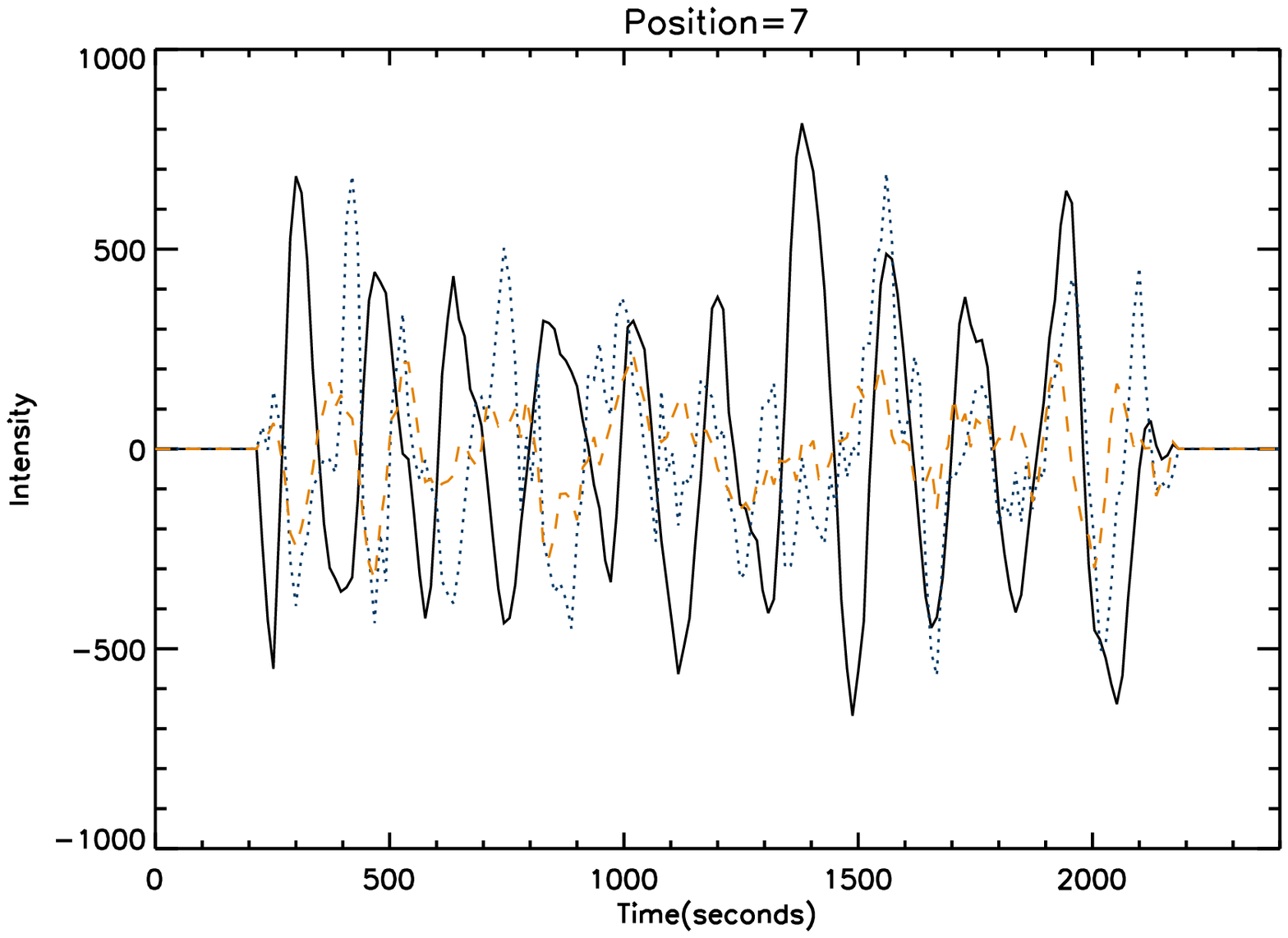}
              }
     \vspace{-0.32\textwidth}   
     \centerline{\Large \bf     
      \hspace{0.0 \textwidth} \color{black}{\small(c)}
      \hspace{0.400\textwidth}  \color{black}{\small(d)}
         \hfill}
     \vspace{0.31\textwidth}    
              
\caption{Cuts through de-trended intensities for  171~\AA\ (black solid), 193~\AA\ (orange dashed), and 131~\AA\
(blue dotted) for different positions along the tube, for 22 June 2011. }
   \label{F-cuts_22/06}
   \end{figure}

Figure \ref{F-cuts_22/06} again shows cuts at positions 1, 3, 5, and 7 for the three different wavelengths.  At position 1 the three
wavelengths match quite well for most of the time.  The 131~\AA\ and 193~\AA\ lines match very
well throughout, but the 171~\AA\ line seems to have a greater frequency than the 193~\AA\ and
131~\AA\ lines.  It is only in phase for times greater than 1200 seconds.  The 193~\AA\ and 131~\AA\
lines continue to match at positions 3 and 5, with the 171~\AA\ line remaining
out of phase with the others.  At position 7 the 131~\AA\ and 193~\AA\ signals have now
drifted slightly out of phase with one another and the 171~\AA\ line has remained approximately half a wavelength out
of phase.

\begin{figure}[!h]
 \centering
\scalebox{0.32}{\includegraphics{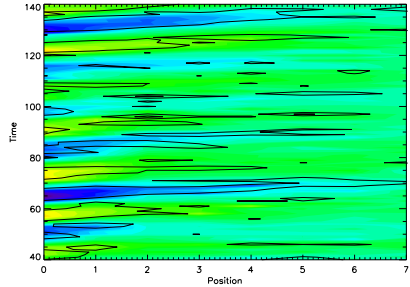}}
\caption{Contour plot of the 193~\AA\ running-difference with 131~\AA\ overplotted (thick black lines) for 22 June 2011 }
\label{F-contour2_22/06}
\end{figure}

Figure \ref{F-contour2_22/06}  shows a contour plot of the 193~\AA\ running
difference with the 131~\AA\ contours overplotted.  The bands in the 193~\AA\ contour plot
are outlined quite well by the 131~\AA\ contours.  There is evidence of the gradients of the bands increasing in the 131~\AA\ contours.   The gradient of the 131~\AA\ contours
may be slightly greater than the 193~\AA\ contours but not by a large amount. The similarities between the 193~\AA\ and 131~\AA\ passbands at sunspot locations is discussed further in Section \ref{S-Cool}.

If we consider the velocities calculated from the running-difference images, 
there does seem to be a systematic dependence on the temperature, which fits with the slow magneto-acoustic wave
interpretation.  From Figure \ref{F-cuts_22/06} it appears that the 171~\AA\ PDs are
travelling at a different velocity than the  others, which only drift slightly out
of phase as they travel further along the loop. 

\subsection{Alternative Methods for Calculating the Velocities}
   \label{S-velocities}

The velocities in Tables \ref{T-One} and \ref{T-Two} are calculated from manually measuring the gradient of the PDs in the running-difference images (Figures \ref{F-run_diff_22/09} and \ref{F-run_diff_22/06}).  Although this method is known to give a reasonable estimate of the velocities, it is subjective ({\it i.e.} user dependent), and the errors associated with it can be substantial (\inlinecite{yuan12}).  This method will be referred to as method 1 (M1).  We have used a further two methods to calculate the velocity. Method 2 (M2): for each intensity band we find the location of the maximum for each position of the band.  The positions of the maximum are then plotted against position along the loop and the gradient of a line fitted to these points is taken to be an estimate of the velocity.  Method 3 (M3): we find the correlation and the time lag between the signals at each position.  In this case, the lag gives an estimate of the velocity.  This is the same method used by \inlinecite{tian11b} and  \inlinecite{mcintosh12} to calculate the velocities.  The errors associated with M2 and M3 range between 5\,--\,15km$s^{-1}$.  The velocities are calculated for both the sunspot (22 June 2011) and non-sunspot (22 September 2011) examples using M2 and M3 and are displayed in Tables \ref{T-method2_22_09} and \ref{T-method2_22_06}.

\begin{table}[!h]
\caption{Calculated velocities using methods 2 and 3 (M2/M3) for PDs associated with running-difference images, for all
wavelengths (in
km$s^{-1}$), for 22 June 2011.}
\label{T-method2_22_09}
\begin{tabular}{c c c c}
\hline
Band & 131~\AA\ & 171~\AA\ & 193~\AA\ \\ 
\hline
1 & 87/113 & 140/138  & 120/168 \\
2 & 90/90 & 124/128  & 147/170 \\
3 & 86/62 & 116/132 & 141/126\\
4 & 92/117 & 105/123 & 120/151 \\
5 & 51/82 & 100/121 & 124/131 \\ 
\hline
\end{tabular}
\end{table}

\begin{table}[!h]
\caption{Calculated velocities using method 2 and 3 (M2/M3) for PDs associated with running-difference images, for all
wavelengths (in
km$s^{-1}$), for 22 September 2011.}
\label{T-method2_22_06}
\begin{tabular}{c c c c}
\hline
Band & 131~\AA\ & 171~\AA\ & 193~\AA\ \\ 
\hline
1 & 30/39 & 27/60  & 27/59 \\
2 & 91/100& 98/63  & 66/65 \\
3 & 58/62 & 71/63 & 61/66 \\
4 & 45/41 & 40/60 & 53/86 \\
5 & 36/41 & 49/61 & 26/40 \\ 
\hline
\end{tabular}
\end{table}

The velocities calculated using method 2 are similar to those found using M1, with the mean values across the five bands within 5\,--\,20 km$s^{-1}$ of each other. Our earlier results concerning the temperature dependence are still present; for the sunspot example (22 June 2011) we still find a temperature dependence but no clear dependence is present in the non-sunspot example (22 September 2011) using M2. The velocities calculated using M3 are greater than the velocities calculated by M1 and M2 for both examples, but not by a significant amount.  Even with this increase in the velocities we again confirm the results found using M1.  We can be confident that the results found in Sections \ref{S-22/09/11} and \ref{22_06_2011}  are not dependent on the way that we have measured the gradient. 

\subsection{Other Examples}

In total we have identified 41 loops over eight active regions.  Information on the eight data sets we have considered is displayed in Table \ref{T-data}.

\begin{table}[!h]
\caption{The eight data sets that contain the 41 examples studied.}
\label{T-data}
\begin{tabular}{c lc c}
\hline
Data Set & Date & Start Time  \\ 
\hline
A & 16 September 2010 & 12:05 UT    \\
B & 19 March 2011 & 12:55 UT   \\
C & 22 September 2011 & 00:35 UT   \\
D & 3 April 2011 & 15:20 UT   \\
E & 1 October 2011 & 13:35 UT   \\
F & 28 March 2011 & 14:45 UT      \\
G & 24 August 2011 & 09:40 UT     \\
H & 22 June 2011 & 15:13 UT    \\ 
\hline
\end{tabular}
\end{table}

The same analysis has been undertaken for the 39 other examples and characteristic velocities using the three methods and temperature dependence are displayed in Table
\ref{T-Three}. Speeds in brackets correspond to intensity bands that are less clear in the respective running-difference images.  For the examples we have investigated here there are two categories of less clear: i) in some cases the PDs did not persist for the entire time interval (and hence we only had a limited number of bands to measure) or ii) the PDs only showed up near the footpoint of the loops (and hence the slope became difficult to measure).  It is interesting to note these cases are always non-sunspot examples, again highlighting the more  intermittent, varying nature of non-sunspot PDs.  Each example has been categorised in one of two categories: the
velocity of the PDs are dependent on temperature or they are independent of
temperature.  Two of the three methods need to show a temperature dependence for that example to be defined as temperature dependent.  For the majority of cases the three methods are consistent and the examples that are not, are explained by the superscript.  The solar co-ordinates given in the table correspond to the
footpoints of the loop, in arcseconds.  Whether or not the loop footpoints are located in a sunspot
is also indicated in the S column in Table \ref{T-Three}.

\begin{table}[!ht]
\caption{The location, characteristic velocities for the three methods (M1/M2/M3) and temperature dependence (TD) of all the examples considered.  Parentheses indicate that the intensity bands are less clear in the running-difference image. $^1$ velocities calculated using M2 show a TD. $^2$ velocities calculated using M1 show a TD. $^3$ velocities calculated using M1 does not show a TD. $^4$ velocities calculated using M3 show a TD.} 
\label{T-Three}
\begin{tabular}{c c c c c c c}
\hline
Data & Loop& 131~\AA\ & 171~\AA\ &193~\AA\ & Temp& S \\
Set& Coords& & & & Depen. & \\
\hline
A & (-309,-377) & 135/66/99 & 109/81/92 & 109/61/87 & N  & N  \\
  &(-289,-375) & (124/88/113) & 98/90/98 & 121/120/127 & N$^1$ & N\\
  &(-288,-361) & (142/155/151) & 146/86/124& 142/81/97 & N & N  \\
  &(-110,-499) & 85/59/68 & 57/49/59 & 128/118/132 & N & N\\
  &(-94,-500) & (114/85/115) & 125/165/157 & 119/108/122 & N & N\\
  &(-83,-501) & (70/57/64) & 78/41/53 & 83/76/81 & N$^2$ & N\\ 
 &(-69,-499) & (60/52/100) & 53/49/70 & 82/94/105 & N & N \\ 
 &(-69,-442) & 68/58/78 & 88/78/82 & 105/92/119 & Y & Y \\ 
 &(-67,-431) & 53/53/79 & 67/92/108 & 72/104/125 & Y & Y \\ 
B & (-73,-383) & (85/62/111) & 91/74/61 & 85/60/75 & N & N \\ 
 &(-174,-416)& (66/65/117) & 103/102/108 & 65/62/109 & N & N \\ 
 &(-72,-346) & (107/73/95) & 62/76/81 & 80/75/111 & N & N \\ 
C &(-670,204) & 39/50/69 & 36/40/38 & 39/55/58 & N & N \\ 
 &(-672,189) & 87/53/73 & 58/50/58 & 87/85/100 & N & N \\ 
 &(-673,154) & 75/85/76 & 73/125/101 & 81/96/110 & N & N \\
 &(-662,133) & 80/108/124 & 84/102/109 & 94/89/85 & N$^2$ & N \\
D & (289,329) & 29/24/22 & 44/44/37 & 43/48/54 & Y$^3$ & Y \\
 &(289,339) & 38/43/42 & 49/46/60 & 57/81/75 & Y & Y \\
 &(282,341) & 40/40/53 & 44/47/71 & 63/96/102 & Y & Y \\
E & (496,95) & (96/77/77) & 73/47/64 & 94/95/113 & N & N \\
 & (479,92) & (98/71/110) & 105/61/77 & 108/91/104 & N$^2$ & N \\
 & (459,90) & (96/83/88) & 66/62/75 & 80/99/105 & N & N \\
 & (442,100) & (75/113/99) & 102/60/88 & 127/87/97 & N$^2$ & N \\
 & (397,159) & 96/92/131 & 66/69/79 & 84/111/114 & N & N \\
 & (398,167) & 125/111/87 & 126/85/89 & 115/75/120 & N & N \\
 & (436,147) & 39/70/75 & 37/54/55 & 44/61/38 & N & Y \\
F & (-208,-206) & (63/66/111) & 65/98/90 & 78/67/112 & N$^2$ & N \\
 & (-156,-157) & (118/90/106) & 71/108/134 & 99/67/90 & N & N \\
 & (-203,-212) & 95/57/75 & 57/92/110 & 77/153/120 & Y$^3$ & N \\
 & (-424,-204) & (106/130/132) & 107/129/126 & 111/154/128 & N$^2$ & N \\
G & (555,174) & 68/80/64 & 71/93/117 & 77/101/127 & Y & Y \\
 & (557,180) & 71/58/63 & 76/76/64 & 95/101/111 & Y & Y \\
 & (552,184) & 94/78/93 & 111/90/99 & 116/113/119 & Y & Y \\
H& (559,178) & 97/64/81 & 124/106/125 & 129/96/121 & N$^2$ & N \\
 & (538,154) & 131/97/72  & 140/114/138 & 130/107/112 & N & N \\
 & (530,141) & 135/83/117 & 142/126/133 & 123/84/109 & N & N \\
 & (461,263) & 66/39/51 & 83/69/95 & 76/61/96 & N$^4$ & Y \\
 & (463,260) & 62/44/53 & 91/91/117 & 97/95/125 & Y & Y \\
 & (446,259) & 71/77/75 & 86/103/100 & 93/114/129 & Y & Y \\
\hline
\end{tabular}
\end{table}

The characteristic speeds displayed in Table \ref{T-Three} are the mean of the
average  velocities calculated from the intensity bands. In 38 out of 41 of the
examples, whether their velocities are dependent on temperature depends on
whether they are located at sunspot or non-sunspot location, {\it i.e.} double Ys or
Ns.  PDs that are dependent on temperature are mainly found in sunspots and
PDs whose velocity are not dependent on temperature are mainly found in non
sunspot regions.  In 11 of the 13 (85\%) sunspot examples the PDs are dependent on the temperature,
and at non-sunspot locations 27 out of the 28 (96\%) examples the PDs are not dependent on temperature.  Of these 27, eight examples showed a temperature dependence in one of the methods for calculating velocity. Hence, for the examples analysed here, the dependence of the PD velocity on temperature seems to correlate with the location (sunspot or non-sunspot region).

\section{Properties of PDs Across an Active Region}
     \label{S-Active}

\begin{figure}[!h]
 \centering
\scalebox{0.45}{\includegraphics{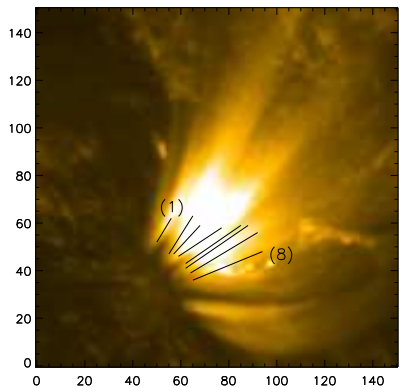}}
\caption{171 ~\AA\ intensity image showing where the eigth arcs  are defined for  AR11236 at 15:13UT on the 22 June 2011.}
\label{F-outline}
\end{figure}

These PDs are thought to arise from the leakage of global {\it p}-modes
into the solar atmosphere \cite{depontieu05,demoortel07,malins07}.  This is known to lead to periods of approximately five
minutes for non-sunspot locations and three minutes above sunspots.  We will now
investigate how properties such as period and velocity change across a smaller
scale, {\it i.e.} how they change across a single set of sunspot and non-sunspot loops.  Again we consider the two
primary active regions analysed in Section \ref{S-Propspeed} and focus on the 171~\AA\ passband. Eight arcs are
identified in this set of sunspot loops and we study over which scale the properties of
the PDs change.   Figure \ref{F-outline} shows the area in which the eight arcs
are defined for the first example (22 June 2011).  Arc 1 is located closest to the left side in Figure \ref{F-outline} and arc 8 is located closest to the right-hand side. Running differences are then
calculated in the usual way.

\begin{figure}[!h]    
   \centerline{\hspace*{0.015\textwidth}
            \includegraphics[width=0.400\textwidth,clip=]{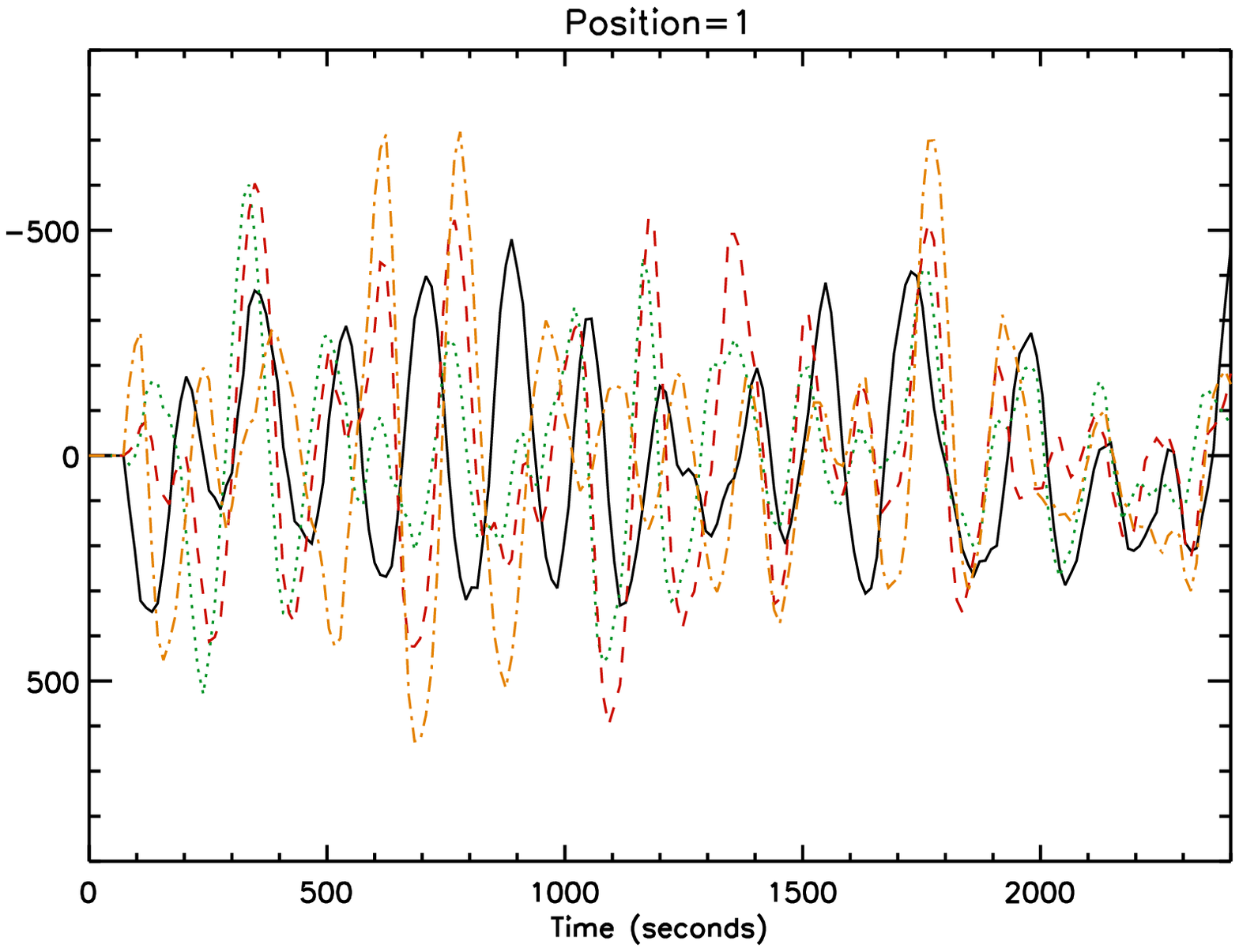}
              \hspace*{-0.03\textwidth}
            \includegraphics[width=0.400\textwidth,clip=]{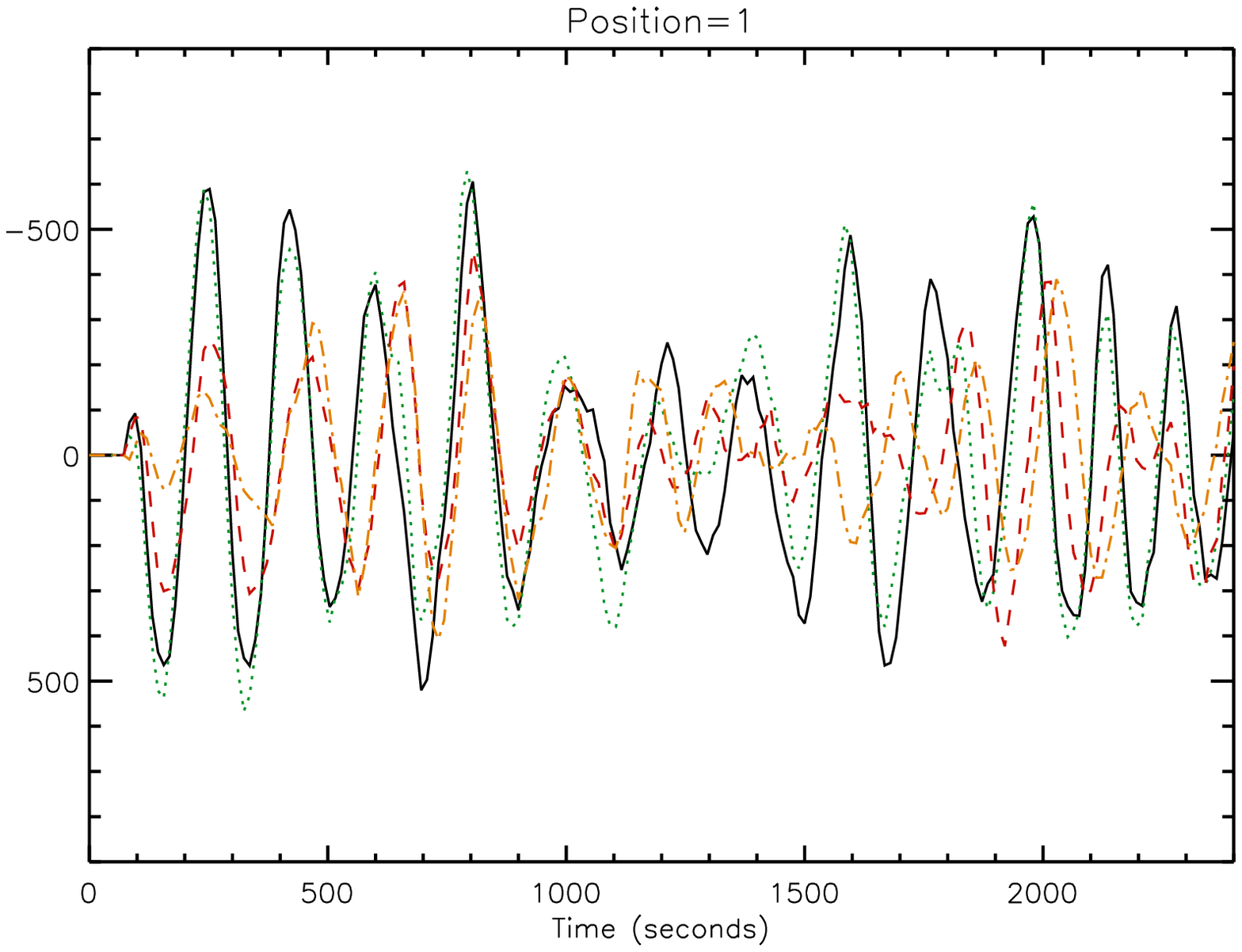}
              }
     \vspace{-0.32\textwidth}   
     \centerline{\Large \bf     
      \hspace{0.07 \textwidth}  \color{black}{\small(a)}
      \hspace{0.415\textwidth}  \color{black}{\small(b)}
         \hfill}
     \vspace{0.31\textwidth}    
   \centerline{\hspace*{0.015\textwidth}
            \includegraphics[width=0.400\textwidth,clip=]{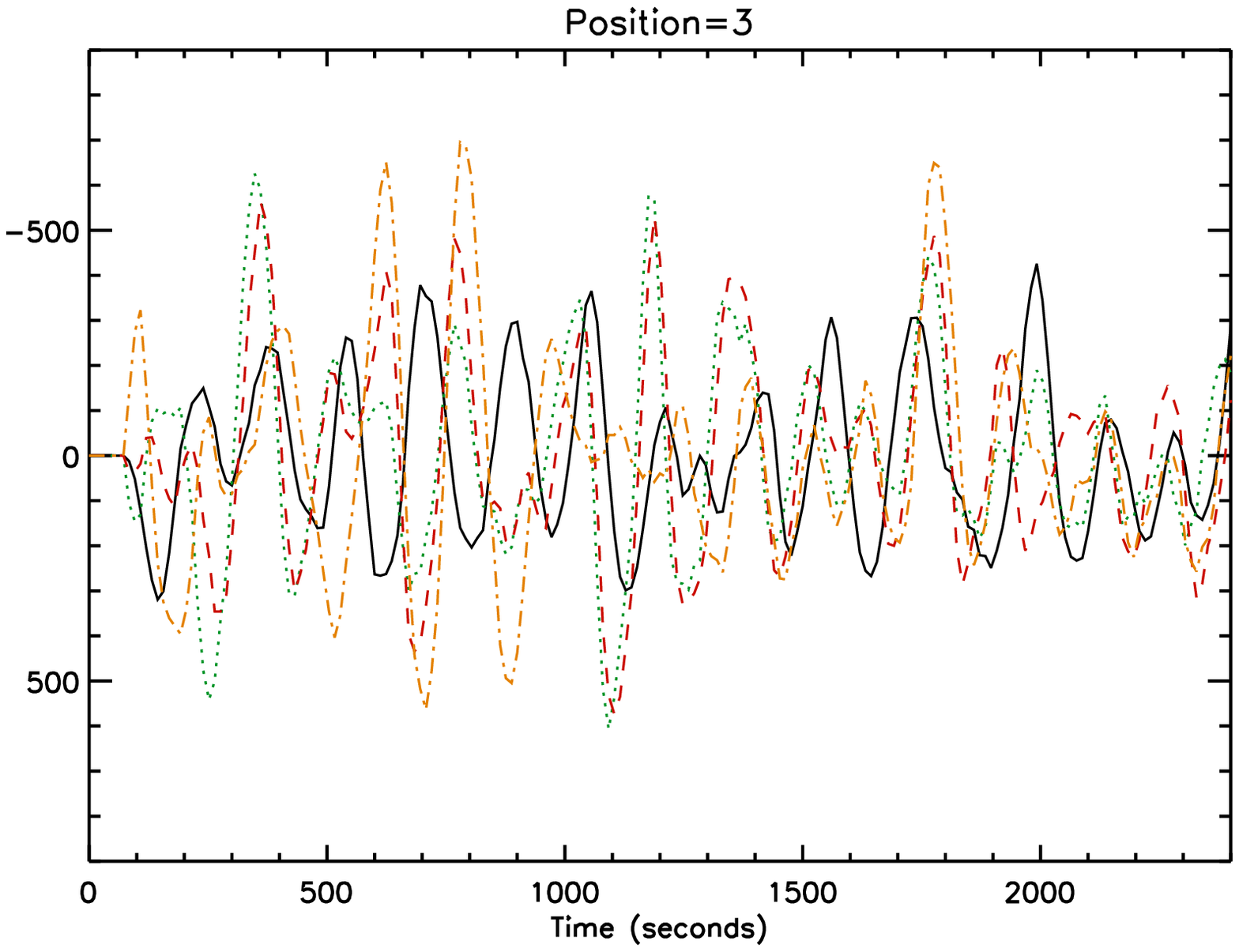}
             \hspace*{-0.03\textwidth}
            \includegraphics[width=0.400\textwidth,clip=]{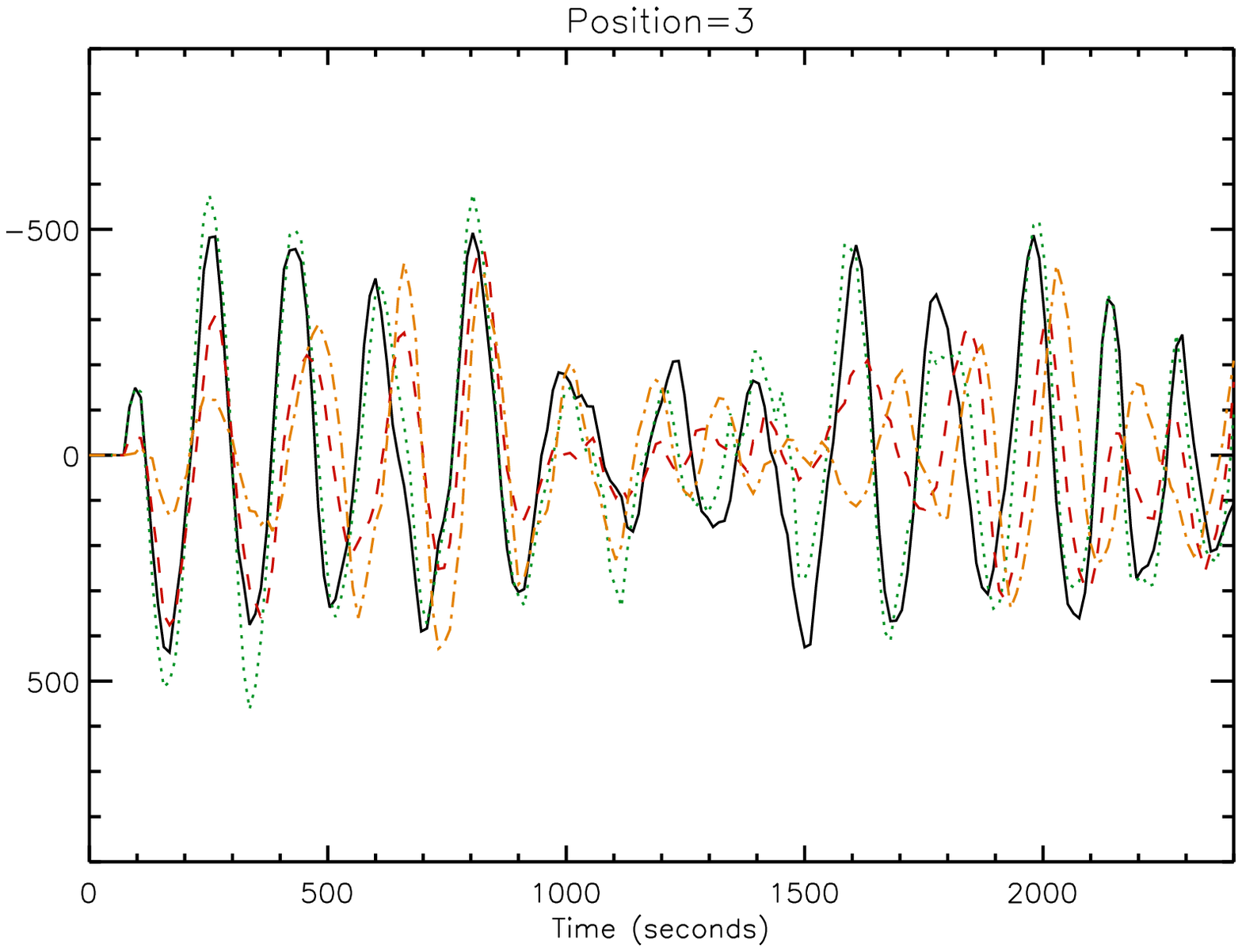}
              }
     \vspace{-0.32\textwidth}   
     \centerline{\Large \bf     
      \hspace{0.07 \textwidth} \color{black}{\small(c)}
      \hspace{0.415\textwidth}  \color{black}{\small(d)}
         \hfill}
     \vspace{0.31\textwidth}    

 \centerline{\hspace*{0.015\textwidth}
          \includegraphics[width=0.400\textwidth,clip=]{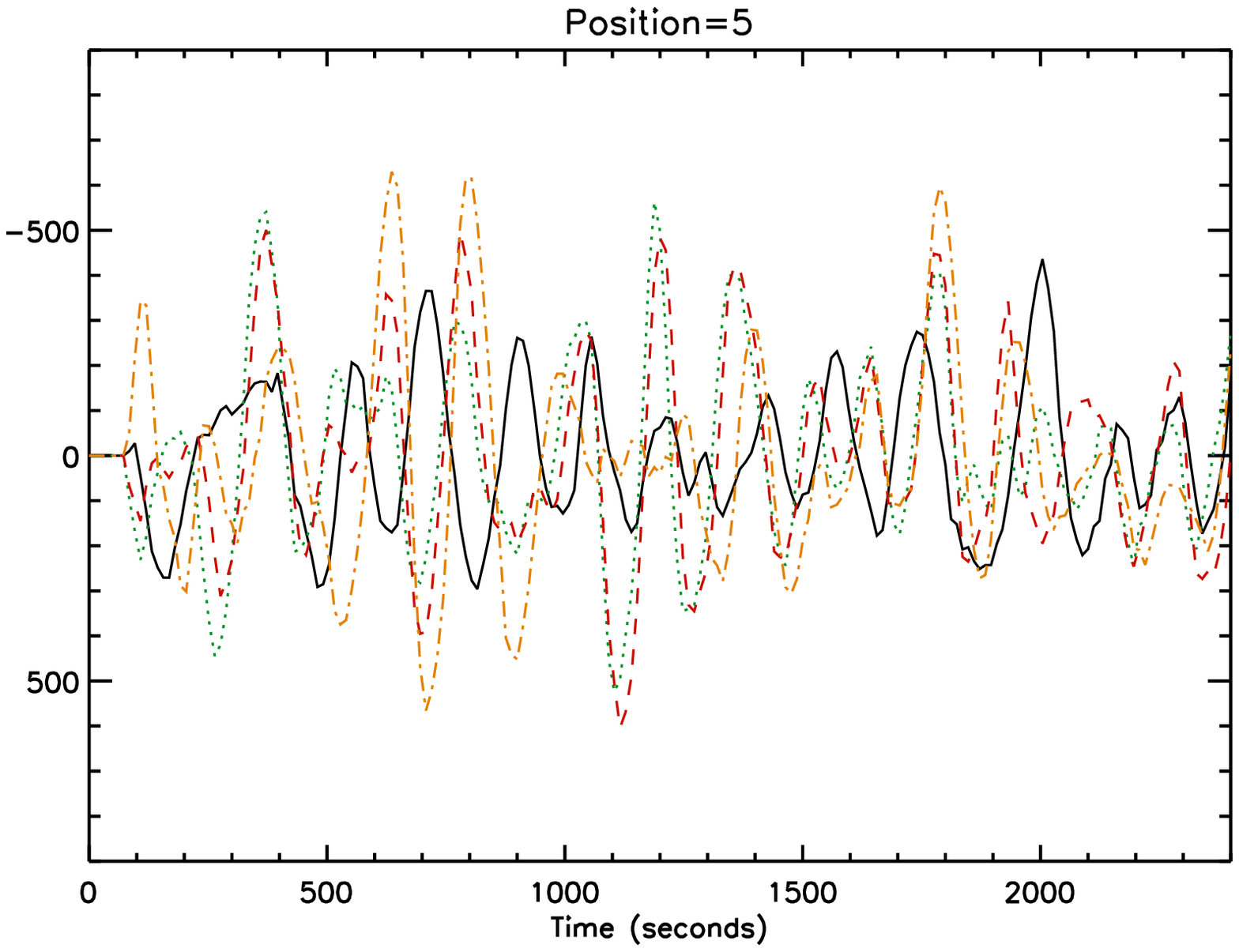}
            \hspace*{-0.03\textwidth}
          \includegraphics[width=0.400\textwidth,clip=]{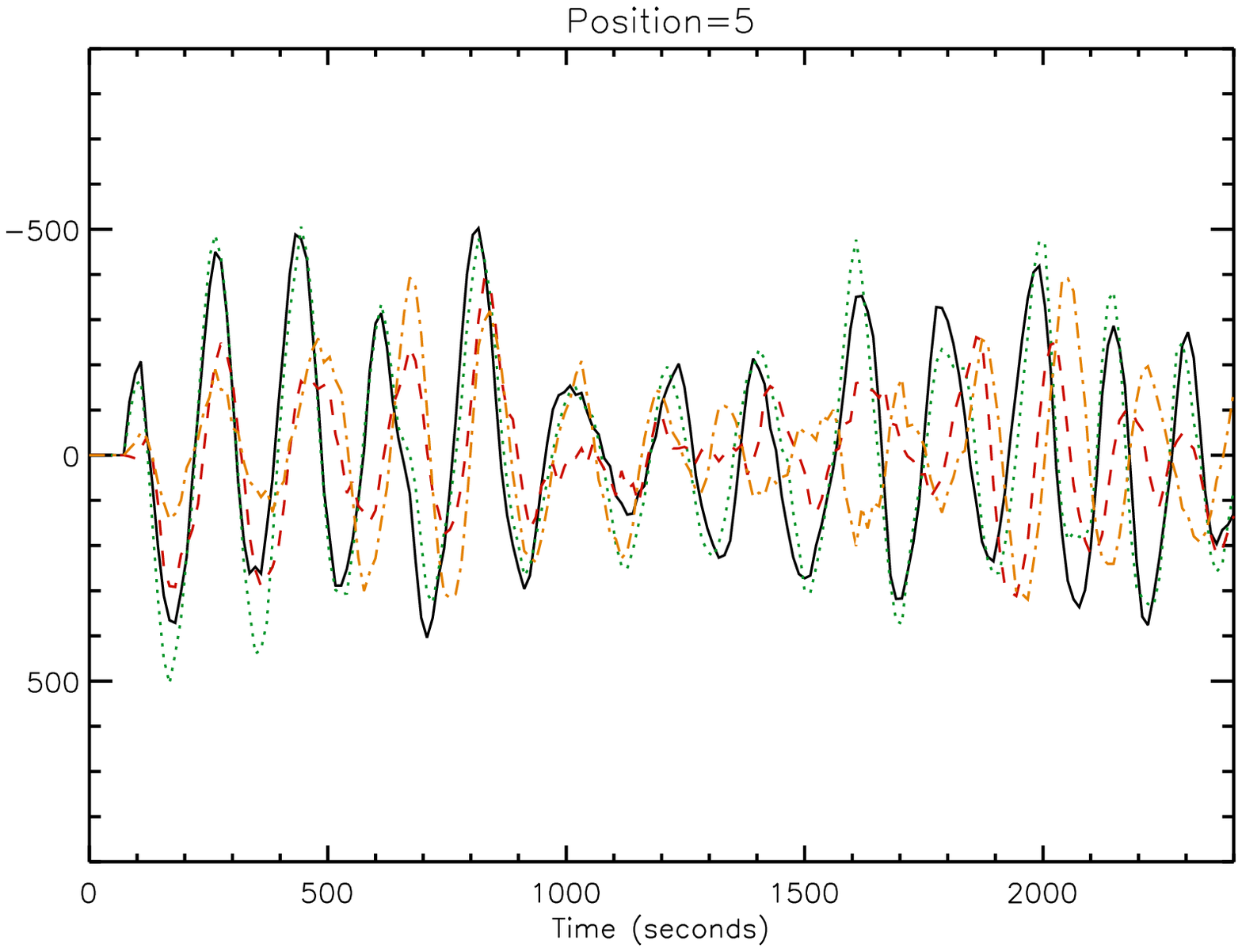}
              }
     \vspace{-0.32\textwidth}   
     \centerline{\Large \bf     
      \hspace{0.07 \textwidth} \color{black}{\small(e)}
      \hspace{0.415\textwidth}  \color{black}{\small(f)}
         \hfill}
     \vspace{0.31\textwidth}    

 \centerline{\hspace*{0.015\textwidth}
          \includegraphics[width=0.400\textwidth,clip=]{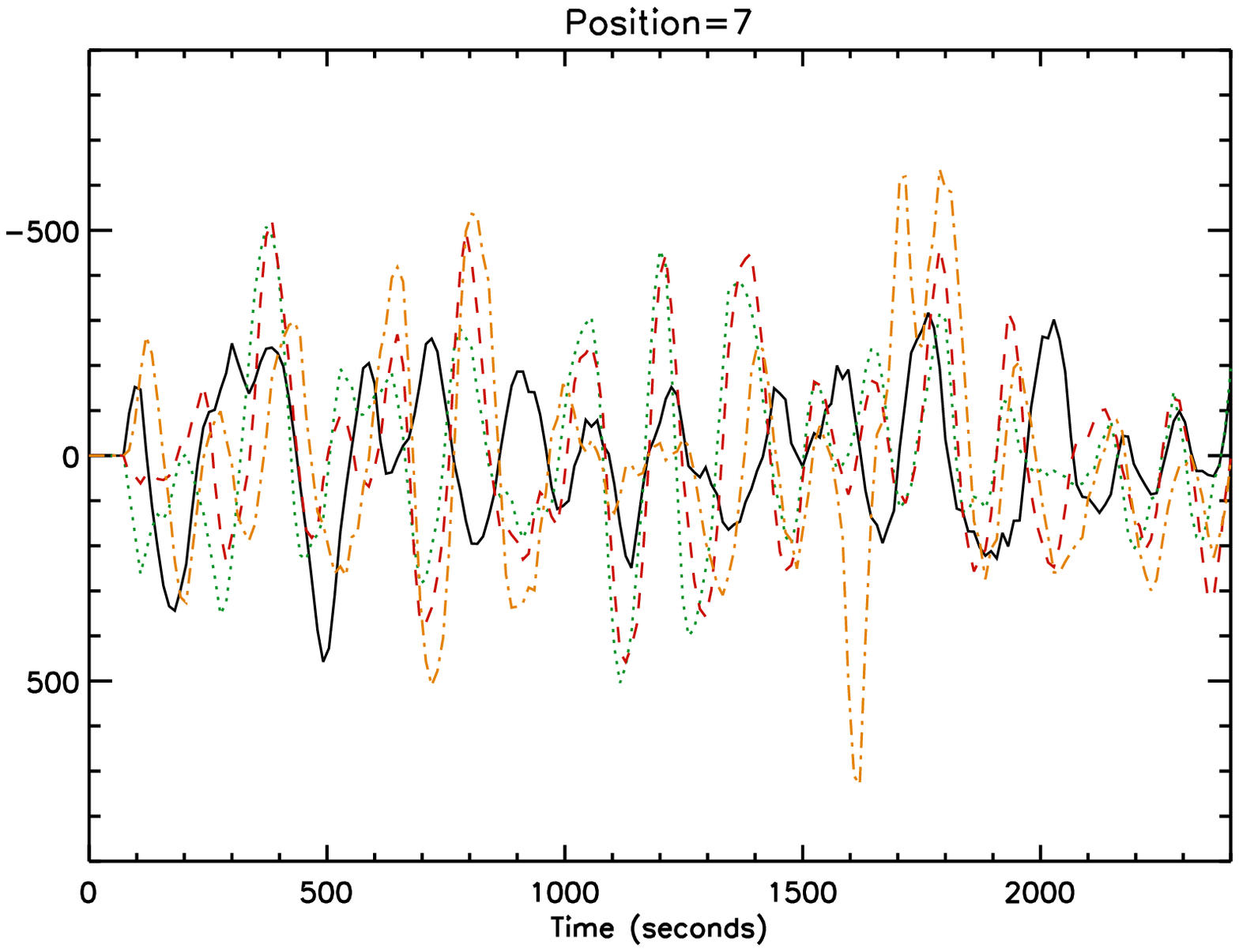}
            \hspace*{-0.03\textwidth}
          \includegraphics[width=0.400\textwidth,clip=]{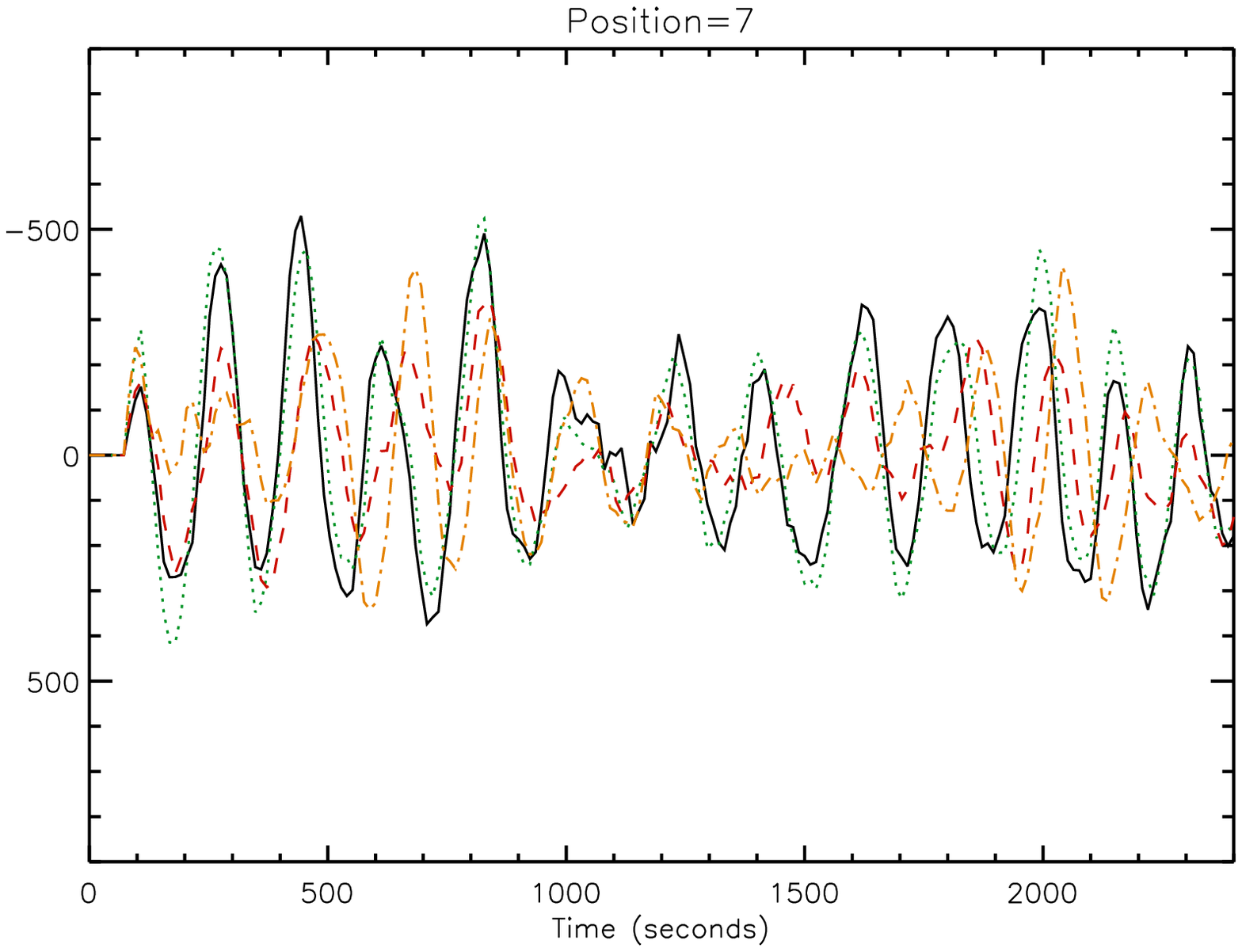}
              }
     \vspace{-0.32\textwidth}   
     \centerline{\Large \bf     
      \hspace{0.07 \textwidth} \color{black}{\small(g)}
      \hspace{0.415\textwidth}  \color{black}{\small(h)}
         \hfill}
     \vspace{0.31\textwidth}    
              
\caption{Cuts through the 171~\AA\ running-difference images for arcs 1\,--\,8 at all positions for 22 June 2011.  (a), (c), (e), and (g) show arcs 1\,--\,4 and  (b), (d), (f), and (h) show arcs 5\,--\,8.  The solid black lines corresponds to arcs 1 and 5, the green dotted lines to arcs 2 and 6, red dashed to 3 and 7, and orange dot--dashed to 4 and 8.   }
   \label{F-phase}
   \end{figure}

Figure \ref{F-phase} shows cuts through running-difference images for the eight
arcs.  Panels (a), (c), (e), and (g) are for arcs 1\,--\,4 and panels (b),
(d), (f), and (h) are for arcs 5\,--\,8.  The black solid lines represents arc 1 in
(a),(c), (e), and (h) and arc 5 in (b), (d), (f), and (h).  The green dotted lines
represent arcs  2 and 6, the red dashed line arcs 3 and 7 and orange dot--dashed lines represents arcs 4
and 8.

It is clear from Figure \ref{F-phase} that arcs 5\,--\,8 are approximately in phase
for all positions along the arc.  Arcs 2\,--\,4 are also in phase for all
positions but arc 1 is slightly out of phase with 2\,--\,4.  To quantify these phase differences, we  calculate the cross
correlation between each loop at all positions as a function of the lag.  Table
\ref{T-correlate} shows the maximum correlation between two loops and the time lag at
which this correlation is achieved. 

\begin{table}[!h]
\caption{Cross correlation between 171 arcs at position 1 along the loop defined on the 22 June 2011. The subscript denotes the lag (i units of 12 seconds) where the maximum correlation is found. The table is symmetric, and the blank spaces would have the same values as their corresponding location, with the sign of the lag changing. }
\begin{tabular}{c c c c c c c c c}
\hline
Arc  & 1 & 2 &  3 & 4 &  5 &  6 & 7 & 8 \\
[0.5ex]
\hline
 1 & 1 & $0.365_{-3}$ & $0.187_{-2}$ & $0.471_5$ & $0.344_{-10}$ & $0.381_{-10}$ &
$0.360_8$ & $0.477_9$ \\
 2 &  & 1 & $0.815_1$ & $0.215_2$ & $0.273_5$ & $0.256_5$ & $0.264_{-8}$ &
$0.181_{12}$ \\
 3 &  &  & 1 & $0.476_1$ & $0.384_3$ & $0.356_3$ & $0.263_{-9}$ & $0.130_{-14}$ \\
 4 &  &  & & 1 & $0.569_1$ & $0.522_1$ & $0.582_3$ & $0.451_4$\\ 
 5 & &  & & & 1 & $0.897_0$ & $0.582_2$ & $0.297_4$\\
 6 & &  & & & & 1 & $0.721_2$ & $0.373_3$\\
 7 & &  & & & & & 1 & $0.687_1$\\
 8 & &  & & & & & & 1 \\ [1ex]
\hline
\end{tabular}
\label{T-correlate}
\end{table}

This analysis is repeated for positions 3, 5, and 7 and the corresponding correlation tables
are in the appendix.  From Table \ref{T-correlate} we can see that
arcs 2\,--\,4 have maximum correlation with each other at relatively small lag
positions, i.e. loops 2\,--\,4 are mostly in phase with each other.  Arcs 5\,--\,8
are also approximately in phase at this position.    The lag where the maximum
correlation occurs becomes greater when you consider two arcs that are not
located next to each other.  The fact that the PDs do not correlate over the entire extent of this ensemble of sunspot loops suggests the underlying driver changes on smaller scales.  However, we have to keep in mind that some of the lag could also be caused by the fact that the starting points of the arcs do not line up perfectly. 

\begin{figure}[!t]
 \centering
\scalebox{0.45}{\includegraphics{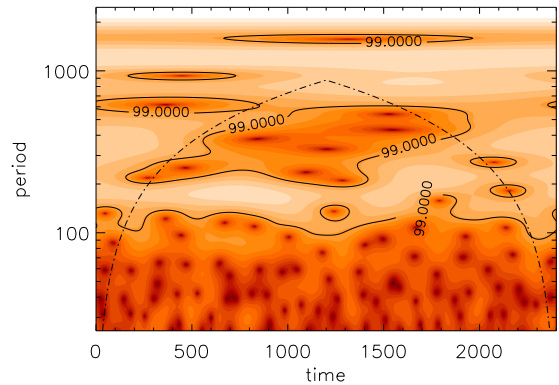}}
\caption{Wavelet analysis for arc 1 on 22 June 2011 at position 1.}
\label{wavelet}
\end{figure}

The periods of these disturbances are calculated using a wavelet transform
(Torrence and Compo 1998) with the Morlet function as the mother wavelet and
are displayed in Table \ref{T-Five}.  The range of periods is calculated using the bottom and top of the band that is above the confidence interval. An example wavelet is shown from arc 1 position 1 in Figure \ref{wavelet}.

\begin{table}[!h]
\caption{Table showing the periods and characteristic 171~\AA\ velocity for arcs 1\,--\,8 for 22 June 2011. }
\label{T-Five}
\begin{tabular}{c c c}     
  \hline                   
Arc & Period[s] & Char. Velocity[km$s^{-1}$]   \\
  \hline
1& 150\,--\,200 & 131  \\
2 & 100\,--\,200 & 128 \\
3 & 120\,--\,190 & 130 \\
4 & 150\,--\,210 & 135   \\
5 & 160\,--\,200 & 152  \\
6 & 160\,--\,200 & 136  \\
7 & 150\,--\,210 & 167  \\
8 & 150\,--\,200 & 143  \\
  \hline
\end{tabular}
\end{table}

From Table \ref{T-Five}  we can see that the PDs associated with all arcs have
approximately the same period  which appears centred around 180 seconds (three minutes) as expected for sunspot loops \cite{demoortel00}.  The velocities of the PDs
as seen in the 171~\AA\ passband are also displayed. Arcs 1\,--\,4 all propagate with
approximately the same speed.  Arcs 5\,--\,8 also propagate with approximately the
same speed as each other but at a slightly greater speed than arcs 1\,--\,4.

\subsection{22 September 2011 (Non Sunspot)}

\begin{figure}[!h]
 \centering
\scalebox{0.45}{\includegraphics{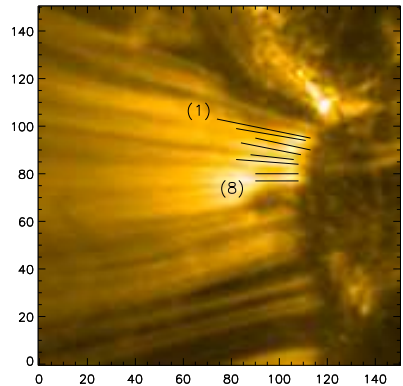}}
\label{F-out}
\caption{171~\AA\ intensity image showing where eight arcs are defined for AR11301 at 00:35UT on 22 September 2011.}
\end{figure}

\begin{figure}[!h]    
   \centerline{\hspace*{0.015\textwidth}
            \includegraphics[width=0.400\textwidth,clip=]{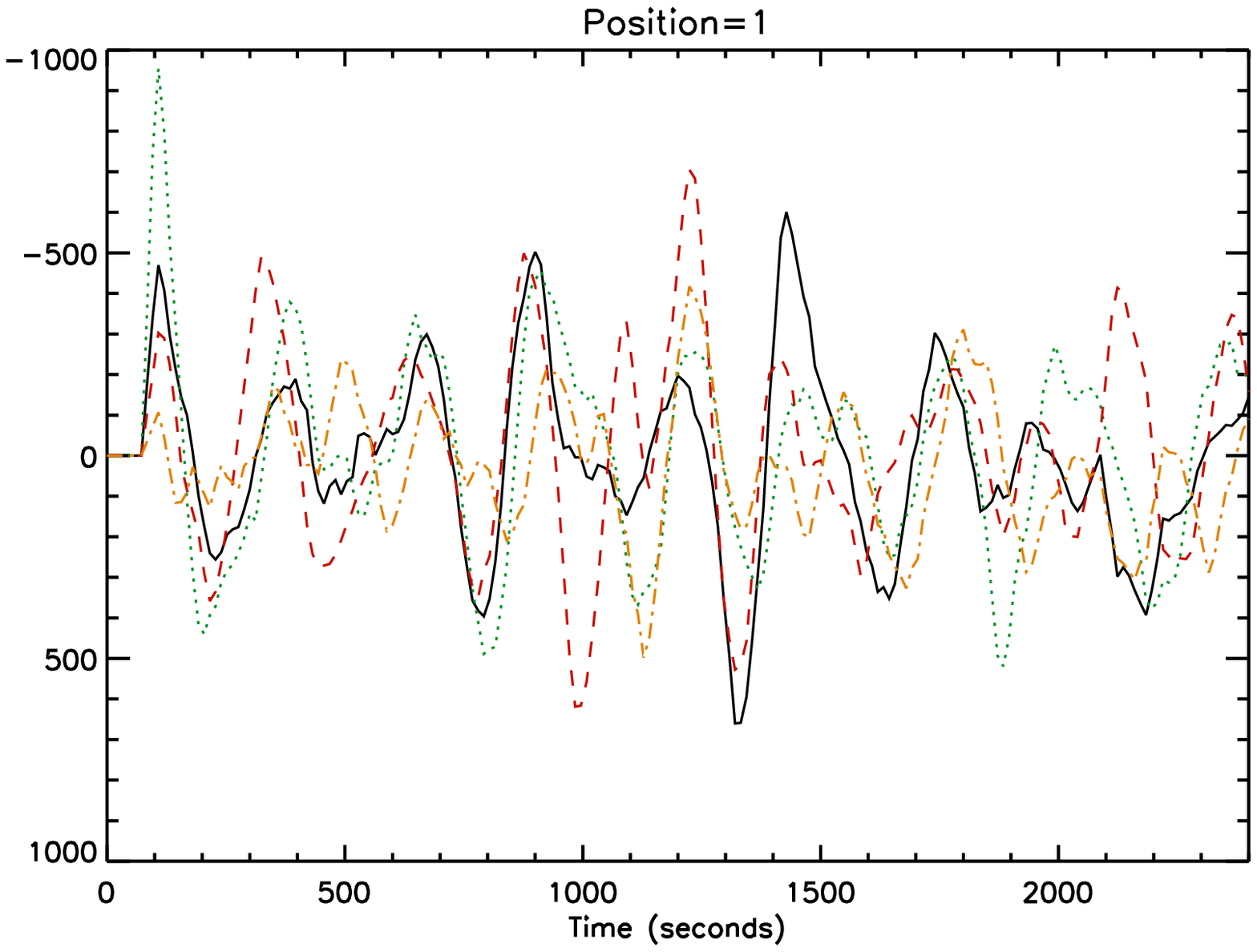}
              \hspace*{-0.03\textwidth}
            \includegraphics[width=0.400\textwidth,clip=]{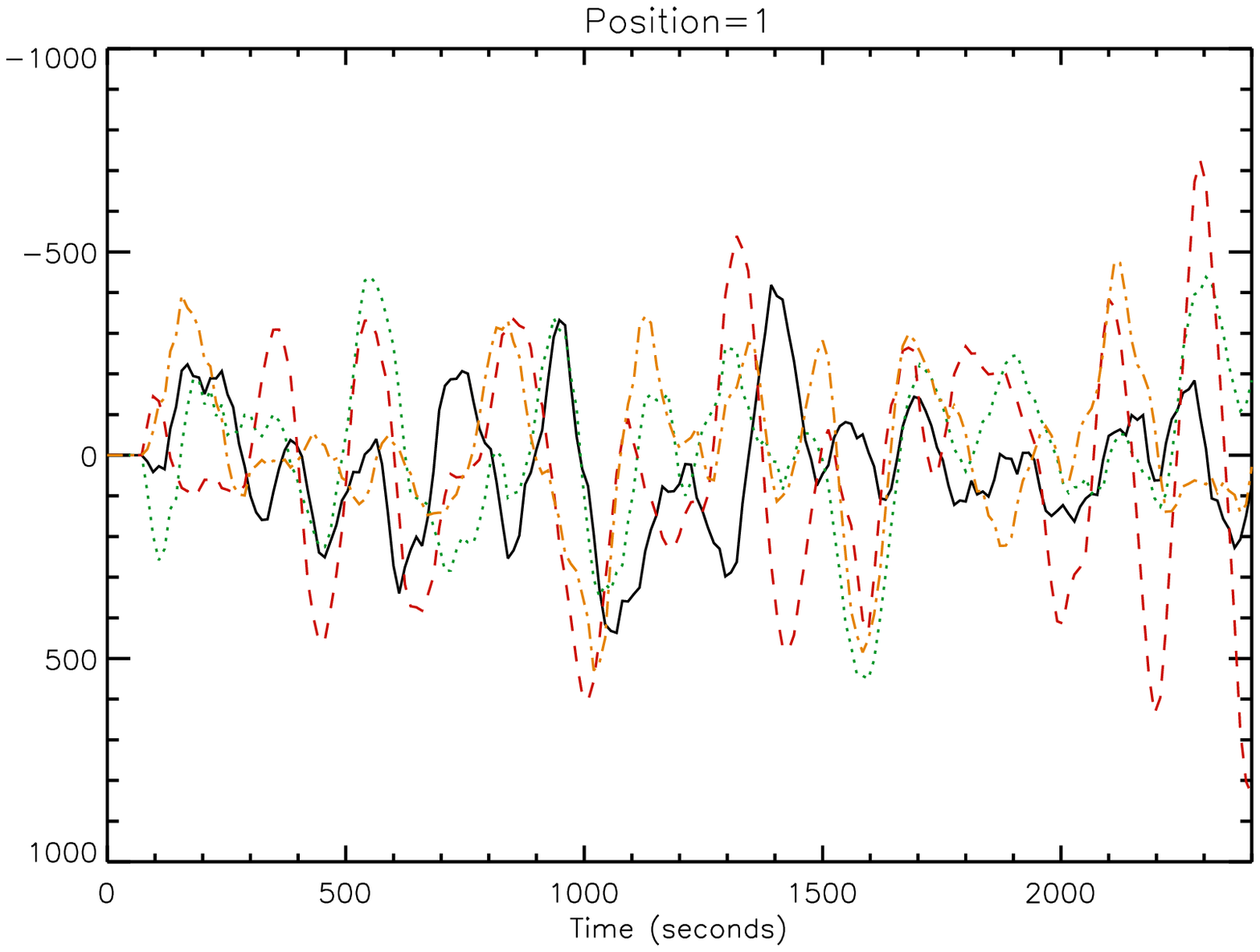}
              }
     \vspace{-0.32\textwidth}   
     \centerline{\Large \bf     
      \hspace{0.07 \textwidth}  \color{black}{\small(a)}
      \hspace{0.415\textwidth}  \color{black}{\small(b)}
         \hfill}
     \vspace{0.31\textwidth}    
   \centerline{\hspace*{0.015\textwidth}
    \includegraphics[width=0.400\textwidth,clip=]{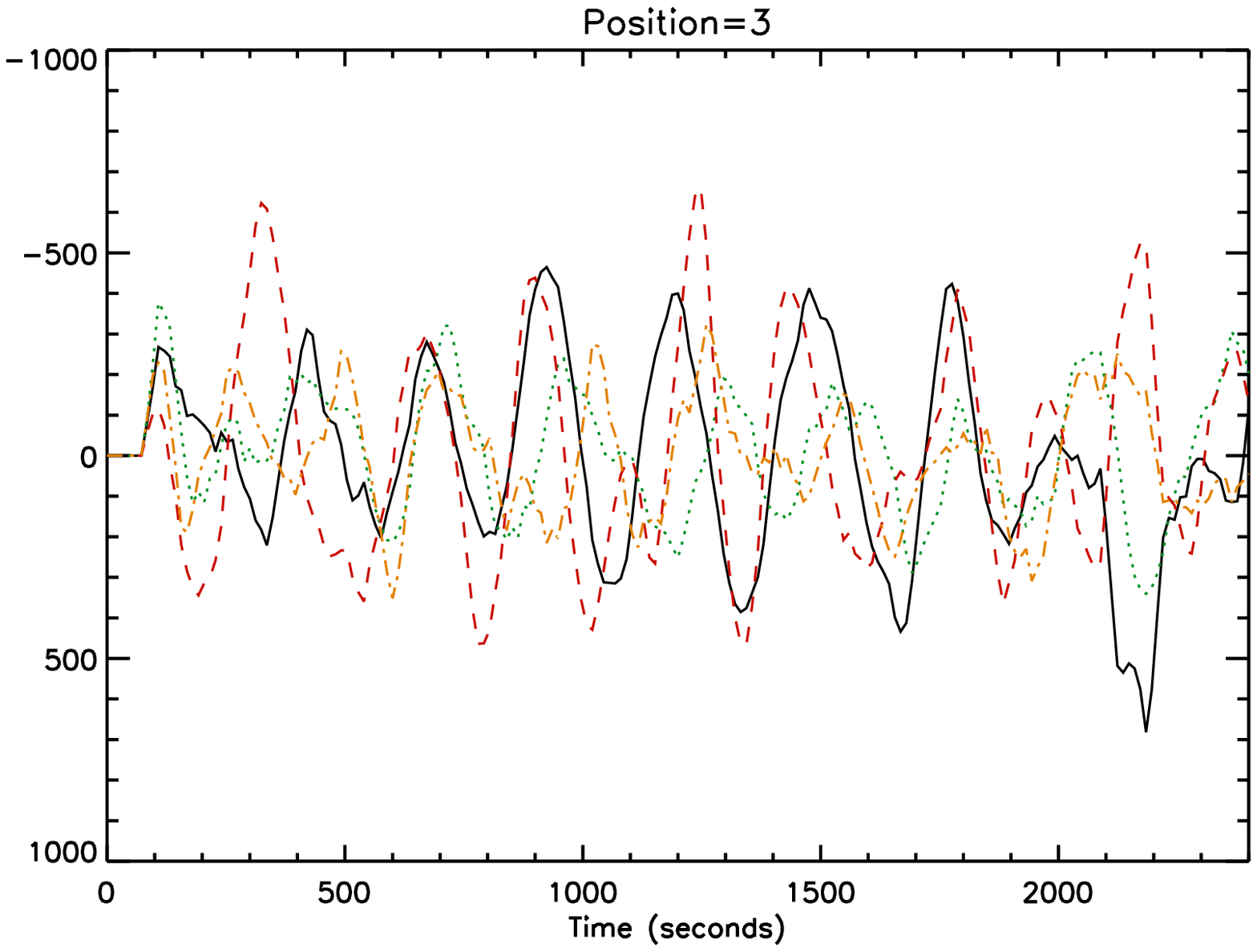}
               \hspace*{-0.03\textwidth}
            \includegraphics[width=0.400\textwidth,clip=]{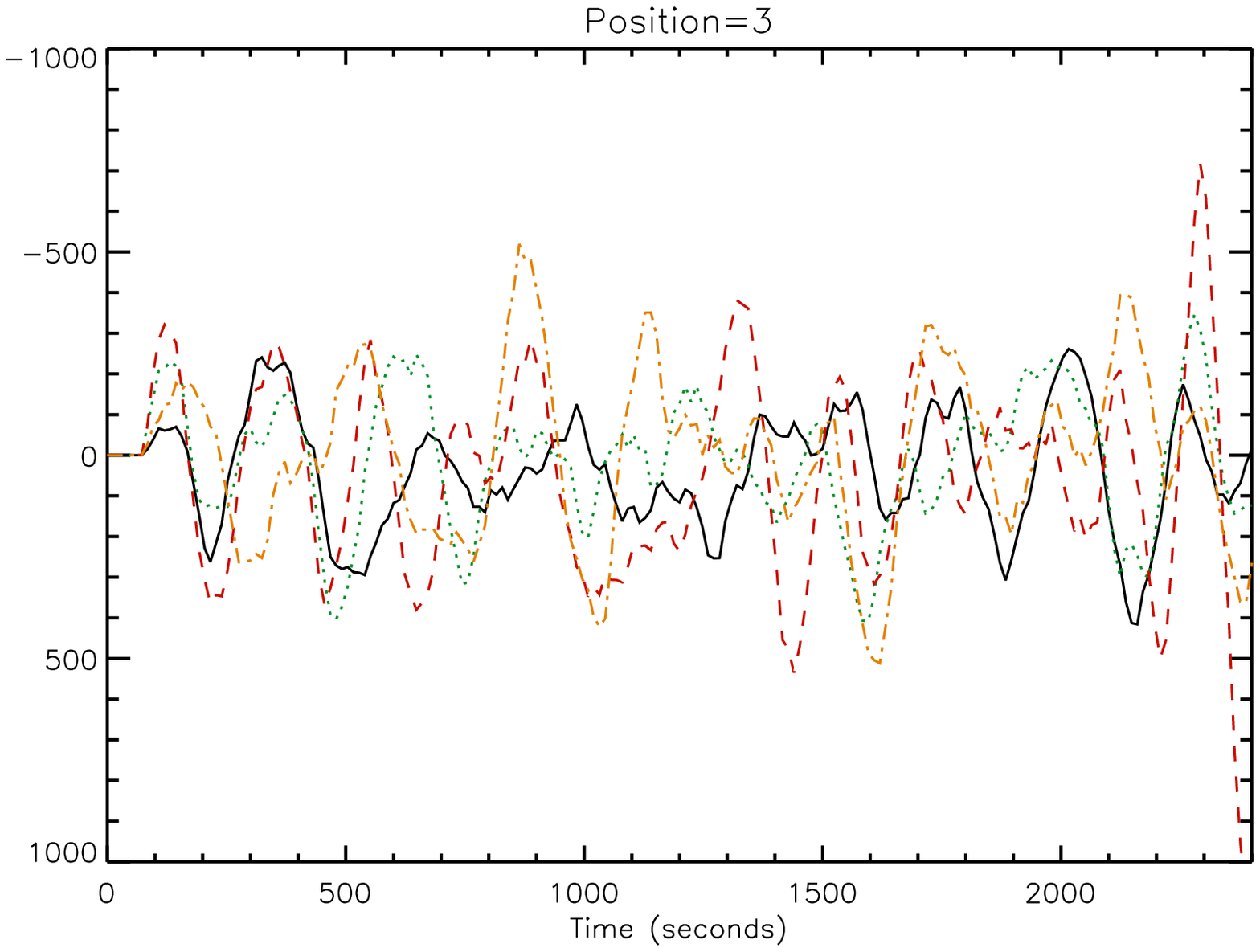}
              }
     \vspace{-0.32\textwidth}   
     \centerline{\Large \bf     
      \hspace{0.07 \textwidth} \color{black}{\small(c)}
      \hspace{0.415\textwidth}  \color{black}{\small(d)}
         \hfill}
     \vspace{0.31\textwidth}    

 \centerline{\hspace*{0.015\textwidth}
          \includegraphics[width=0.400\textwidth,clip=]{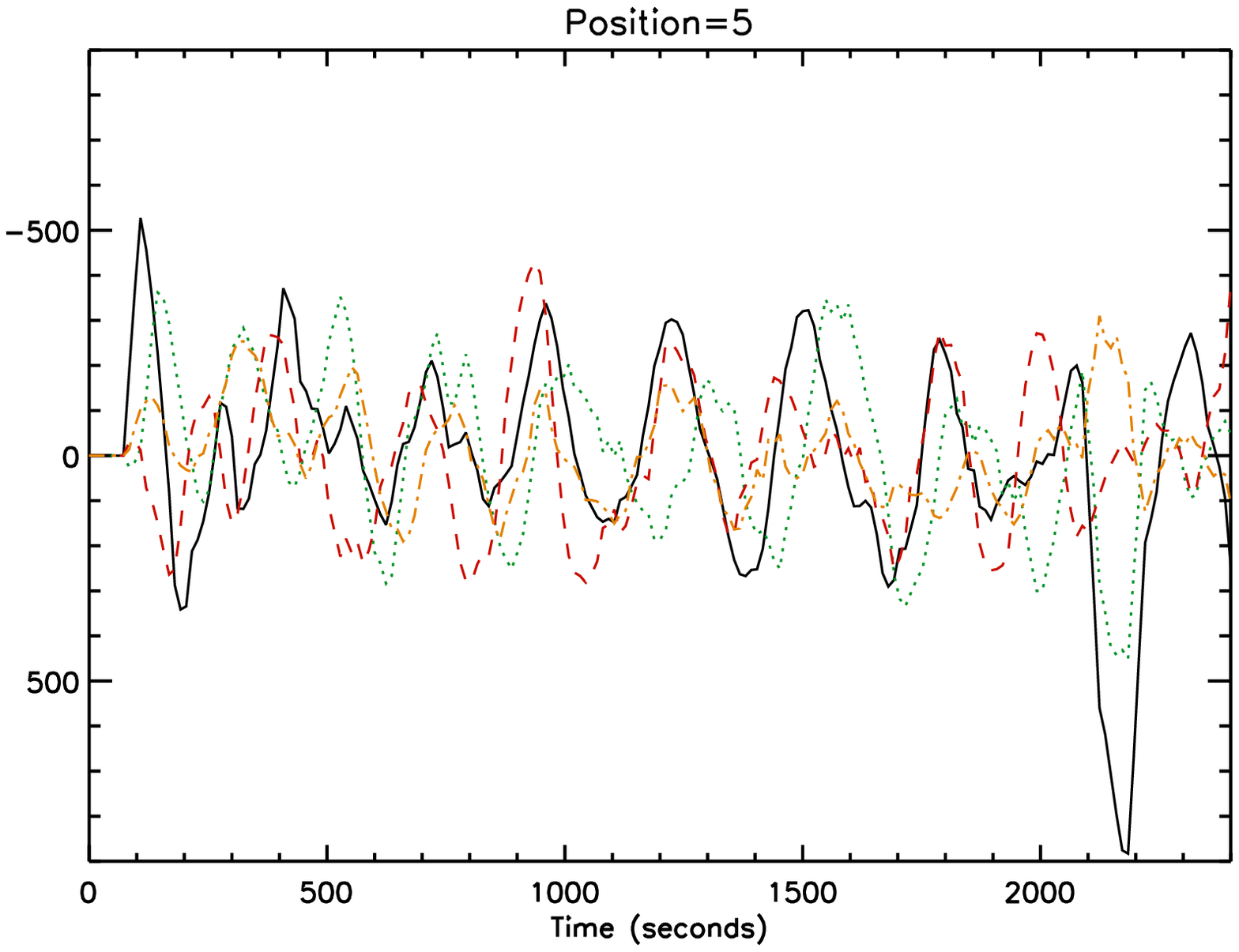}
            \hspace*{-0.03\textwidth}
          \includegraphics[width=0.400\textwidth,clip=]{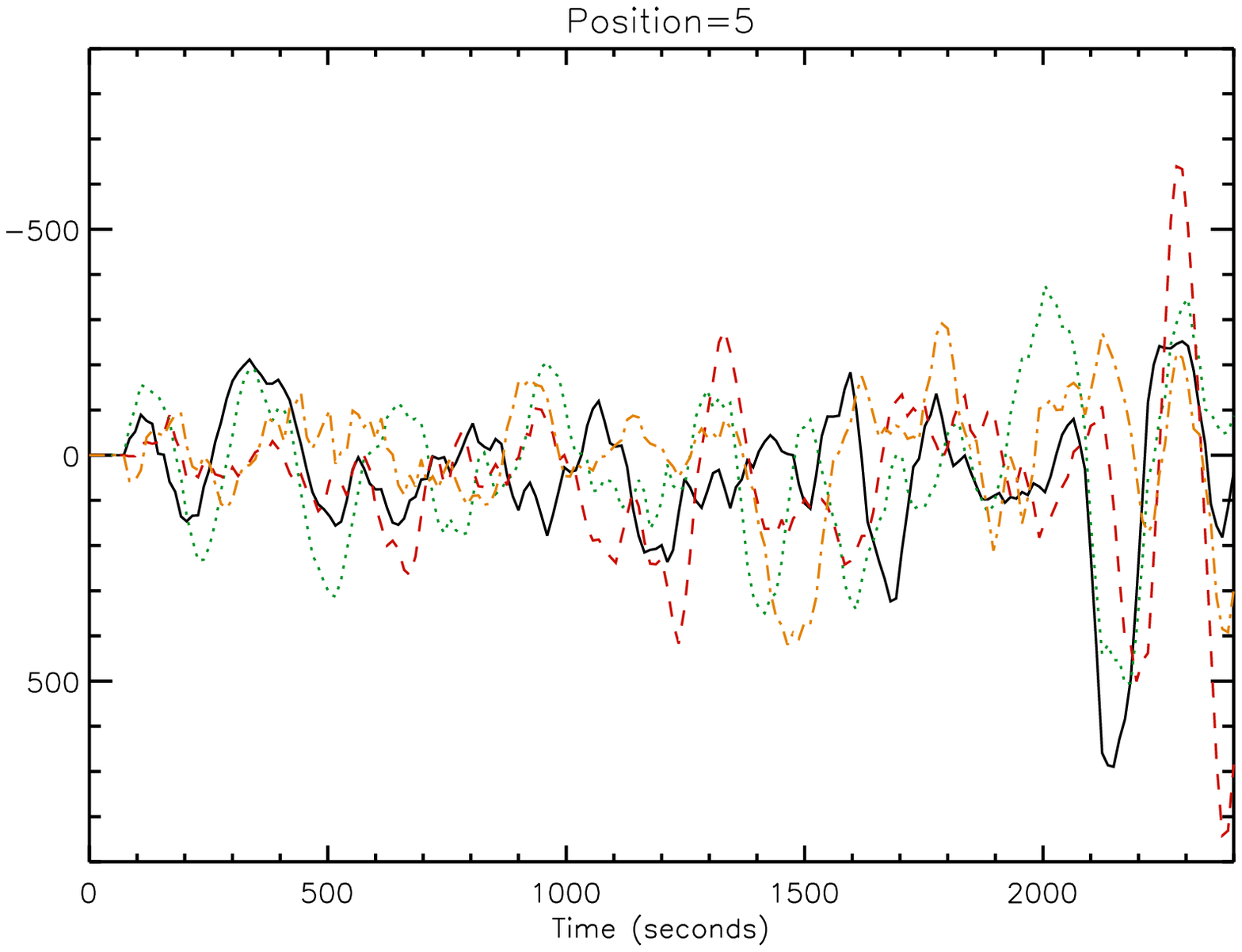}
              }
     \vspace{-0.32\textwidth}   
     \centerline{\Large \bf     
      \hspace{0.07 \textwidth} \color{black}{\small(e)}
      \hspace{0.415\textwidth}  \color{black}{\small(f)}
         \hfill}
     \vspace{0.31\textwidth}    

 \centerline{\hspace*{0.015\textwidth}
          \includegraphics[width=0.400\textwidth,clip=]{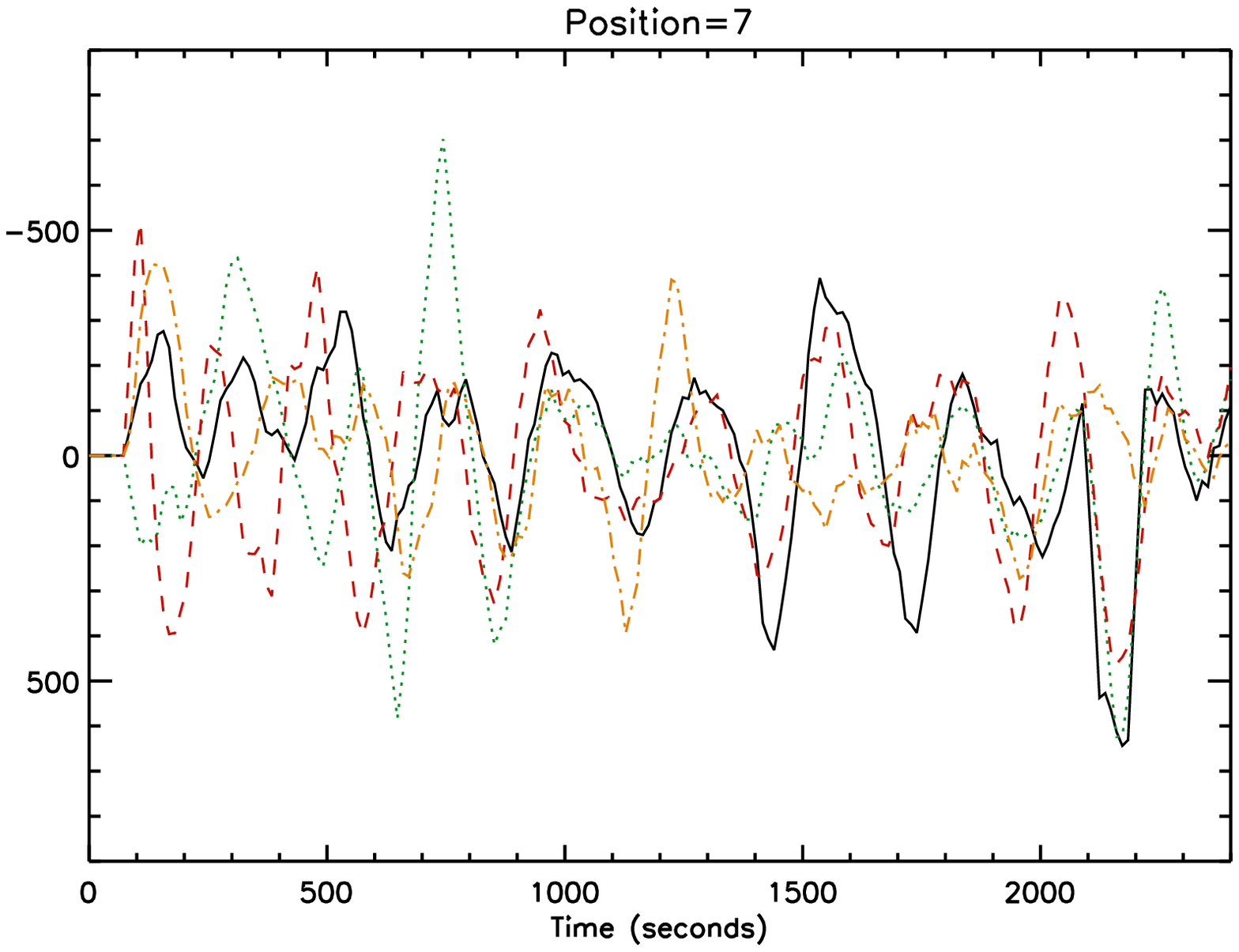}
            \hspace*{-0.03\textwidth}
          \includegraphics[width=0.400\textwidth,clip=]{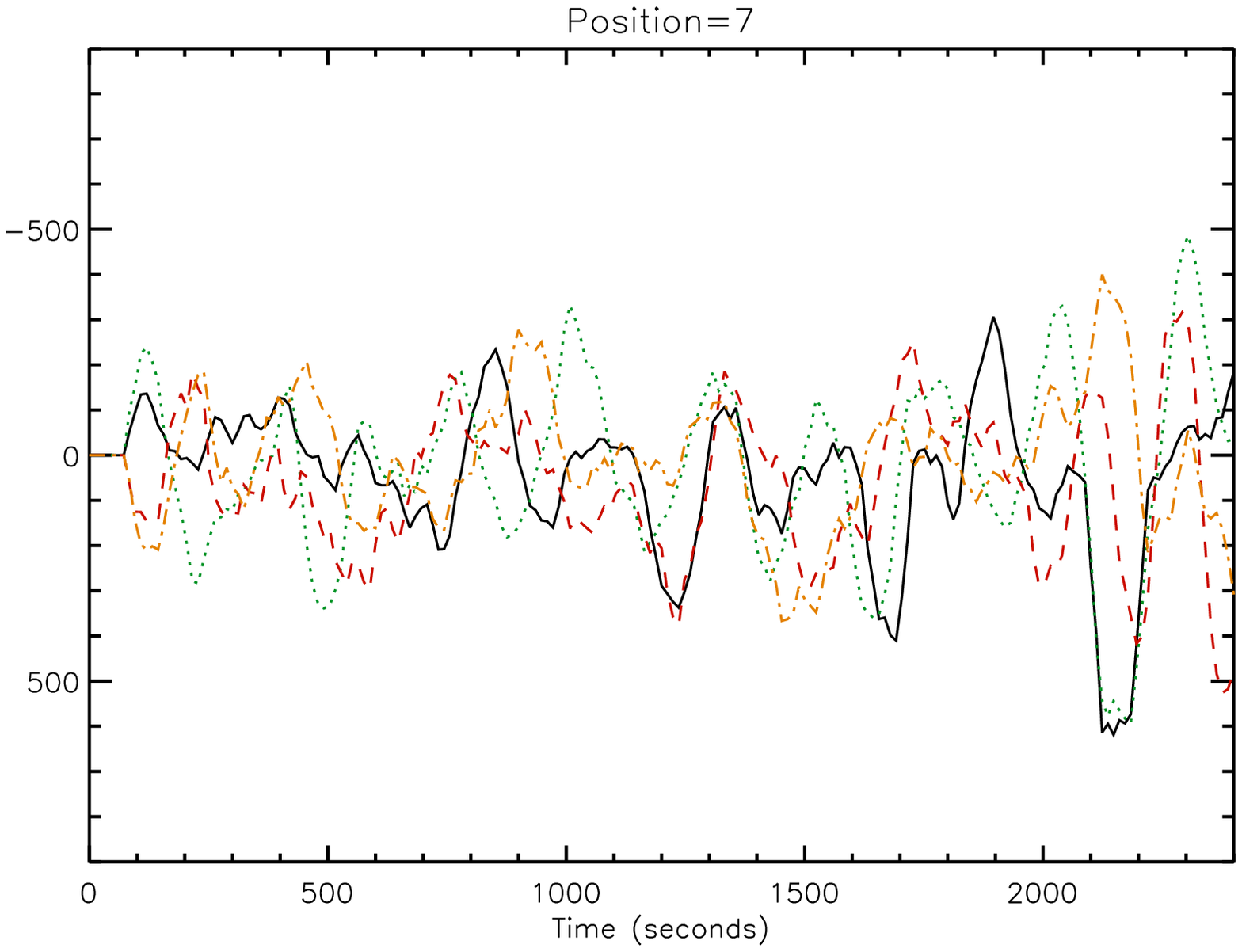}
              }
     \vspace{-0.32\textwidth}   
     \centerline{\Large \bf     
      \hspace{0.07 \textwidth} \color{black}{\small(g)}
      \hspace{0.415\textwidth}  \color{black}{\small(h)}
         \hfill}
     \vspace{0.31\textwidth}    
              
\caption{Cuts through the 171~\AA\ running-difference images for arcs 1\,--\,8 at all positions for 22 September 2011.  (a), (c), (e), and (g) show arcs 1\,--\,4 and  (b), (d), (f), and (h) show arcs 5\,--\,8.  The solid black lines correspond to arcs 1 and 5, the green dotted line to arcs 2 and 6, red dashed to 3 and 7, and orange dot--dashed to 4 and 8.}
   \label{F-phase2}
   \end{figure}

This analysis is also done on the other primary data set (22 September 2011).  We define
eight arcs (Figure \ref{F-out}) between the two lines displayed in Figure \ref{F-out}, with arc 1 again closest to the top of the image and arc 8 defined closest to the bottom.  Running
differences are constructed in the usual way and cuts are taken at several positions and are displayed in Figure \ref{F-phase2}.

Each line in Figure \ref{F-phase2} is defined in the same way as Figure
\ref{F-phase}. We can see that arcs 6\,--\,8 (green dotted/red dashed/orange dot--dashed in the right-hand column of graphs) match very well for all positions. Arc
5 appears to match well at some times but is completely out of  phase at
others.  Arcs 1, 2, and 4 are almost in phase at position 1, whilst the red dashed line (arc 3) is out
of phase at certain times.  At positions 3 and 5, the arc 2 line (green dotted line in left hand column) is approximately
in phase with the others and at position 7 there is some evidence of them
starting to become out of phase.  We have calculated the cross correlation for this example and the results are
displayed in Table \ref{T-correlate2}.

\begin{table}[!h]
\caption{ Cross correlation between 171~\AA\ arcs at position 1 along the loop for 22 September 2011. The subscript denotes the lag (in units of 12 seconds) where the maximum correlation is found.}
\begin{tabular}{c c c c c c c c c}
\hline
 Arc & 1 & 2 & 3 & 4 & 5 & 6 & 7 & 8 \\
[0.5ex]
\hline
1 & 1 & $0.659_1$ & $0.506_0$ & $0.409_{-29}$ & $0.095_0$ & $0.252_{-6}$ &
$0.274_{-9}$ & $0.170_{-6}$\\
2 &  & 1 & $0.406_{-1}$ & $0.406_1$ & $0.081_{11}$ & $0.393_{-7}$ & $0.173_{-9}$ &
$0.098_{27}$ \\
3 &  &  & 1 & $0.244_9$ & $0.106_{11}$ & $0.161_{-7}$ & $0.229_{-3}$ & $0.229_{-1}$ \\
4 &  &  & & 1 & $0.156_{14}$ & $0.398_{-9}$ & $0.316_{-11}$ & $0.476_{-8}$\\ 
5 & &  & & & 1 & $0.144_1$ & $0.226_{12}$ & $0.340_{-5}$\\
6 & &  & & & & 1 & $0.403_{-2}$ & $0.267_0$\\
7 & &  & & & & & 1 & $0.389_1$\\
8 & &  & & & & & & 1 \\ [1ex]
\hline
\end{tabular}
\label{T-correlate2}
\end{table}

There is no clear pattern in Table \ref{T-correlate2}. On average, arcs correlate better
with arcs located close to them.  There are exceptions though.  For example, arc 4 only has a high correlation
with arc 2, arcs 6\,--\,8 on average are well correlated with most maximum
correlations occurring within one or two time frames.  However, the correlation values are overall
lower than in the previous case, which could be due to the arc footpoints not lining up very well. The lags associated with arcs 1 and 4 and arcs 2 and 8 are large but these do not give a true reflection of the correlation between the two arcs as a lag in this range corresponds to approximately a full period.  These arcs are actually in phase as seen in Figure \ref{F-phase2}. The dominant periods and characteristic velocities for this example are
displayed in Table \ref{T-Seven}.

\begin{table}[!h]
\caption{Table showing the periods and characteristic 171~\AA\ velocity for arcs 1\,--\,8 for 22 September 2011. }
\label{T-Seven}
\begin{tabular}{c c c}     
  \hline                   
Arc & Period(s) & Char. Velocity(km$s^{-1}$)   \\
  \hline
1& 250\,--\,300 & 74  \\
2 & 270\,--\,320 & 76 \\
3 & 270\,--\,320 & 61 \\
4 & 260\,--\,310 & 93   \\
5 & 200\,--\,290 & 88  \\
6 & 300\,--\,360 & 84  \\
7 & 270\,--\,300 & 109  \\
8 & 270\,--\,320 & 91  \\
  \hline
\end{tabular}
\end{table}

As in the previous example the dominant periods are approximately constant for
each of the arcs.  As we would expect given this example is not a sunspot region, the
dominant periods are longer and closer to five minutes (300 seconds).  The velocities for this
example seem to increase slightly as we move down the arcs. 

The periods of the PDs are constant across the two active regions in both cases.  The velocities stay approximately constant with some variation, where the small variations could possibly be due to changes in the inclination angles.
     
\section{Removing the Cool Emission from the 193~\AA\ Passband}
\label{S-Cool}

As shown by Del Zanna \etal (2011) using simultaneous {\it Hinode}/EIS spectra and SDO/AIA images, AR loop legs produce strong  Fe \textsc{viii} and Fe \textsc{ix}
``cool'' emission dominating the 131 and 171 ~\AA\ bands.
The 193~\AA\ band is multithermal, in that
strong emission from  Fe \textsc{viii} and Fe \textsc{ix}  lines alongside
Fe \textsc{xi} and Fe \textsc{xii} is observed. Weak  emission from
a range of even lower temperature lines (mostly from O \textsc{v} and Fe \textsc{vii})
is also present.
As described by Del Zanna \etal (2011), the atomic data for
Fe \textsc{ix}, Fe \textsc{xi}, and Fe \textsc{xii} are relatively well understood,
while the Fe \textsc{viii} data are more uncertain. The Fe \textsc{vii} data are
very uncertain and have not yet been included in the CHIANTI database
\cite{landi12}.

We have devised a rough method to
 estimate the main cool contribution (from  Fe \textsc{viii} and  Fe \textsc{ix})
to the 193~\AA\ passband, in order to subtract it, and study the
properties of the hot (T $>$ 1 MK) emission in the band.
 The loop legs we have chosen have strong emission in the 131~\AA\  and 171~\AA\
passband, formed in the log T[K]= 5.5\,--\,5.9 range.
There is observational evidence based on spectroscopy that at each location
the plasma distribution in loop legs is nearly isothermal
({\it e.g.}  \inlinecite{delzanna03},\inlinecite{delzanna11},\inlinecite{tripathi09}).
As a first approximation, it is therefore reasonable to assume that
at each location the plasma is isothermal.
With this assumption, we then estimate the isothermal temperature
and emission measure of the main cool component for each pixel using the
observed  171~\AA\  and 131~\AA\ count rates and the
respective response functions calculated using CHIANTI v.7 \cite{landi12} (The 193 response function has been calculated using the method outlined in \inlinecite{delzanna11}).
We simply divide the observed counts by the responses and take the
intersection of the curves (see Figure 12) as the estimate of the isothermal temperature.

\begin{figure}[!h]
 \centering
\scalebox{0.32}{\includegraphics{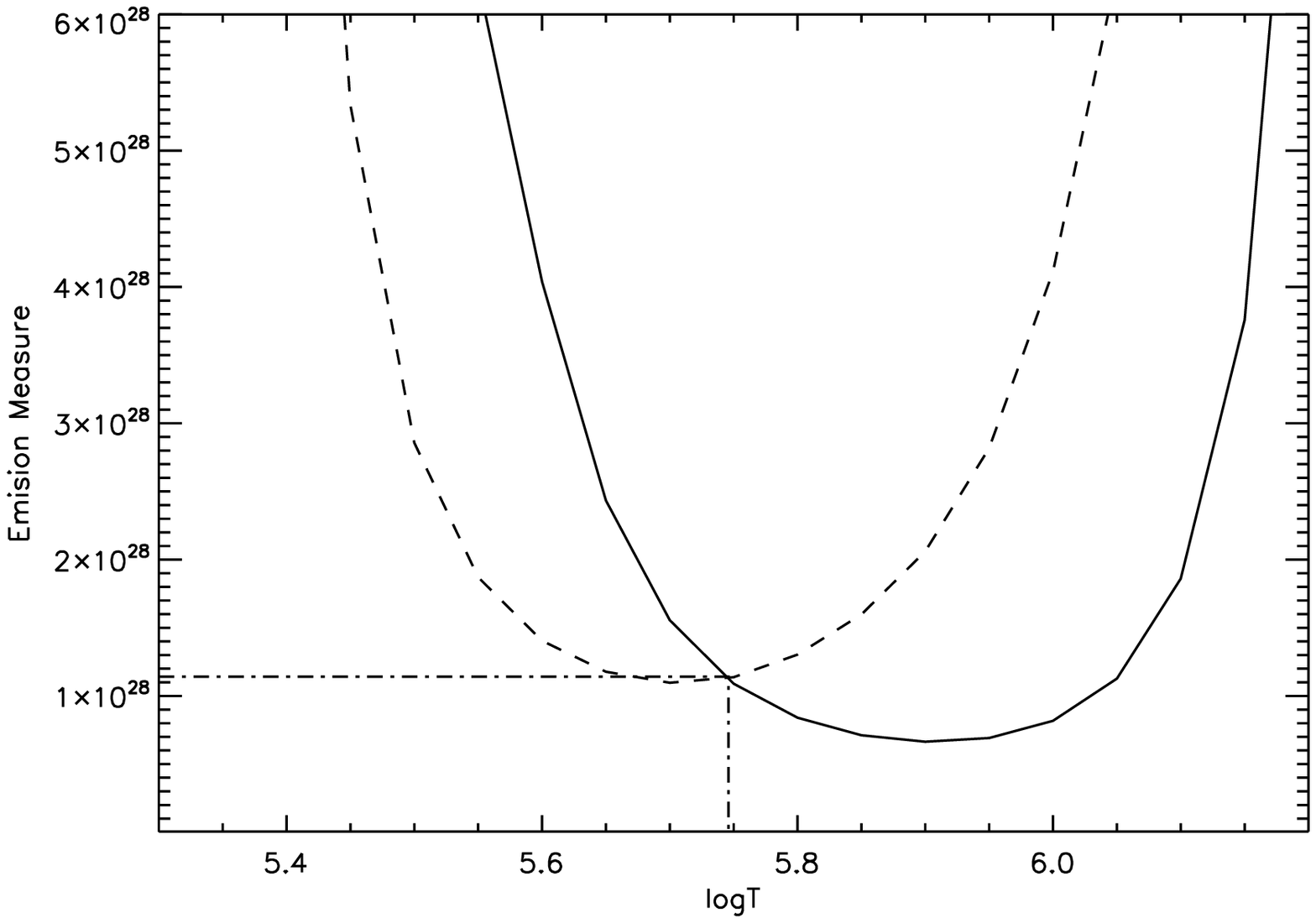}}
\scalebox{0.32}{\includegraphics{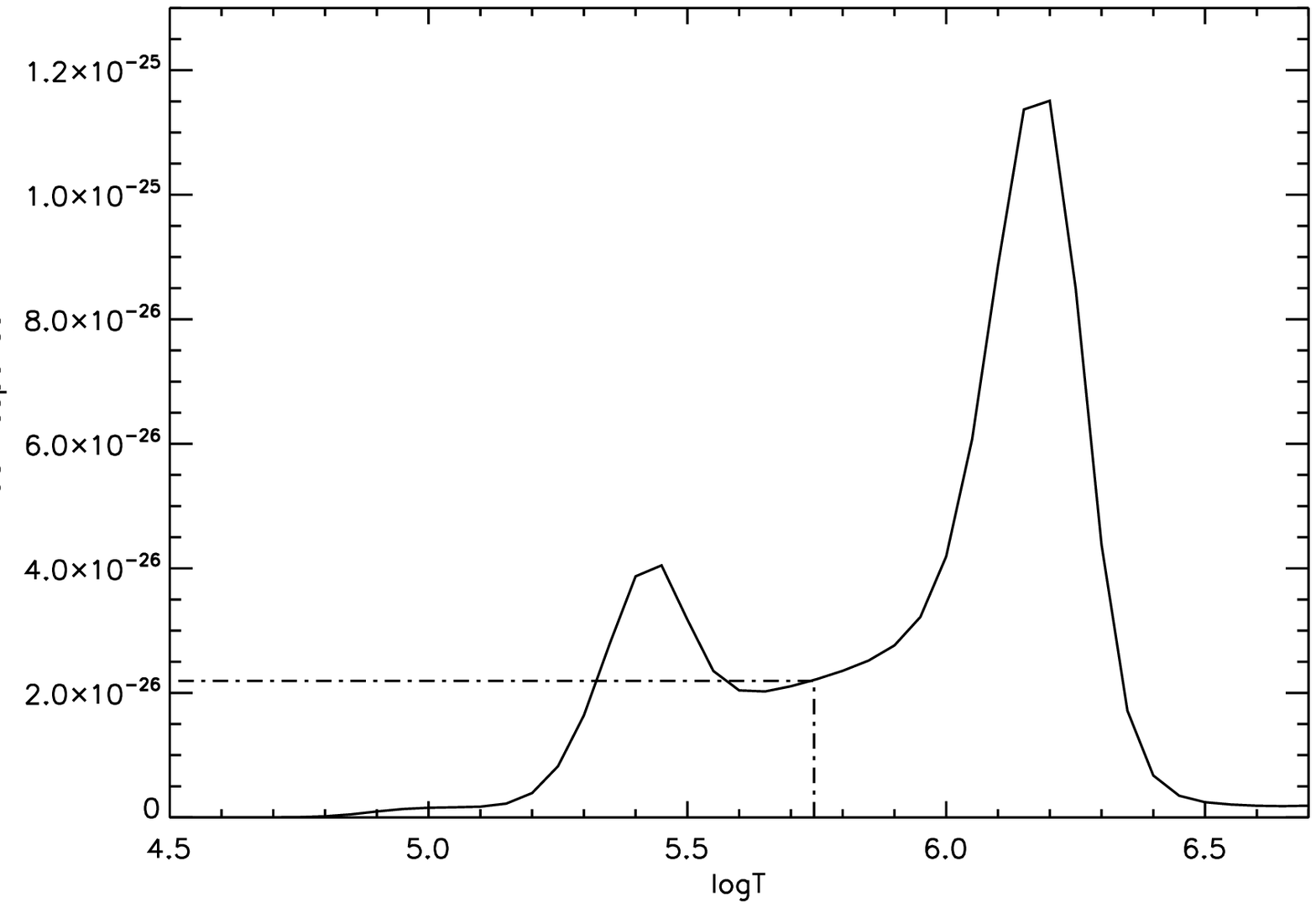}}
\label{emiss}
\caption{Left:Emission measure loci curves for a pixel in 171~\AA\ (solid line) and 131~\AA\ (dashed line).  The dot-dashed line from the {\it x}-axis indicates the value of the isothermal temperature.  The dot--dashed line that crosses the {\it y}-axis indicates the value of the emission measure at the isothermal temperature.  The pixel is located at the loop footpoint and is from the sunspot example (22 June 2011).  Right: the temperature response function for the 193 ~\AA\ passband. The dot--dashed line from the {\it x}-axis indicates the value of the isothermal temperature.  The dot dashed line that crosses the {\it y}-axis indicates the value of the response function at the isothermal temperature } 
\end{figure}

The method is basically the Emission Measure Loci
one (\inlinecite{delzanna03}) applied to the AIA  bands.
We define the value where the dot--dashed line in the left plot in Figure 12 cross the {\it x}-axis as the isothermal temperature [$T_i$].  The value of the emission measure at the isothermal temperature  (the value where the dot--dashed line in the left plot crosses the {\it y}-axis) as $E(T_i)$ and the value of the 193~\AA\ response function at the isothermal temperature (the value where the dot--dashed line in the right plot crosses the {\it y}-axis) as $R(T_i)$. We can now estimate
the contribution (in DN$s^{-1}$) to the 193~\AA\ band due to the Fe \textsc{viii} and Fe \textsc{ix}
lines [C] by $C=E(T_i) R(T_i).$ for a given pixel.
This gives us an estimate of the main cool contribution for a given pixel to the 193~\AA\ emission.
This is a lower estimate, given that it does not take into account lower temperature
emission.  \inlinecite{delzanna11} measured the cool emission in the 193~\AA\ band in
loop legs and footpoints to be as large as 40\%.
The present estimates provide a range of somewhat lower
(but still significant) values, from about 10 to 40\%.
Once the cool emission is subtracted, we expect the dominant emission in the
193~\AA\ band to be originating from Fe \textsc{xi} and Fe \textsc{xii} lines, i.e.
from 1\,--\,2 MK plasma. We refer to this as the ``hot'' emission in the 193~\AA\ band.
The procedure was  automated for all pixels in all  193~\AA\
images and the cool contribution subtracted.

\subsection{22 June 2011 (Sunspot)}

\begin{figure}[!h]
\centering
\scalebox{0.32}{\includegraphics{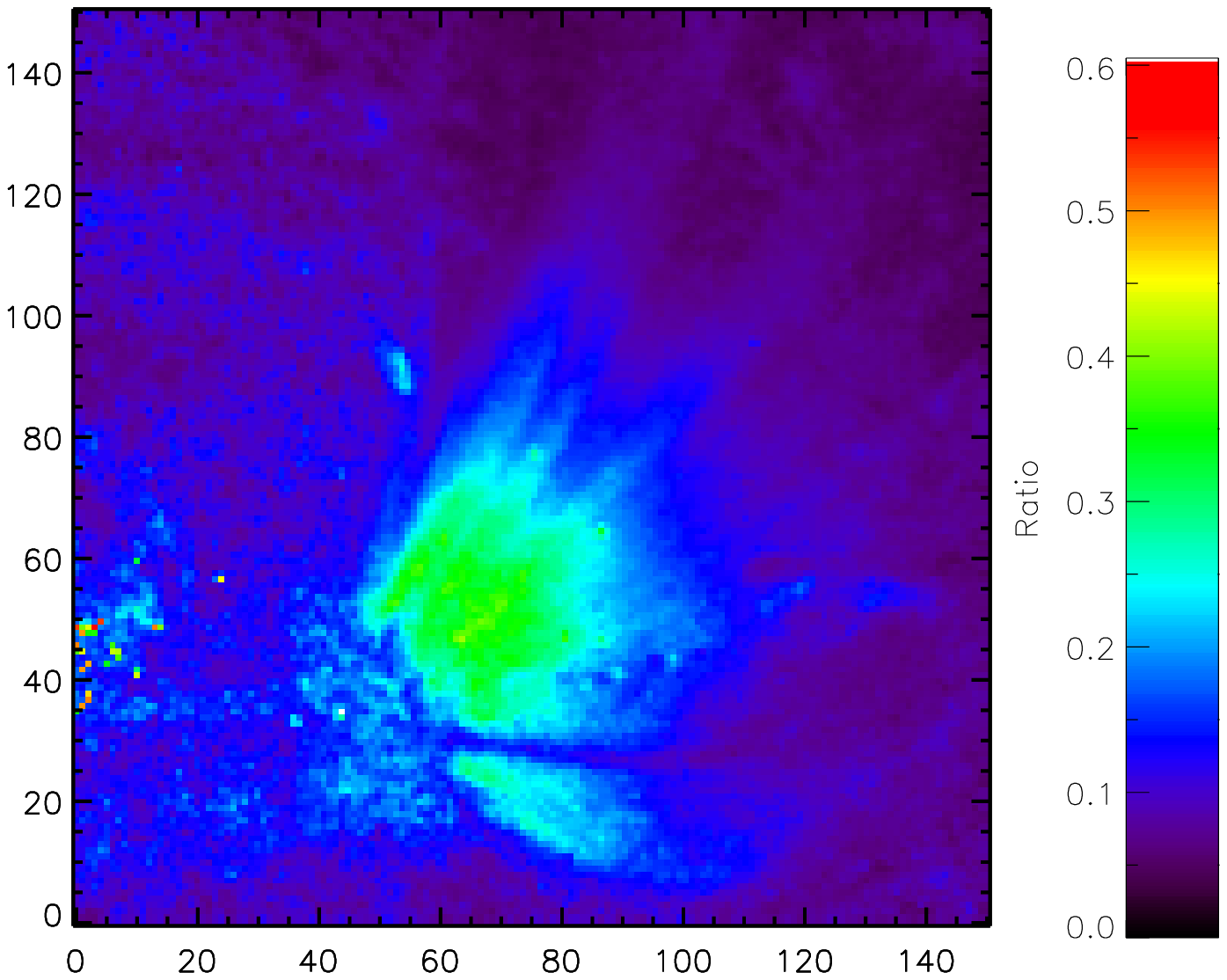}}
\scalebox{0.32}{\includegraphics{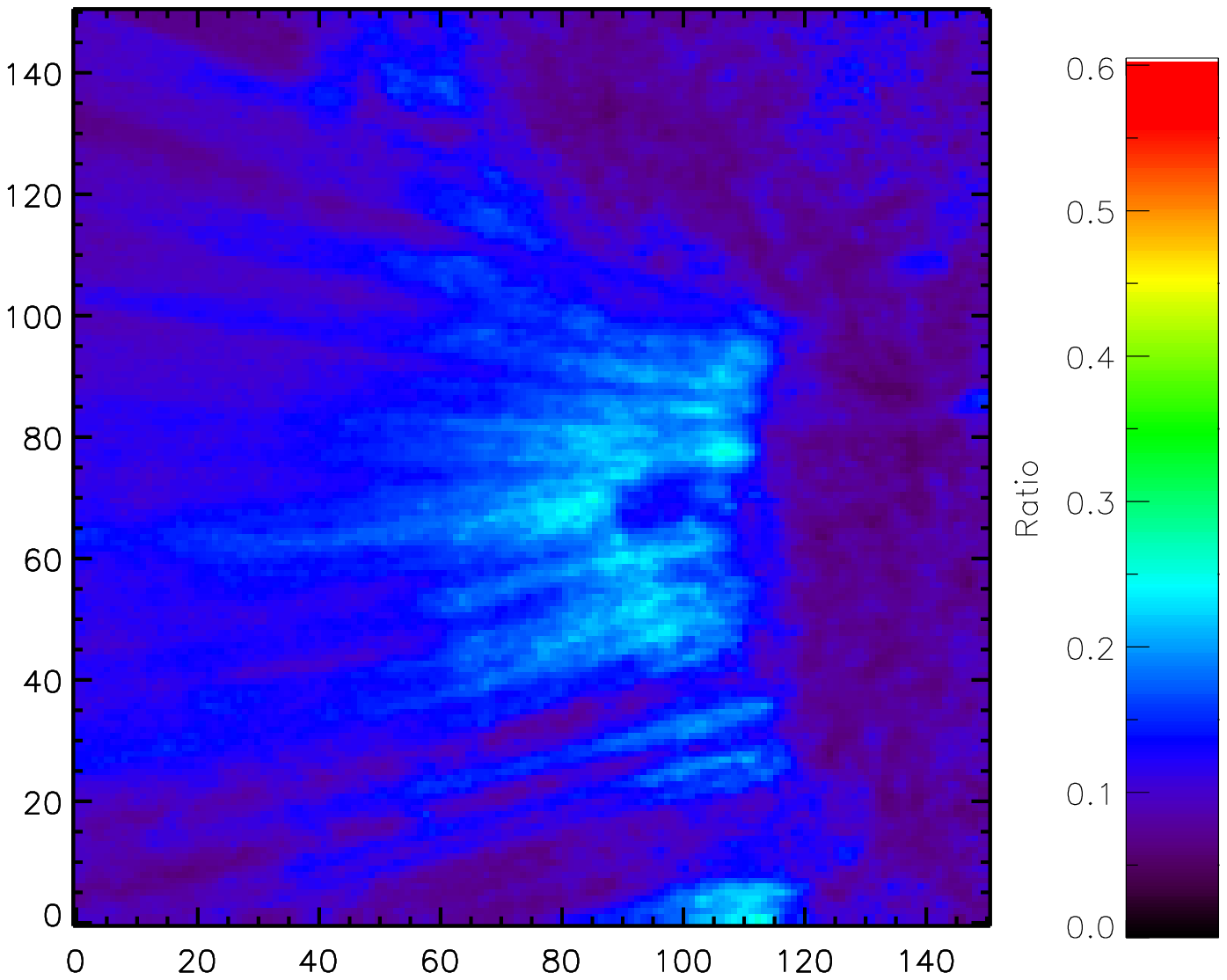}}
\caption{The ratio of the calculated cool emission to the full 193~\AA\ emission for (left) the sunspot region (22 June 2011) and (right) the non-sunspot region (22 September 2011).}
\label{F-ratio}
\end{figure}

We have applied this technique to the two primary data sets analysed in
Sections \ref{S-Propspeed} and \ref{S-Active} to  investigate how the
properties of the PDs change in the 193~\AA\ passband.  We first plot the ratio of the cool contribution to the full emission for both examples (Figure \ref{F-ratio}).

We can see from Figure \ref{F-ratio} that there is a greater percentage of the cool emission at the sunspot example than the non-sunspot one.  At the sunspot example the cool contribution accounts for about 30\,--\,40\% of the full emission,  compared to 15 - 25\% in the non-sunspot example.    

\begin{figure}[!h]
\centering
\scalebox{0.32}{\includegraphics{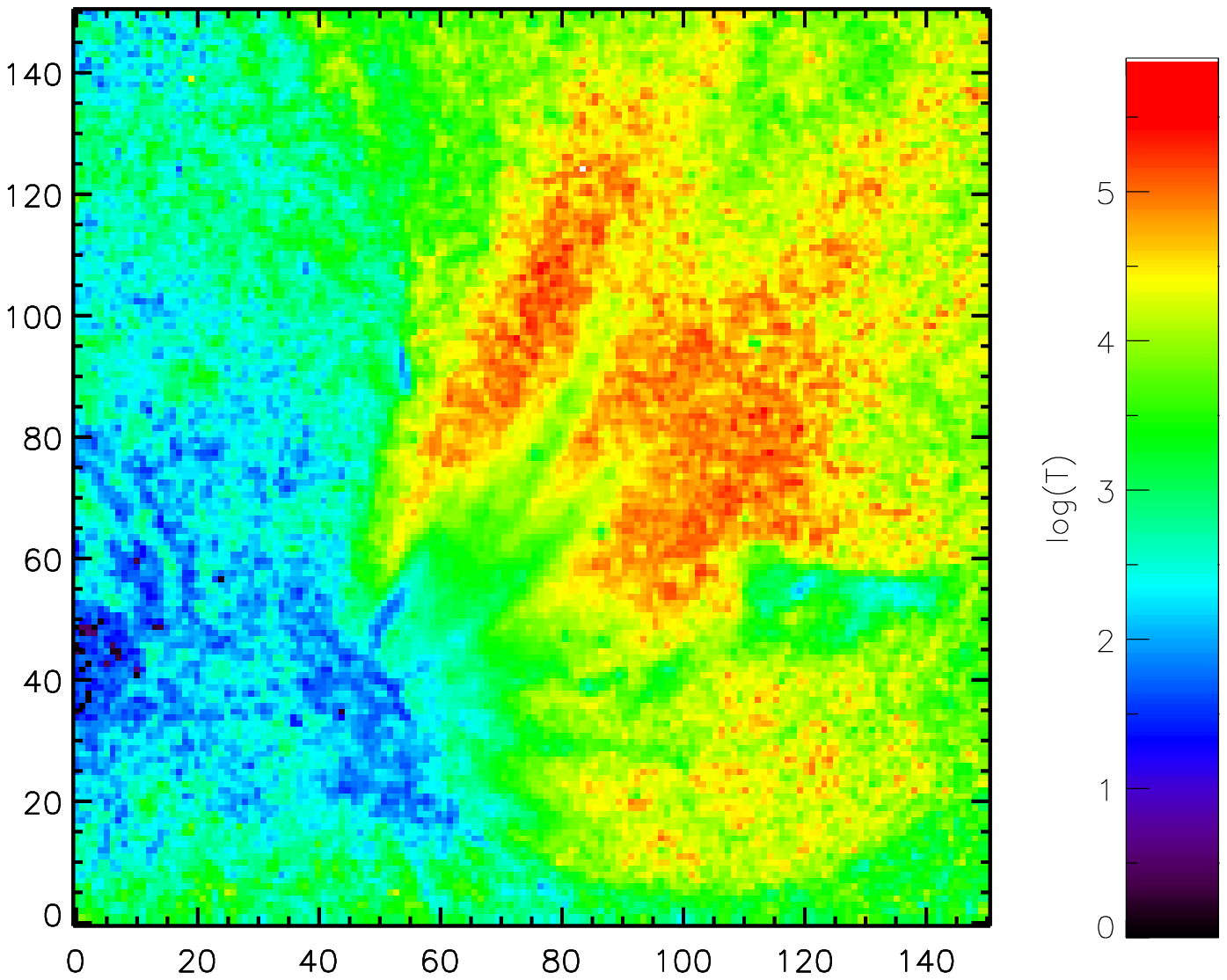}}
\scalebox{0.32}{\includegraphics{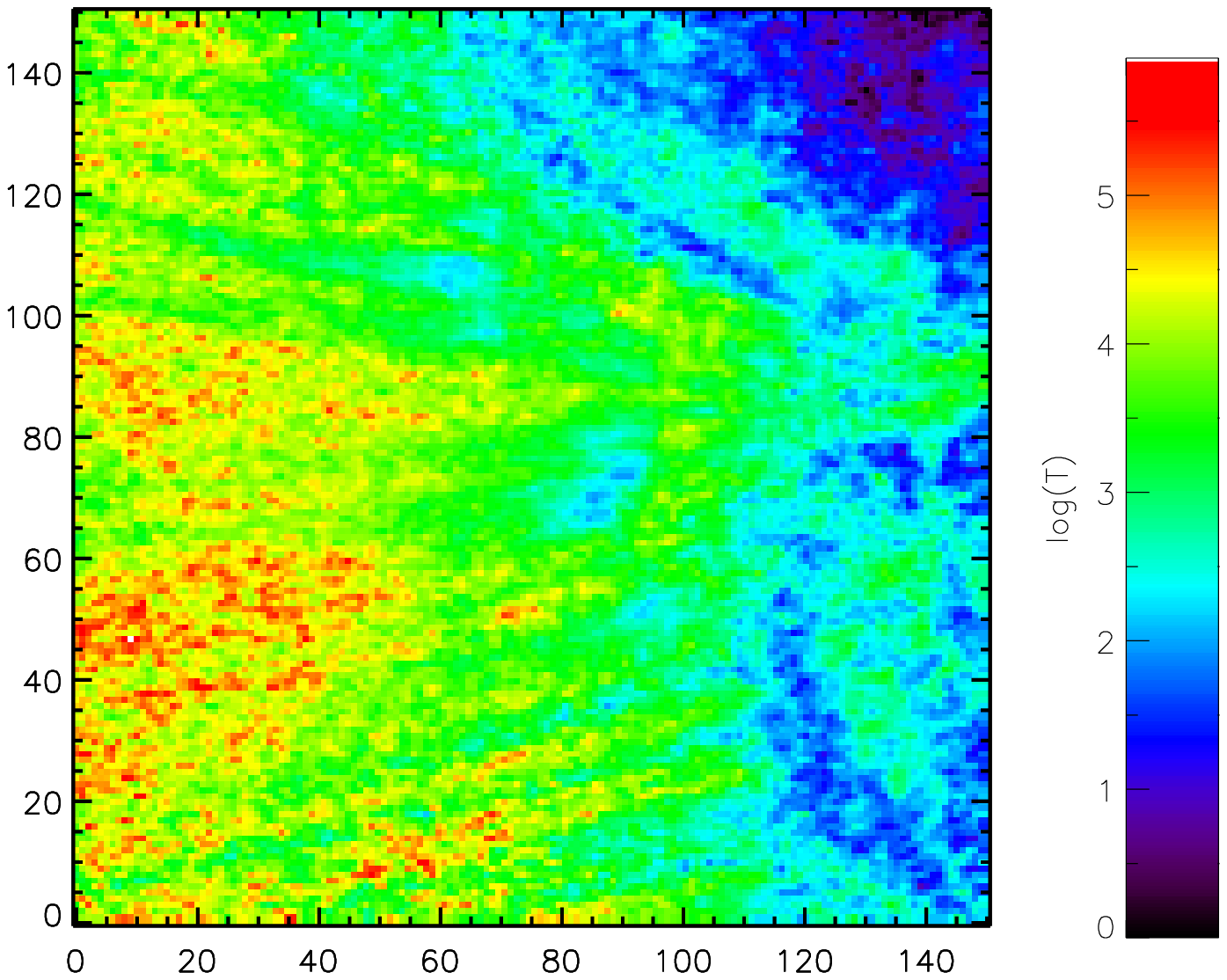}}
\caption{The calculated isothermal temperature for the sunspot region (22 June 2011) and (right) the non-sunspot region (22 September 2011).}
\label{F-isotherm}
\end{figure}

The isothermal temperature is plotted in a similar way (Figure \ref{F-isotherm}).

From Figure \ref{F-isotherm} it is clear that there is a general increase in the (isothermal) temperature along the loops.  This trend is observed in both the sunspot and non-sunspot examples.

\begin{figure}[!h]
\centering
\scalebox{0.32}{\includegraphics{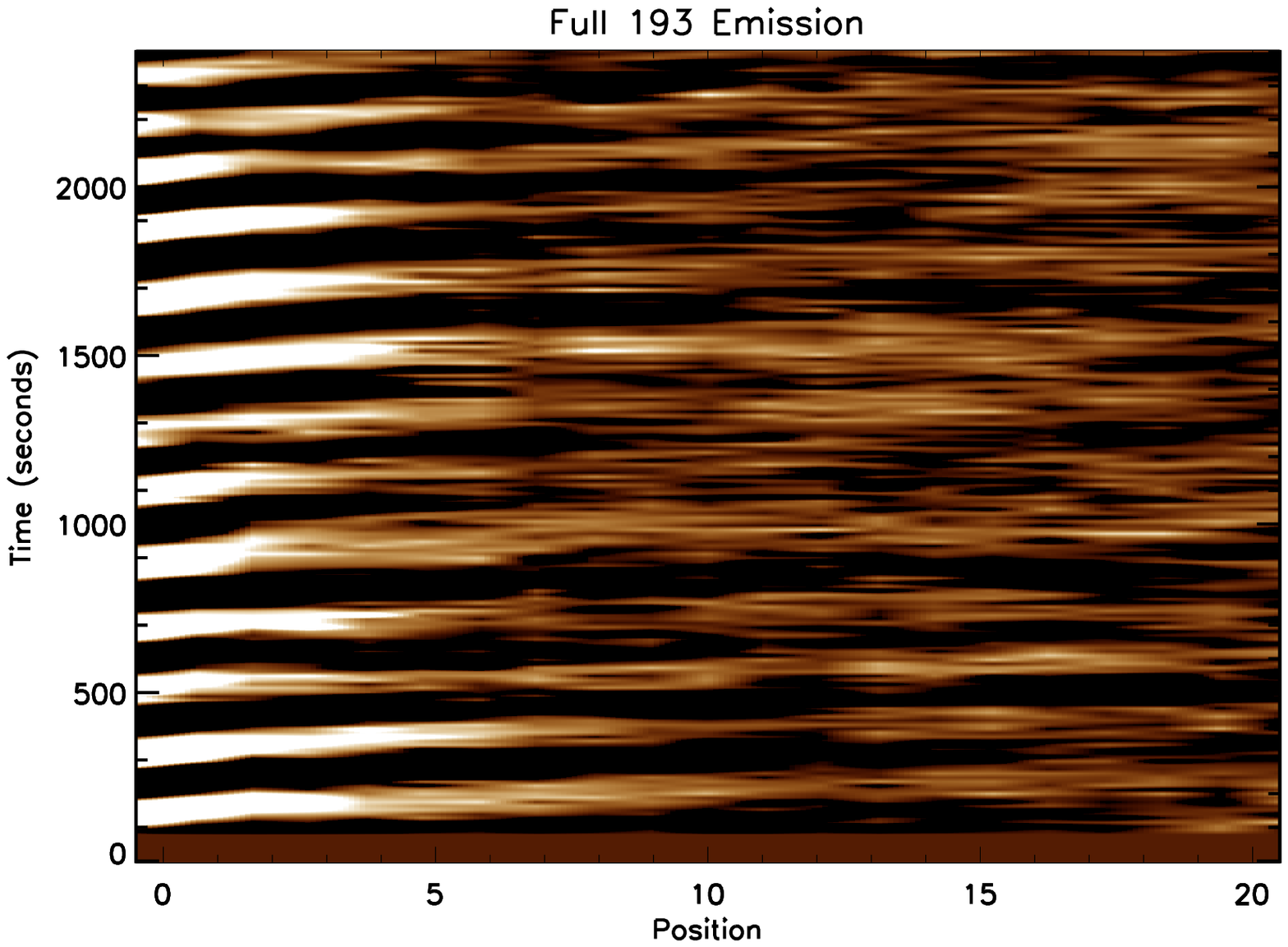}}
\scalebox{0.32}{\includegraphics{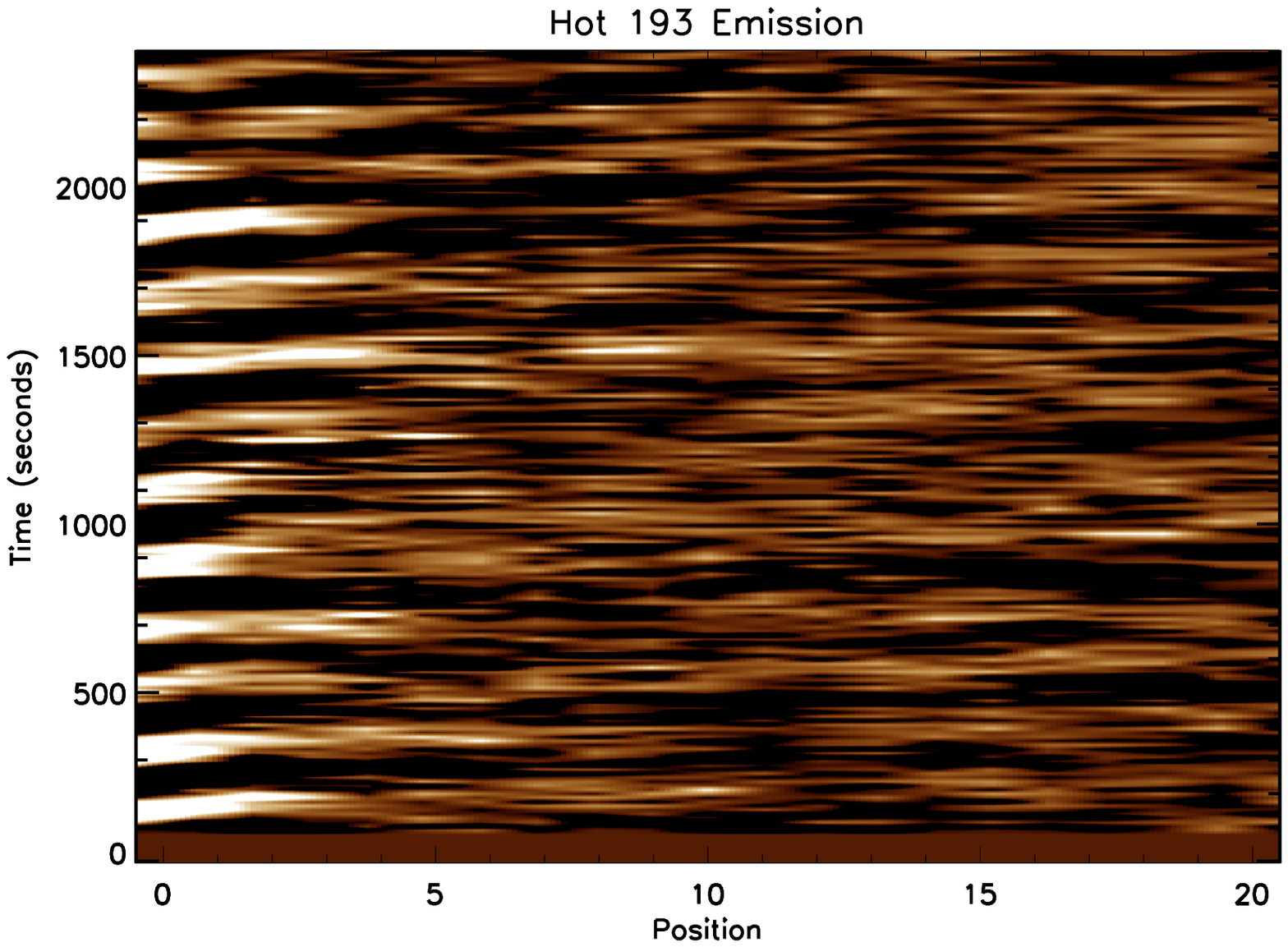}}
\caption{Running-difference images for the loop outlined in Figure \ref{F-run_diff_22/06}.  The left shows the running-difference associated with the full
193~\AA\ emission.  The right hand plot shows the running-difference associated
with the hot 193~\AA\ component only.}
\label{F-compare}
\end{figure}

We now compare the properties of PDs in the hot emission to those in the full emission.  Figure \ref{F-compare} shows running-difference images created using the same data set
as Figure \ref{F-run_diff_22/06}, associated with the full
emission (left) and the hot component only (right).

We can see from Figure \ref{F-compare} that there are clear differences in the PDs.  PDs associated with the hot emission only propagate to
positions 5\,--\,7 before they are no longer distinguishable compared to positions 12\,--\,14 in the
full emission case.  We calculated the velocities of the PDs in the cool, full, and hot data sets using the same methods as Section \ref{S-velocities} and they are displayed in Table \ref{T-Four}.

\begin{table}[!h]
\caption{Characteristic velocities associated with running-difference images for the full, hot, and cool 193~\AA\ emission for 22 June 2011, calculated using methods 1,2, and 3.}

\begin{tabular}{c c c c}
\hline
Method & Cool & Full & Hot \\ 
\hline
1 & 69  & 98 & 118\\
2 & 70  & 89 & 104 \\
3 & 98  & 123 & 143\\

\hline
\end{tabular}
\label{T-Four}
\end{table}

The velocities displayed in Table \ref{T-Four} show that on average the
velocities of the PDs increase from the cool emission to the hot emission and this is consistent between the three methods for calculating the velocities.  



\subsection{22 September 2011 (Non sunspot)}

\begin{figure}[!ht]
\centering
\scalebox{0.32}{\includegraphics{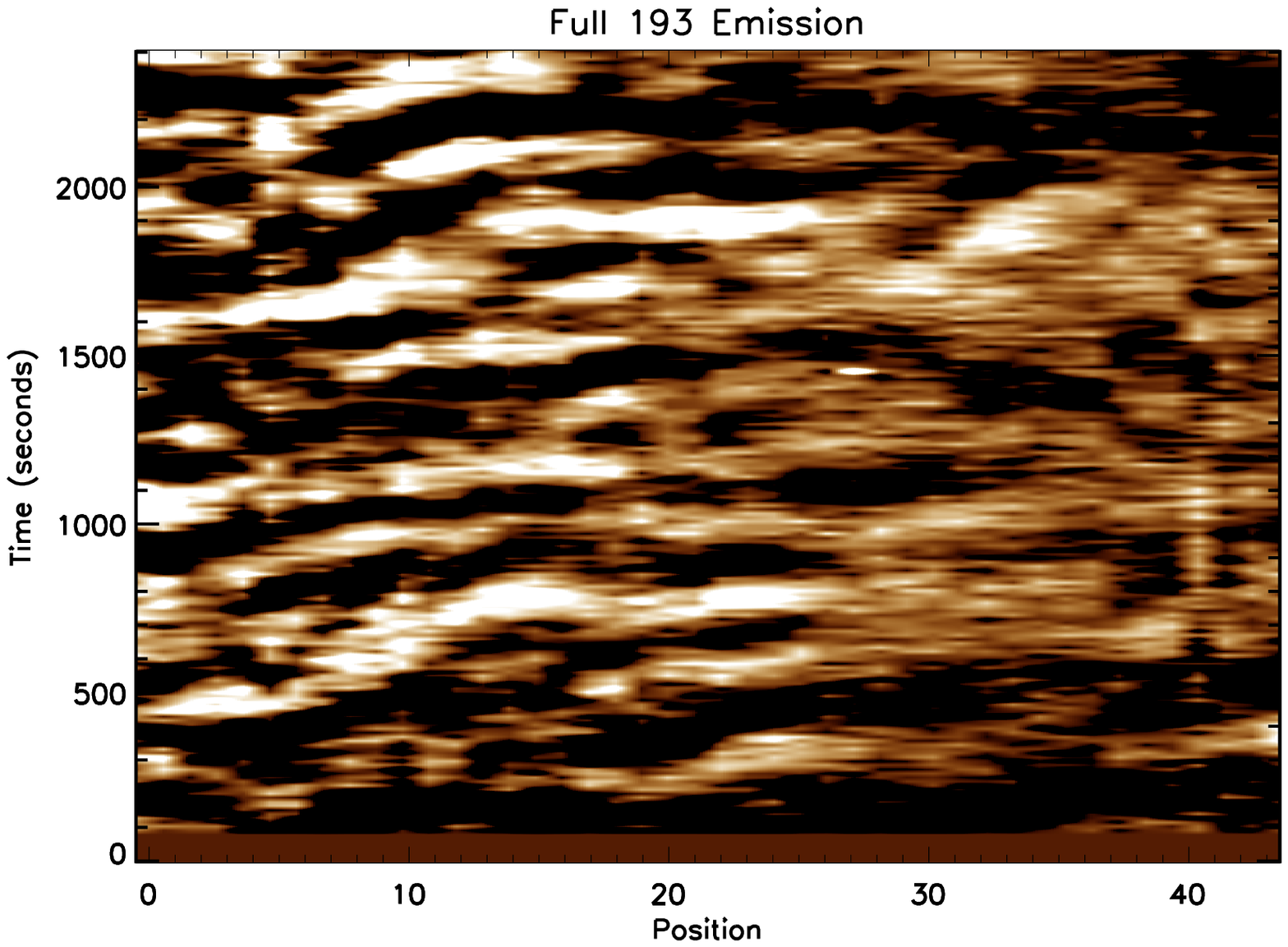}}
\scalebox{0.32}{\includegraphics{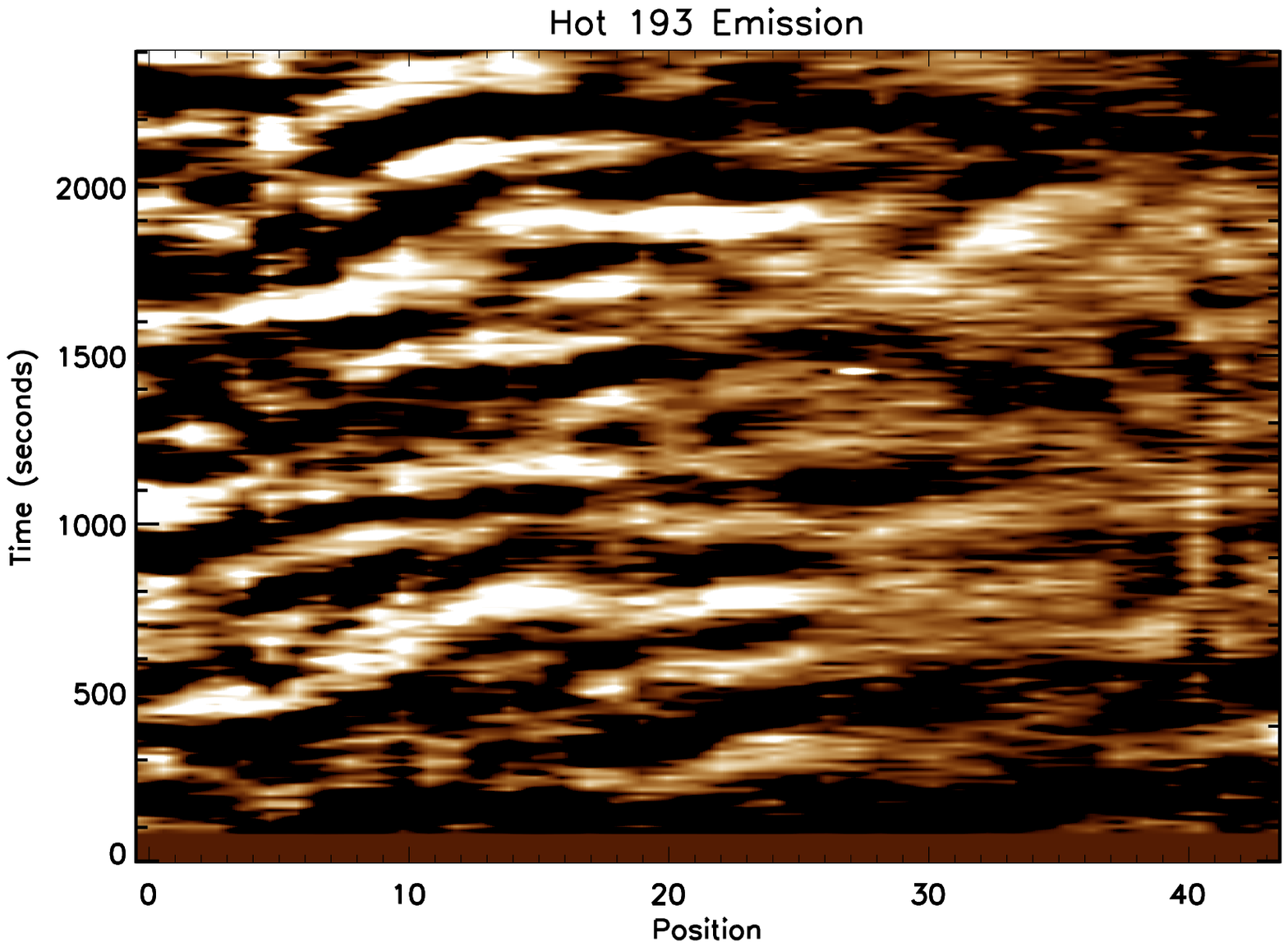}}
\caption{Running difference images for the loop outlined in Figure \ref{F-run_diff_22/09}.  The left shows the running-difference image associated with the full
193 emission.  The right hand plot shows the running-difference image associated
with the hot 193 contribution only.}
\label{F-compare2}
\end{figure}

We now carry out the same analysis on our second primary data set (22 September 2011). The cool contribution to the 193~\AA\ passband has been calculated in the same way as
for the previous example.  Running-difference images for this example are shown
in Figure \ref{F-compare2}.

For this example it is clear that the PDs associated with the hot emission have
almost identical properties as the PDs from the full emission.  They are exactly
in phase, propagate the same distance along the loop, and have the same period.  


In total we have carried out this analysis for seven loops. Four of these loops are
located at sunspots and three at non-sunspot locations.  PDs located at  non-sunspot locations appear to be identical in the hot component and in the full emission.  Also, the PDs seen in the cool emission are more similar
to those seen in the 171~\AA\ than the 131~\AA\ passband. This is the case for all the
non-sunspot examples studied.  PDs seen in the hot emission and located at
sunspots are not identical to the full emission PDs; they propagate a shorter
distance along the loop before they become unidentifiable and travel at a slightly greater
velocity.  PDs associated with the cool contribution at these locations are very
similar to those seen in the 131~\AA\ passband.  

This suggests that the PDs at sunspots are more likely to be slow
magneto-acoustic waves. Indeed, when we have removed the cool contribution from the 193 ~\AA\
line, the PDs have a slightly greater velocity that they have in the full
emission case.  Slow magneto-acoustic waves are expected to travel at the local sound
speed and hence their velocity should increase with temperature.  The main damping mechanism
of slow magneto-acoustic waves is thought to be thermal conduction \cite{demoortel04}, which is consistent with the fact that the PDs in the hot emission case appear to damp quicker than in cooler lines.  At non-sunspot locations the PDs
associated with the hot emission are identical to those in the full emission. 
Along with the lower-intensity bands seen in the cool emission we can conclude
that removing the cool contribution at non-sunspot regions has little to no
effect on the properties of the PDs.        

\section{Discussions and Conclusions}
     \label{S-Conclusions}

The aim of this paper was to undertake a detailed investigation of the propagation speed of observed  PDs to gain a greater insight into the temperature dependence of the PD properties. 

In Section \ref{S-Propspeed} we considered the velocities of PDs across the 131, 171, and 193~\AA\ passbands. We studied two main examples, one located at a sunspot (22 June 2011) and one above a non-sunspot (plage) region (22 September 2011). The velocities calculated for the sunspot example displayed a temperature dependence, where velocities increased when the PDs were propagating in hotter plasmas.  This velocity difference was found to be consistent with an interpretation in terms of slow magneto-acoustic  waves, especially when the effect of removing the cooler contribution from the 193~\AA\ emission is taken into account.  The velocities found  in the non-sunspot example  did not display a clear temperature dependence.  The  velocities found were approximately constant across the three wavelengths. These results were confirmed by recalculating the velocities from a further two methods.  This analysis was then used on a further 39 examples.  Our total sample of 41 cases included 13 sunspot and 28 non-sunspot locations.  The results suggest a strong relationship between whether or not the PDs are temperature dependent and whether the defined loops are located at a sunspot.  Of the 13 loops located at sunspots, 11 showed PD propagation speeds that are temperature dependent, and for the 28 not located at sunspots, 27 did not show a temperature dependence.   This suggests that PDs found at sunspots are far more likely to be temperature dependent and hence fit the slow-wave interpretation. For non-sunspot loops, the results are less clear. PDs located at these regions are less likely to be temperature dependent.  This fact makes the slow magneto-acoustic wave interpretation less likely, as a slow wave would adjust its velocity to the local sound speed. 

In Section \ref{S-Active} we studied how the properties of these PDs change across a set of active region loops and a set of sunspot loops.  It was found that the velocities of the PDs can change across an active region but the periods stay constant across the active region.  This trend was found for both the sunspot and non-sunspot examples.    

In the final section we considered the effect of removing the cool contribution in the 193~\AA\ passband on the properties of the PDs.  A rough method was devised to remove the contribution due to the cooler ions and this technique was used on seven examples.  We found  that the properties of the PDs do not change when the cool contribution is removed when considering loops at non-sunspot locations. Properties of PDs associated with loops located at sunspots did change when the cool contribution is removed.  The PDs seen in the hot emission damp more rapidly than the PDs seen in the full emission cases.  The velocities of the PDs in the hot emission are found to be slightly greater than those in the full emission.  This analysis further suggests that PDs  seen at sunspots agree with the wave interpretation, as this interpretation explains the increase of the PDs velocity and the rapid damping (as thermal conduction is more efficient at higher temperatures).  Removing the cool contribution had little or no effect on the PDs at non-sunspot areas.  Plasma is  hotter at non-sunspot regions, and it is not surprising that there is less of an effect due to the cool emission at these regions.

\begin{acks}
GK acknowledges the financial support of the STFC.  IDM acknowledges support of a Royal Society University Research Fellowship.  The authors would like to thank  H. Tian for useful comments on the article.
\end{acks}

\appendix

Correlation results for positions 3, 5, and 7 for the 22 June 2011 are displayed in Tables \ref{T-A1}, \ref{T-A2}, and \ref{T-A3}.

\begin{table}[!h]
\caption{Cross correlation between 171 arcs at position 3 along the loop defined on the 22 June 2011. The subscript denotes the lag where the maximum correlation is found.}
\begin{tabular}{c c c c c c c c c}
\hline
Arc  & 1 & 2 &  3 & 4 &  5 &  6 & 7 & 8 \\
[0.5ex]
\hline
 1 & 1 & $0.302_{-3}$ & $0.209_{10}$ & $0.415_6$ & $0.284_{3}$ & $0.242_{-10}$ &
$0.205_{-10}$ & $0.404_{10}$ \\
 2 &  & 1 & $0.831_1$ & $0.350_2$ & $0.324_5$ & $0.307_6$ & $0.321_{-9}$ &
$0.141_{-11}$ \\
 3 &  &  & 1 & $0.511_1$ & $0.429_3$ & $0.378_4$ & $0.343_{6}$ & $0.159_{8}$ \\
 4 &  &  & & 1 & $0.600_0$ & $0.536_-15$ & $0.554_3$ & $0.535_4$\\ 
 5 & &  & & & 1 & $0.891_0$ & $0.661_2$ & $0.478_4$\\
 6 & &  & & & & 1 & $0.746_1$ & $0.471_3$\\
 7 & &  & & & & & 1 & $0.683_1$\\
 8 & &  & & & & & & 1 \\ [1ex]
\hline
\end{tabular}
\label{T-A1}
\end{table}

\begin{table}[!h]
\caption{Cross correlation between 171~\AA\  arcs at position 5 along the loop defined on the 22 June 2011. The subscript denotes the lag where the maximum correlation is found. }
\begin{tabular}{c c c c c c c c c}
\hline
Arc  & 1 & 2 &  3 & 4 &  5 &  6 & 7 & 8 \\
[0.5ex]
\hline
 1 & 1 & $0.190_{9}$ & $0.165_{9}$ & $0.374_7$ & $0.209_{-10}$ & $0.207_{-10}$ &
$0.171_{10}$ & $0.394_{10}$ \\
 2 &  & 1 & $0.832_1$ & $0.444_2$ & $0.369_3$ & $0.364_4$ & $0.339_{6}$ &
$0.118_{-9}$ \\
 3 &  &  & 1 & $0.566_1$ & $0.481_3$ & $0.448_3$ & $0.411_{6}$ & $0.156_{9}$ \\
 4 &  &  & & 1 & $0.631_0$ & $0.525_-15$ & $0.535_3$ & $0.433_4$\\ 
 5 & &  & & & 1 & $0.913_0$ & $0.673_2$ & $0.435_5$\\
 6 & &  & & & & 1 & $0.717_2$ & $0.346_4$\\
 7 & &  & & & & & 1 & $0.575_1$\\
 8 & &  & & & & & & 1 \\ [1ex]
\hline
\end{tabular}
\label{T-A2}
\end{table}

\begin{table}[!h]
\caption{Cross correlation between 171 ~\AA\ arcs at position 7 along the loop defined on the 22 June 2011. The subscript denotes the lag where the maximum correlation is found. }
\begin{tabular}{c c c c c c c c c}
\hline
Arc  & 1 & 2 &  3 & 4 &  5 &  6 & 7 & 8 \\
[0.5ex]
\hline
 1 & 1 & $0.154_{-2}$ & $0.114_{10}$ & $0.143_6$ & $0.192_{3}$ & $0.152_{-11}$ &
$0.164_{-10}$ & $0.288_{10}$ \\
 2 &  & 1 & $0.849_1$ & $0.445_3$ & $0.412_4$ & $0.390_4$ & $0.363_{6}$ &
$0.157_{10}$ \\
 3 &  &  & 1 & $0.554_2$ & $0.521_3$ & $0.492_3$ & $0.394_{6}$ & $0.186_{6}$ \\
 4 &  &  & & 1 & $0.488_0$ & $0.416_1$ & $0.444_3$ & $0.369_4$\\ 
 5 & &  & & & 1 & $0.889_0$ & $0.662_2$ & $0.481_5$\\
 6 & &  & & & & 1 & $0.719_2$ & $0.466_4$\\
 7 & &  & & & & & 1 & $0.547_1$\\
 8 & &  & & & & & & 1 \\ [1ex]
\hline
\end{tabular}
\label{T-A3}
\end{table}

Correlation results for positions 3, 5, and 7 for the 22 September 2011 are displayed in Tables \ref{T-A4}, \ref{T-A5}, and \ref{T-A6}.

\begin{table}[!h]
\caption{Cross correlation between 171 ~\AA\ arcs at position 3 along the loop defined on the 22 September 2011. The subscript denotes the lag where the maximum correlation is found. }
\begin{tabular}{c c c c c c c c c}
\hline
Arc  & 1 & 2 &  3 & 4 &  5 &  6 & 7 & 8 \\
[0.5ex]
\hline
 1 & 1 & $0.467_{4}$ & $0.276_{-2}$ & $0.406_{-15}$ & $0.389_{-2}$ & $0.311_{-3}$ &
$0.114_{-9}$ & $0.316_{-4}$ \\
 2 &  & 1 & $0.378_{-6}$ & $0.330_0$ & $0.480_{-4}$ & $0.500_{-8}$ & $0.132_{-10}$ &
$0.365_{14}$ \\
 3 &  &  & 1 & $0.290_{-15}$ & $0.248_{5}$ & $0.199_{-1}$ & $0.209_{9}$ & $0.198_{-3}$ \\
 4 &  &  & & 1 & $0.243_{-2}$ & $0.405_{-9}$ & $0.226_7$ & $0.222_9$\\ 
 5 & &  & & & 1 & $0.351_{-2}$ & $0.111_0$ & $0.267_{14}$\\
 6 & &  & & & & 1 & $0.320_1$ & $0.204_{-11}$\\
 7 & &  & & & & & 1 & $0.403_0$\\
 8 & &  & & & & & & 1 \\ [1ex]
\hline
\end{tabular}
\label{T-A4}
\end{table}

\begin{table}[!h]
\caption{Cross correlation between 171 ~\AA\ arcs at position 5 along the loop defined on the 22 September 2011. The subscript denotes the lag where the maximum correlation is found. }
\begin{tabular}{c c c c c c c c c}
\hline
Arc  & 1 & 2 &  3 & 4 &  5 &  6 & 7 & 8 \\
[0.5ex]
\hline
 1 & 1 & $0.476_{2}$ & $0.451_{-2}$ & $0.194_{4}$ & $0.524_{0}$ & $0.521_{0}$ &
$0.196_{2}$ & $0.074_{6}$ \\
 2 &  & 1 & $0.484_{-6}$ & $0.169_{-3}$ & $0.445_{0}$ & $0.177_{-2}$ & $0.273_{3}$ &
$0.127_{4}$ \\
 3 &  &  & 1 & $0.218_{4}$ & $0.157_{4}$ & $0.266_{3}$ & $0.211_{7}$ & $0.083_{-15}$ \\
 4 &  &  & & 1 & $0.163_{14}$ & $0.194_{-14}$ & $0.248_{12}$ & $0.217_{12}$\\ 
 5 & &  & & & 1 & $0.428_{0}$ & $0.356_2$ & $0.210_{-14}$\\
 6 & &  & & & & 1 & $0.310_1$ & $0.230_{-15}$\\
 7 & &  & & & & & 1 & $0.428_0$\\
 8 & &  & & & & & & 1 \\ [1ex]
\hline
\end{tabular}
\label{T-A5}
\end{table}

\begin{table}[!h]
\caption{Cross correlation between 171 ~\AA\ arcs at position 7 along the loop defined on the 22 September 2011. The subscript denotes the lag where the maximum correlation is found. }
\begin{tabular}{c c c c c c c c c}
\hline
Arc  & 1 & 2 &  3 & 4 &  5 &  6 & 7 & 8 \\
[0.5ex]
\hline
 1 & 1 & $0.508_{0}$ & $0.499_{-3}$ & $0.130_{1}$ & $0.474_{0}$ & $0.353_{0}$ &
$0.244_{-13}$ & $0.212_{8}$ \\
 2 &  & 1 & $0.347_{-2}$ & $0.371_{-15}$ & $0.221_{2}$ & $0.346_{2}$ & $0.239_{2}$ &
$0.177_{14}$ \\
 3 &  &  & 1 & $0.224_{5}$ & $0.187_{0}$ & $0.400_{0}$ & $0.122_{2}$ & $0.073_{-11}$ \\
 4 &  &  & & 1 & $0.209_{10}$ & $0.194_{-1}$ & $0.066_{11}$ & $0.278_{-10}$\\ 
 5 & &  & & & 1 & $0.489_{-1}$ & $0.313_{-13}$ & $0.182_{-15}$\\
 6 & &  & & & & 1 & $0.171_{-15}$ & $0.262_{-14}$\\
 7 & &  & & & & & 1 & $0.275_1$\\
 8 & &  & & & & & & 1 \\ [1ex]
\hline
\end{tabular}
\label{T-A6}
\end{table}


 
\newpage

\bibliographystyle{spr-mp-sola}




\end{article}

\end{document}